\documentclass[fleqn,usenatbib]{mnras}
\usepackage{newtxtext,newtxmath,natbib}
\usepackage[T1]{fontenc}
\usepackage{ae,aecompl}

\usepackage{eso-pic}

\AddToShipoutPictureBG*{%
  \AtPageUpperLeft{%
    \hspace{0.75\paperwidth}%
    \raisebox{-3.5\baselineskip}{%
      \makebox[0pt][l]{\textnormal{DES 2016-0187}}
}}}%

\AddToShipoutPictureBG*{%
  \AtPageUpperLeft{%
    \hspace{0.75\paperwidth}%
    \raisebox{-4.5\baselineskip}{%
      \makebox[0pt][l]{\textnormal{FERMILAB-PUB-17-025-AE}}
}}}%

\usepackage{graphicx}	
\usepackage{amsmath}	
\usepackage{amssymb}	

\newcommand*{\expectation}[1]{\ensuremath{\left\langle #1 \right\rangle}}
\newcommand*{\logTen}{\ensuremath{\log_{10}}}
\newcommand*{\Msun}{\ensuremath{\, M_{\odot}}}

\newcommand*{\unit}[1]{\ensuremath{\mathrm{\, #1}}}
\newcommand*{\erg}{\unit{erg}}

\newcommand*{\second}{\unit{s}}

\newcommand*{\LamMCMF}{$\lambda_{\mathrm{MCMF}}$}
\newcommand*{\LamRM}{$\lambda_{\mathrm{RM}}$}
\newcommand*{\ProbLam}{$P_\lambda$}
\newcommand*{\ProbS}{$P_\mathrm{s}$}
\newcommand*{\ProbCut}{$P_{\mathrm{cut}}$}
\newcommand*{\ProbStar}{$P_{*}$}

\DeclareMathOperator\arctanh{arctanh}

\title[MCMF Applied to RASS and DES-SV]{A Multi-component Matched Filter Cluster Confirmation Tool for eROSITA:  Initial Application to the RASS and DES-SV Datasets}

\author[Klein et al.]{M.~Klein$^{1,2}$\thanks{E-mail: mklein@usm.uni-muenchen.de},
J.~J.~Mohr$^{1,2,3}$,
S.~Desai$^{4}$,
H.~Israel$^{1}$,
\newauthor
S.~Allam$^{5}$,
A.~Benoit-L{\'e}vy$^{6,7,8}$,
D.~Brooks$^{7}$,
E.~Buckley-Geer$^{5}$,
A. Carnero Rosell$^{9,10}$,
\newauthor
M.~Carrasco~Kind$^{11,12}$,
C.~E.~Cunha$^{13}$,
L.~N.~da Costa$^{9,10}$,
J.~P.~Dietrich$^{1,3}$,
T.~F.~Eifler$^{14}$,
\newauthor
A.~E.~Evrard$^{15,16}$,
J.~Frieman$^{5,17}$,
D.~Gruen$^{13,18}$,
R.~A.~Gruendl$^{11,12}$,
G.~Gutierrez$^{5}$,
\newauthor
K.~Honscheid$^{19,20}$,
D.~J.~James$^{21,22}$,
K.~Kuehn$^{23}$,
M.~Lima$^{24,9}$,
M.~A.~G.~Maia$^{9,10}$,
\newauthor
M.~March$^{25}$,
P.~Melchior$^{26}$,
F.~Menanteau$^{11,12}$,
R.~Miquel$^{27,28}$,
A.~A.~Plazas$^{14}$,
K.~Reil$^{18}$,
\newauthor
A.~K.~Romer$^{29}$,
E.~Sanchez$^{30}$,
B.~Santiago$^{31,9}$,
V.~Scarpine$^{5}$,
M.~Schubnell$^{16}$,
\newauthor
I.~Sevilla-Noarbe$^{30}$,
M.~Smith$^{32}$,
M.~Soares-Santos$^{5}$,
F.~Sobreira$^{9,33}$,
E.~Suchyta$^{34}$,
\newauthor
M.~E.~C.~Swanson$^{12}$,
G.~Tarle$^{16}$
and the DES Collaboration
\\
}

\date{Accepted XXX. Received YYY; in original form ZZZ}
\pubyear{2016}
\begin{document}
\label{firstpage}
\pagerange{\pageref{firstpage}--\pageref{lastpage}}
\maketitle


\begin{abstract}

{\small We describe a multi-component matched filter cluster confirmation tool (MCMF) designed for the study of large X-ray source catalogs produced by the upcoming X-ray all-sky survey mission eROSITA.  We apply the method to confirm a sample of 88 clusters with redshifts $0.05<z<0.8$ in the recently published 2RXS catalog from the ROSAT all-sky survey (RASS) over the 208\,deg$^2$ region overlapped by the Dark Energy Survey (DES) science verification (DES-SV) dataset.  In our pilot study, we examine all X-ray sources, regardless of their extent. 
Our method employs a multi-color red sequence (RS) algorithm that incorporates the X-ray count rate and peak position in determining the region of interest for followup and extracts the positionally and color weighted optical richness \LamMCMF\ as a function of redshift for each source.  Peaks in the \LamMCMF--redshift distribution are identified and used to extract photometric redshifts, richness and uncertainties. The significances of all optical counterparts are characterized using the distribution of richnesses defined along random lines of sight. These significances are used to extract cluster catalogs and to estimate the contamination by random superpositions of unassociated optical systems. The delivered photometric redshift accuracy is $\delta z / (1+z)=0.010$.  We find a well defined X-ray luminosity--\LamMCMF\ relation with an intrinsic scatter of $\delta \ln(\lambda_\mathrm{MCMF}| L_\mathrm{x})=0.21$.  Matching our catalog with the DES-SV redMaPPer catalog yields good agreement in redshift and richness estimates; comparing our catalog with the South Pole Telescope (SPT) selected clusters shows no inconsistencies. SPT clusters in our dataset are consistent with the high mass extension of the RASS based \LamMCMF--mass relation.}

\end{abstract}



\section{Introduction}

The abundance of galaxy clusters over a range of mass and redshift is a powerful probe of the amplitude of mass fluctuations in the Universe \citep{white93b} that is well suited for studies of the cosmic acceleration \citep{wang98,haiman01,vikhlinin09,mantz10a,dehaan16} and consistency tests of the $\Lambda$CDM and wCDM paradigms \citep{rapetti10,bocquet15}.  In addition, galaxy clusters provide the tightest constraints on the dark matter self interaction cross section to date \citep{Sartoris14,robertson17}, and offer many other insights into plasma physics and galaxy evolution. To use clusters for these various purposes, the clusters have to be first identified and their redshifts have to be measured. 

One common challenge for most recent and upcoming cluster surveys is the large number of cluster candidates. One of these surveys is the eROSITA \citep{predehl10} all sky X-ray survey, which will enable the selection of clusters using the X-ray emission from the hot plasma that fills the gravitational potential wells. The expected number of cluster candidates is $\sim10^5$ \citep{merloni12}.  Only a fraction of these candidates will have enough detected X-ray photons to enable a reliable redshift 
from X-rays alone. The majority, especially at high redshift and low mass, will require optical 
photometric redshift estimation.  The Dark Energy Survey \citep[][DES]{DESC16} dataset will provide this information for the majority of the overlapping eROSITA cluster candidates.  Given the size of the sample and the need to precisely understand the selection for a cosmological analysis of the clusters, an automated and objective follow-up method is therefore needed.

In this analysis we present a matched filter method for finding the optical counterparts of an X-ray detected cluster sample.  The method leverages the simple time evolution and clustering of the passive galaxy population in clusters.  These so-called red sequence (RS) galaxies have been used previously to create large samples of clusters and groups for studies of cosmology \citep{gladders00,gladders07,rozo10,rykoff14}.  Our method adopts the prior positional and flux information from the candidate selection in the X-ray when searching for a candidate.  It allows for multiple optical counterparts for each candidate cluster, ranking them by the strength of their signal, given the prior X-ray information.  Finally, we quantify for each counterpart the probability that it is a random superposition of a physically unassociated structure along the line of sight toward an X-ray cluster candidate.  This information can be incorporated in cosmological analyses of the sample.

\begin{figure}
\vskip0.05in
\includegraphics[width=0.95\linewidth]{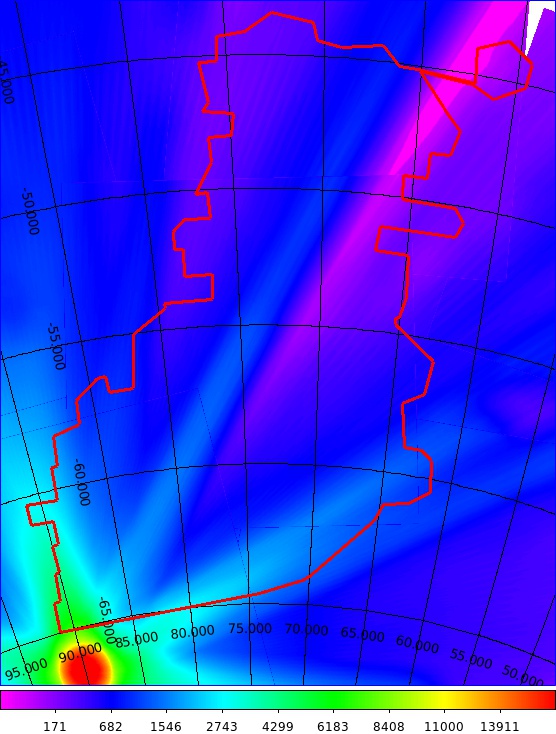}
\includegraphics[width=0.95\linewidth]{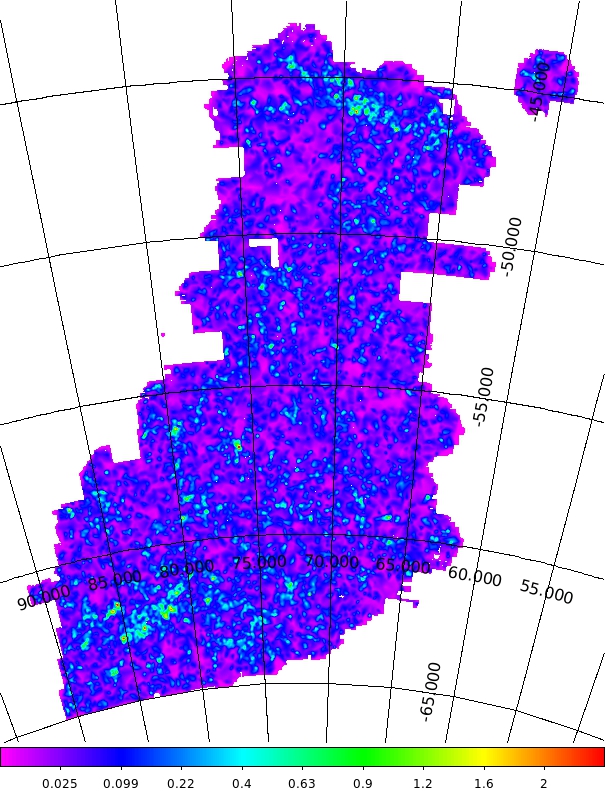}
\vskip-0.05in
\caption{Illustration of the DES-SV footprint. Top panel: DES-SV footprint overlaid on RASS exposure time map. The average exposure time increases from north-west to south-east. The southern ecliptic pole is visible as high exposure time region southern of the SV area. Color coded is the exposure time in seconds. Bottom panel: Galaxy density map of galaxies with colors consistent with the red sequence at z=0.12.}
\label{fig:footprint}
\vskip-0.15in
\end{figure}

The ROSAT All-Sky Survey \citep[RASS,][]{truemper82} is currently the best available all-sky X-ray survey. We therefore adopt faint source catalogs extracted from RASS to perform a systematic test of our algorithm.  To do so we focus on a contiguous area of the DES Science Verification dataset (DES-SV) that overlaps the eastern side of the South Pole Telescope (SPT) survey area.  This area was observed during the first commissioning and verification campaign with DECam \citep{flaugher15}. While the DES-SV data are somewhat shallower than the expected DES final depth, the dataset provides the best representation of the final dataset currently available. The DES-SV footprint is shown in Fig.~\ref{fig:footprint}.

In this paper we describe the data (section~\ref{sec:data}) and method (section~\ref{sec:method}) and its current performance (section~\ref{sec:application}) in verifying clusters and measuring their redshifts as part of an automated X-ray follow up. We examine the impact of contaminating stars (section~\ref{sec:stellarcontamination}) and characterize the cluster sample we extract from the RASS+DES-SV analysis (section~\ref{sec:sample}).  In section~\ref{sec:conclusions} we present our conclusions and comment on the expected number of clusters when applying our method to the full DES area.  Throughout this paper we adopt a flat $\Lambda$CDM cosmology with $\Omega_M=0.3$ and $H_0=70$\,km\,s$^{-1}$\,Mpc$^{-1}$.


%

\section{Data}
\label{sec:data}

The data reduction and catalog creation follow closely that described in \citet{hennig16}. We therefore provide only a brief summary of the methods and data quality here.

\subsection{Observations and Data Reduction}

The DES-SV data were acquired between November 1,  2012 and February 2013. The data reduction is performed using our Cosmology Data Management system (CosmoDM), which is an improved version of the system used to process the Blanco Cosmology Survey Data \citep{desai12} and is a prototype for the development of a data management system being prepared within the Euclid collaboration \citep{laureijs11}.  The data are first processed on a night by night basis, where cross-talk, bias and flat corrections are applied. Each CCD is considered separately; by correcting for varying pixel scale and the resulting persistent variation in sensitivity within each band, we reduce the positional variation of the zero point within each detector.  We measure the persistent variation in sensitivity using star flats extracted using a subset of photometric exposures over the DES-SV period \citep{regnault09,schlafly12,hennig16}.

A first astrometric calibration is performed on the individual exposures using SCAMP \citep{bertin06} with 2MASS \citep{skrutskie06} as the absolute reference catalog. We employ a high quality distortion map that is derived from a large set of overlapping exposures.  The residual scatter between our data and 2MASS is about 200\,mas and is dominated by the positional accuracy of the reference catalog.  Within our system a second astrometric calibration can be performed prior to the coaddition process using all overlapping exposures of a given tile. This second pass internal calibration results in an internal scatter of about 20\,mas (less than a tenth of the pixel scale) around the best astrometric solution.  For the photometry presented here this second correction was not applied, because our tests indicate no significant differences in the galaxy photometry with and without the correction.

For the SPT-East field of the DES-SV dataset  we create coadds with tilings of size 62$'\times$62$'$ on a grid with center offsets of 1$^\circ$.  Neighboring tiles overlap each other by about 2$'$.  The relative calibration of all single epoch images contributing to a tile is carried out for each tile independently, and the repeatability of the resulting calibrated single epoch stellar photometry is excellent \citep[see][]{hennig16}.  

The coadds are created from PSF homogenized single epoch images where the target PSF is described by a Moffat function \citep{moffat72} with FWHM tuned to be the median of all contributing images within a single band.  The coadds we employ for this analysis are median-combined to remove all artifacts.  Following our previous work, the absolute photometric calibration is obtained using the stellar locus, where the absolute zero point comes from the 2MASS $J$ band \citep{desai12,song12b,liu15}.
We use an enhanced version of SExtractor \citep{bertin96} in dual image mode to carry out PSF-corrected model fitting photometric analyses of these images.  The detection image is a coadd of the $i$ and $z$ band images for the tile.  Following this approach we produce an ensemble of several hundred multi-band catalogs over the SPT-East region.  

\begin{figure}
\vskip0.05in
\includegraphics[width=\linewidth]{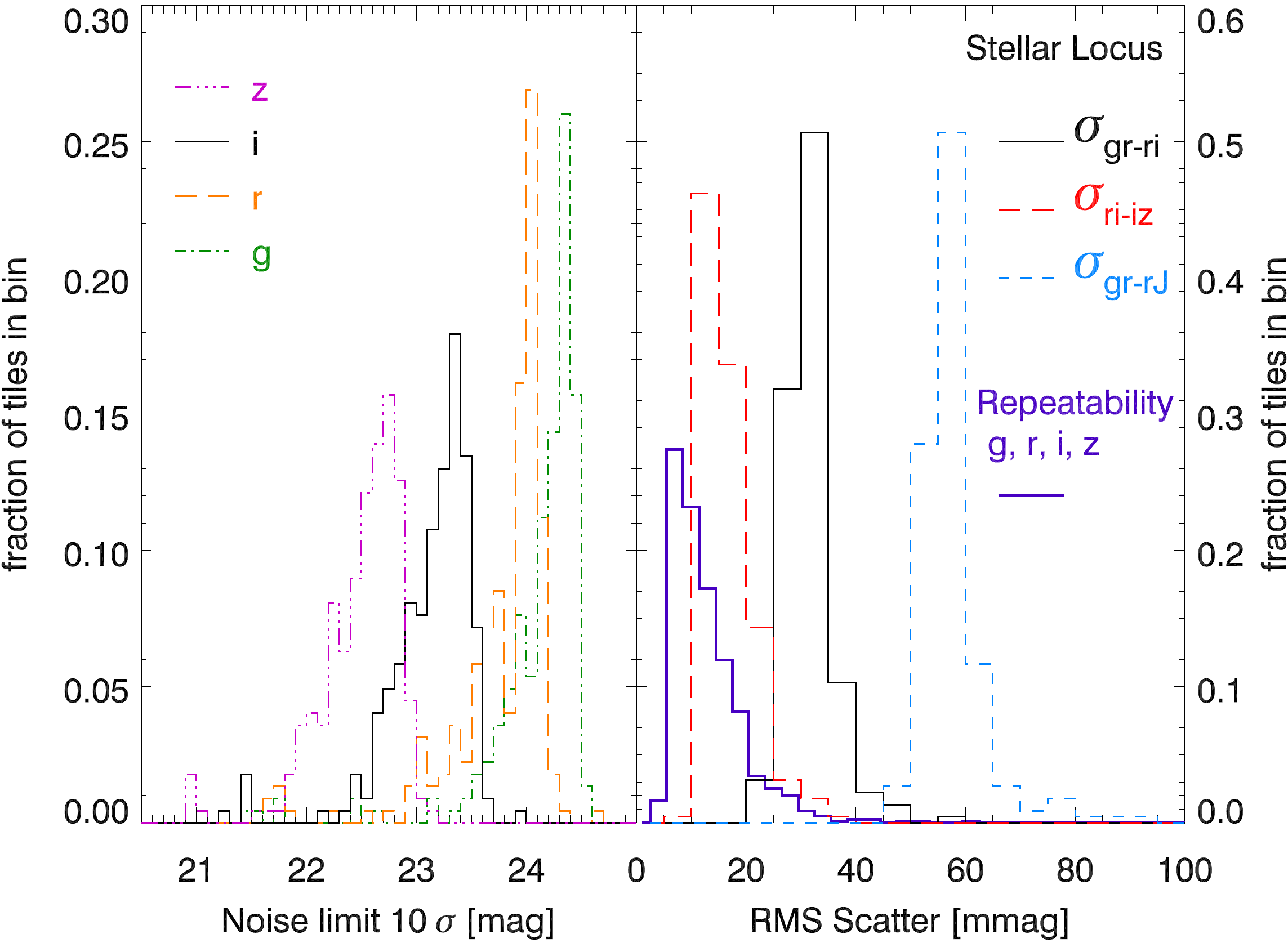}
\vskip-0.05in
\caption{Data quality of the SPT-East region. Left panel: Distribution of $10\sigma$ limiting magnitudes of the individual tiles for the $griz$ bands. Right panel: Distribution of the stellar locus and repeatability scatter.}
\label{fig:sptedq}
\vskip-0.15in
\end{figure}

In the left panel of Fig.~\ref{fig:sptedq}, we show the distributions of the
$10\sigma$ limiting magnitudes measured using the sky noise estimated within 2$''$ diameter apertures in the coadd images.  The $k\sigma$ noise limiting magnitude is defined as
\begin{equation}
m_{\mathrm{lim,k\sigma}} = Z - 2.5\log_{10}{kS_{\mathrm{sky}}}
\end{equation}
where $Z$ is the zeropoint of the stack and $S_{\mathrm{sky}}$ the median
absolute deviation of the measured sky background, which is measured from $1000$ 
randomly positioned apertures that exclude object flux. We find the median $10\sigma$ limiting magnitudes to be
$24.26$, $23.97$, $23.22$ and $22.59$ for the $griz$ bands
of the $223$ SPT-East tiles for which such a measurement could be performed.
We emphasize that because typically only about 60\,percent of the DES-SV exposures meet the DES imaging requirements \citep[see, e.g.][]{hennig16}, that these limiting magnitudes are expected to be $\sim0.3$\,mag shallower than the final DES survey depth.   

In the right panel of Fig.~\ref{fig:sptedq}, the red long-dashed, black solid,
and blue short-dashed histograms denote the measured RMS scatter around the
stellar loci of the same sample of SPT-East coadds.
We show the dispersion in three color-color spaces of objects identified as stars,
obtained from successive fits of the $g\!-\!i$ vs.\ $r\!-\!i$, 
$r\!-\!i$ vs.\ $i\!-\!z$, and $g\!-\!r$ vs.\ $r\!-\!J$ colors.
The 2MASS $J$ band is added to the fit to calibrate our measurements to
an external photometric standard. 
For details of the methods, we refer the reader elsewhere \citep{desai12,hennig16}.
We measure the orthogonal scatter around the stellar loci to be
$\sigma_{\mathrm{gr-ri}}\!=\!15\,\text{mmag}$,
$\sigma_{\mathrm{ri-iz}}\!=\!31\,\text{mmag}$, and
$\sigma_{\mathrm{gr-rJ}}\!=\!56\,\text{mmag}$, in very close agreement to
the $17$, $32$, amd $57\,\text{mmag}$ reported by \citet{hennig16}.  This performance is significantly better than that achieved within the SDSS survey \citep{desai12}.

Overplotted in the right panel of Fig.~\ref{fig:sptedq} as a solid, thick blue
line, we present the RMS \emph{repeatability scatter} distributions of the
SPT-East tiles. This estimator compares measurements of the same source identified
in the different overlapping single epoch images that contribute to the
coadd. Hence it measures the overall photometric stability of the data
set, given the underlying instrumental and atmospheric conditions and the robustness of the calibration procedure.
Following \citet{desai12}, we determine the repeatability scatter,
which trivially increases towards fainter sources, in magnitude bins, from the brightest photometric bin only, where the photon statistics make a vanishingly small contribution to the scatter.  This bin corresponds on average to a magnitude of 15.08 and only includes
those tiles (between $185$ for the $z$-band to $205$ for the $r$-band) which
deviate less than $1\sigma$ ($1.08\,\text{mag}$) from this brightest-bin
magnitude. We find median repeatability scatter of 
$9.6$, $13.7$, $13.8$, and $11.7\,\text{mmag}$ in the $griz$ bands,
with an overall median of $11.7\,\text{mmag}$. 
While this repeatability scatter is larger than the $\sim\!8\,\text{mmag}$
measured by \citet{hennig16} in DES observations of SPT cluster
fields, reflecting the more variable observing depths and conditions, 
our repeatability scatter is significantly smaller compared to the
earlier Blanco Cosmology Survey \citep{desai12} and  Pan-STARRS1 \citep{liu15} surveys.

\subsection{Galaxy Catalog}

In total we use 225 tiles in the SPT-East region and in total have 208.7\,deg$^2$ of unmasked area. We combine all individual tile catalogs and exclude multiple entries in the overlapping regions. Similarly to \citet{hennig16}, we separate galaxies from stars on the bright end by excluding all objects with {\tt spread\_model} less than 0.0025.  This cut allows a clean separation between galaxies and stars down to i$\sim$22\,mag.  Below this magnitude we attempt no stellar exclusion using object morphology.  
In this work we adopt a more conservative limit in $i$ band that is derived for each tile by measuring where the catalog number density per magnitude starts to continuously fall below a best fit power law that was fitted to the region 19$\le$i$\le$21.  Characteristically, this limit lies between i=22.2 and 22.7. 

We note that this rather simplistic approach yields conservative limits and can be improved in future work that will involve larger areas or deeper X-ray data.  We do not expect that the confirmation of RASS selected clusters will be limited by the depth of our DES optical data.  However, we will return to this issue at a later point of this work. 
 
\subsection{The Second ROSAT All-Sky Survey Source Catalog}
The second ROSAT All-Sky Survey source catalog \citep[][2RXS]{boller16} is based on the RASS-3 processed photon event files and uses an improved background determination and detection algorithm, compared to the predecessors: the ROSAT bright source catalog \citep{voges99} and the ROSAT faint source catalog \citep{voges00}. The final source catalog contains about 135000 detections down to the same detection likelihood as the 1RXS faint source catalog but without the additional lower limit on source photon counts.

Further, \citet{boller16} states that this catalog is the largest and most reliable all-sky X-ray catalog available and will likely remain so until the first eROSITA catalogs are created. Thus, the 2RXS catalog is a well suited source for testing the optical follow up of eROSITA cluster candidates and creating cluster catalogs for cosmological studies. Even as the most reliable all-sky X-ray catalog, \citet{boller16} estimates using simulations that the 2RXS contains about 30\,percent spurious detections over the full sky. The fraction of spurious detection show some dependency on the exposure time and a stronger dependency on likelihood threshold \citep{boller16}.

Table~1 in \citet{boller16} lists the spurious fraction in bins of existence likelihood for the average survey exposure time and for exposure times above 4,000 seconds. We use this table to assign a probability to be real to each 2RXS source.  We find a slightly lower spurious fraction of 22\,percent for the SPT-East region we study in this work. This difference is due to the proximity of the south ecliptic pole and the associated increase in exposure time in RASS, as can be seen in Fig.~\ref{fig:footprint}.  Although we assume the fraction of spurious sources is strongly suppressed by our optical confirmation tool, we cannot exclude chance superpositions of a spurious X-ray detection with a real optical counterpart.  

The 2RXS catalog further offers measurements on source extent, source variability and hardness ratio. Due to the large RASS survey PSF with FWHM of  $\sim4$ arcminutes \citep{boese00}, only a few clusters are reliably estimated as extended. Further, the typically low number of source counts of the majority of the sources do not allow precise measurements of hardness ratios and source variability for those sources. Only 0.4\% of the 2RXS sources in our footprint show source variabilities above three sigma. 
Due to the lack of usefulness of those selection parameters we decided not to use them and consequently investigate all 2RXS sources for the possibility of being a cluster. From previous RASS based studies \citep{henry06,Ebeling13} and the estimated spurious fraction, we expect only of the order $\sim 10\%$ of the 2RXS sources to be galaxy clusters. This leads to the fact that cluster confirmation becomes an important factor in this work.

We note that for a potential eROSITA based survey, the source extent can be used to produce very pure cluster catalogs up to high redshifts prior to the optical follow up, partially reducing the requirements in the optical data to provide photometric redshifts only.


%
\begin{figure*}
\begin{minipage}[hbt]{0.33\linewidth}
\centering
\includegraphics[width=0.980\linewidth]{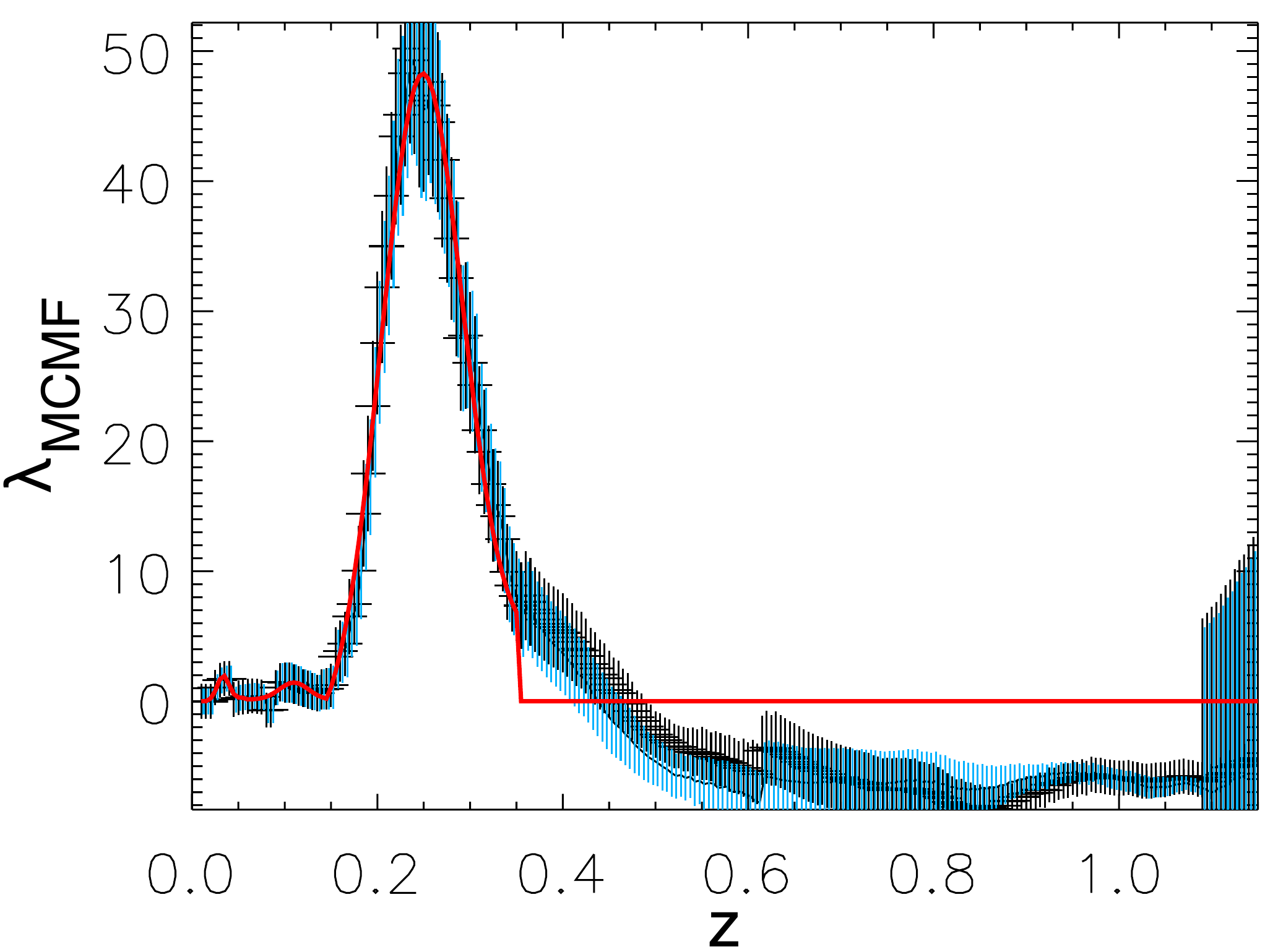}
\includegraphics[keepaspectratio=true,width=1.0\linewidth]{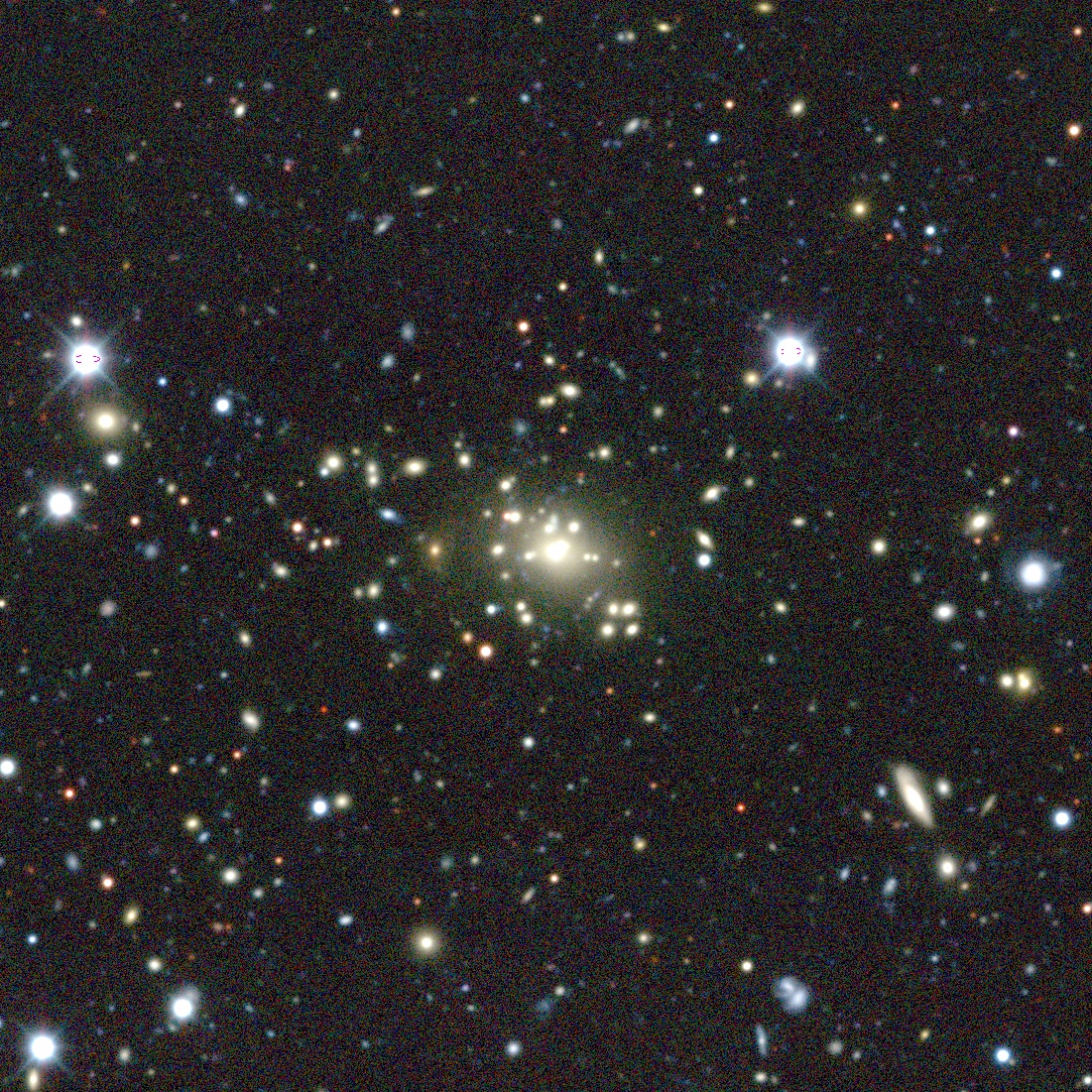}
\textrm{a)}
\end{minipage}
\begin{minipage}[hbt]{0.33\linewidth}
\centering
\includegraphics[width=0.980\linewidth]{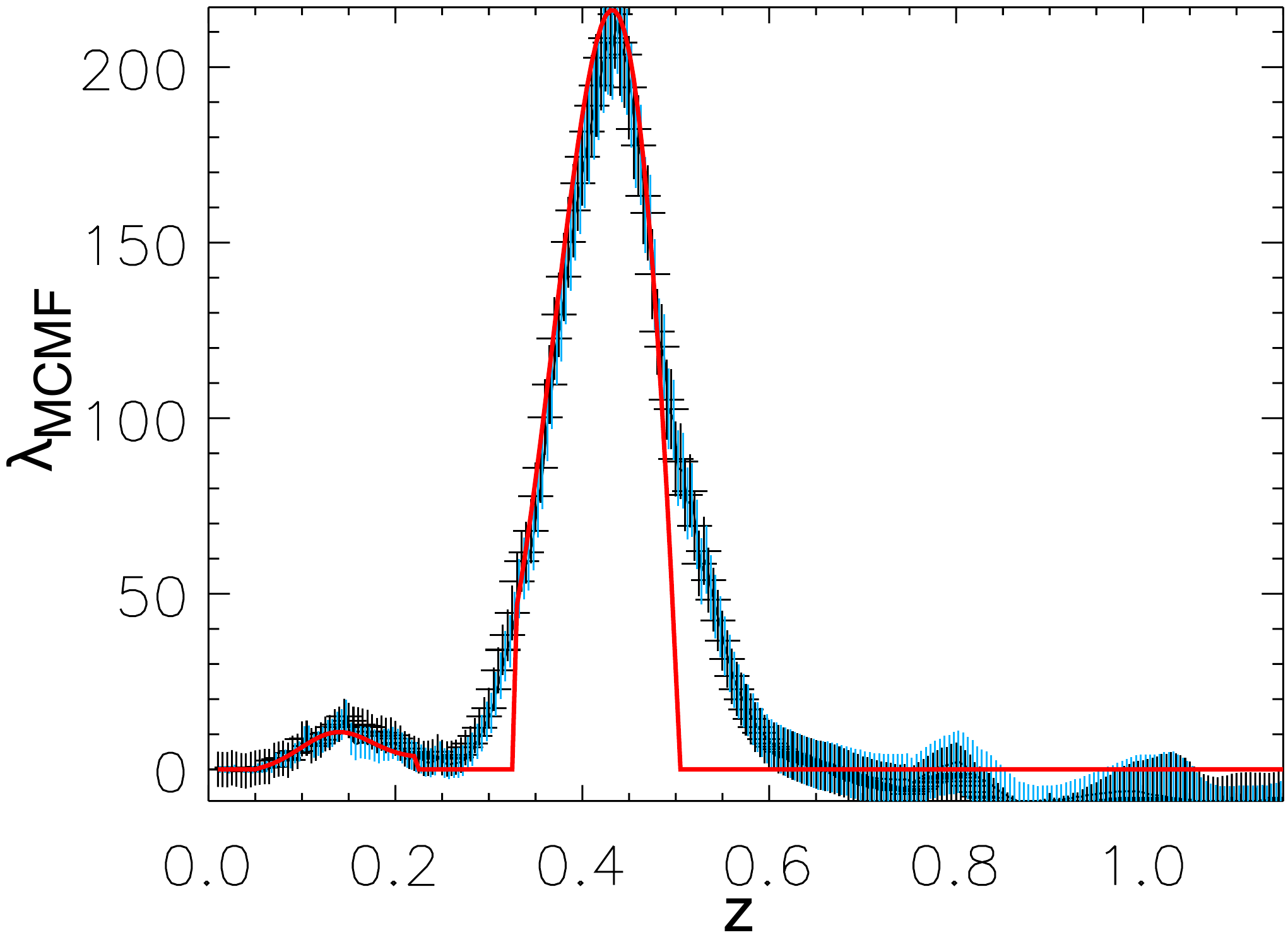}
\includegraphics[keepaspectratio=true,width=1.0\linewidth]{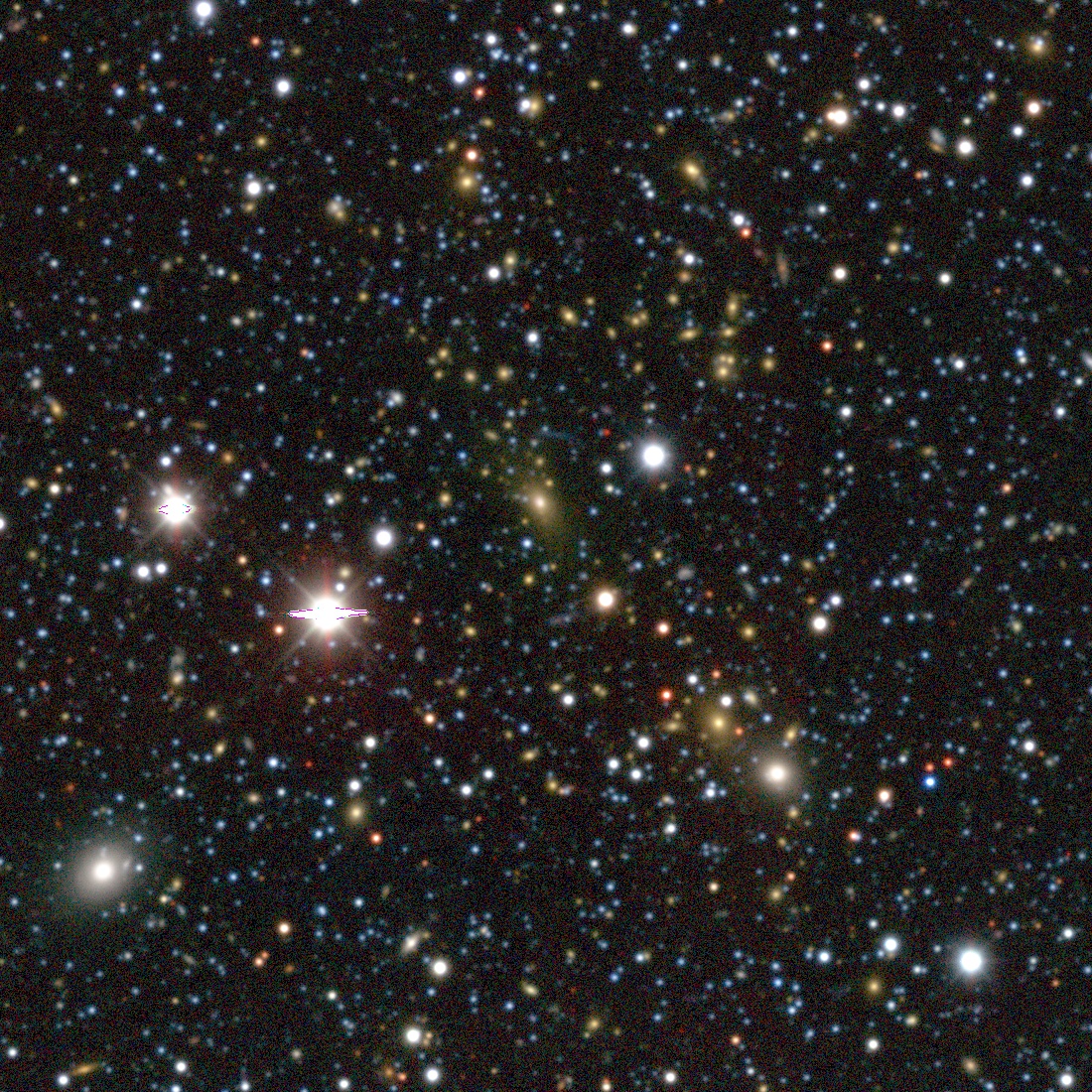} 
\textrm{b)}
\end{minipage}
\begin{minipage}[hbt]{0.33\linewidth}
\centering
\includegraphics[width=0.980\linewidth]{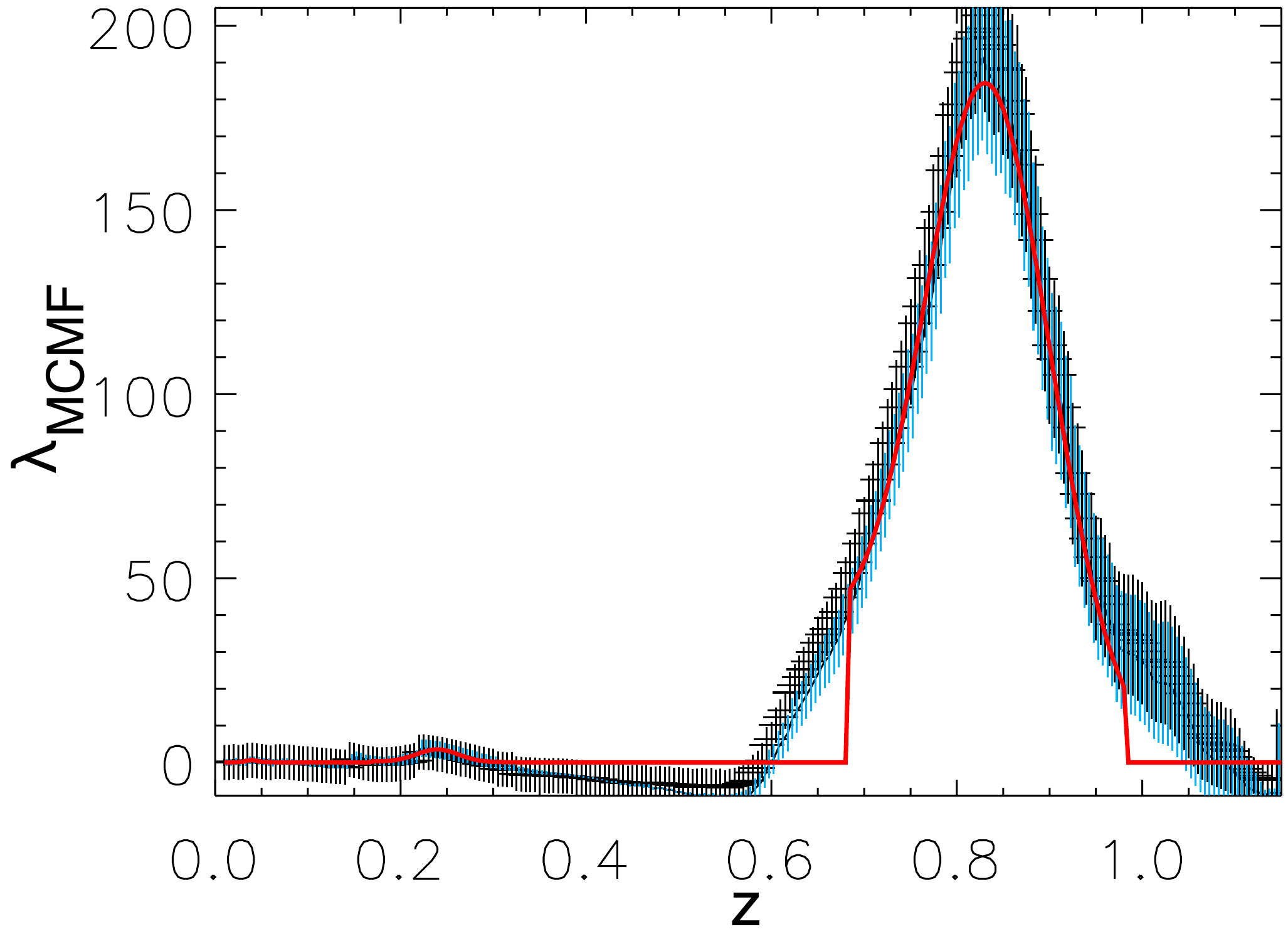}
\includegraphics[keepaspectratio=true,width=1.0\linewidth]{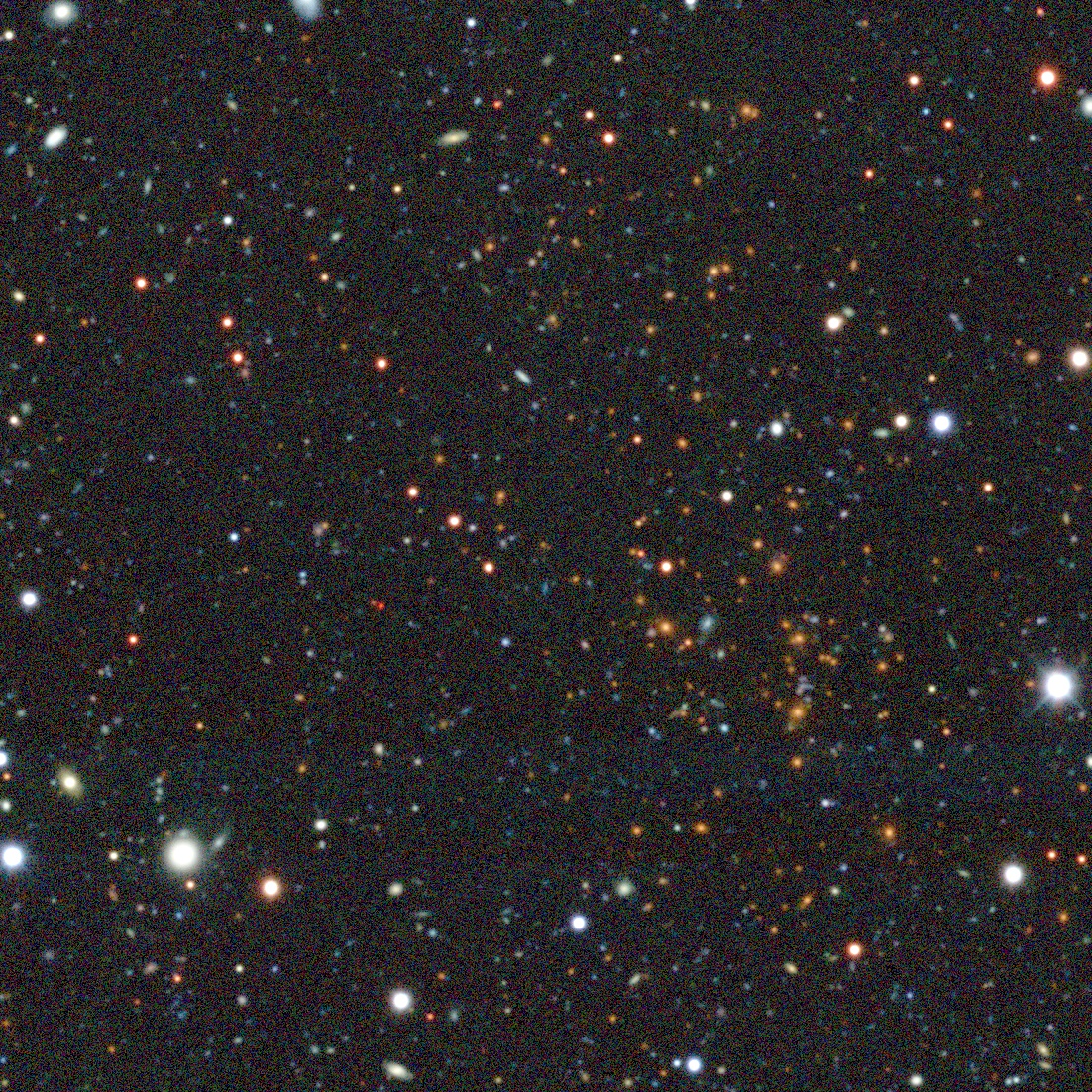}
\textrm{c)}
\end{minipage}
\vskip-0.05in
\caption{Three RASS galaxy clusters identified in the DES-SV region. Top panel: Richness versus redshift plot with Gaussian fit to the most significant peaks. Black points are based on global background estimation, blue points show the results based on the local background estimation. Bottom panel: Optical $grz$ pseudo-color images of the central 5$\times$5 arcmin regions around each of the RASS position. The corrected redshifts found for these clusters are  z=0.22 (a), 0.40 (b) and 0.77 (c).}
\label{fig:clusterexamples}
\vskip-0.15in
\end{figure*}
\begin{figure}
\centering
\vskip0.05in
\includegraphics[keepaspectratio=true,width=\columnwidth]{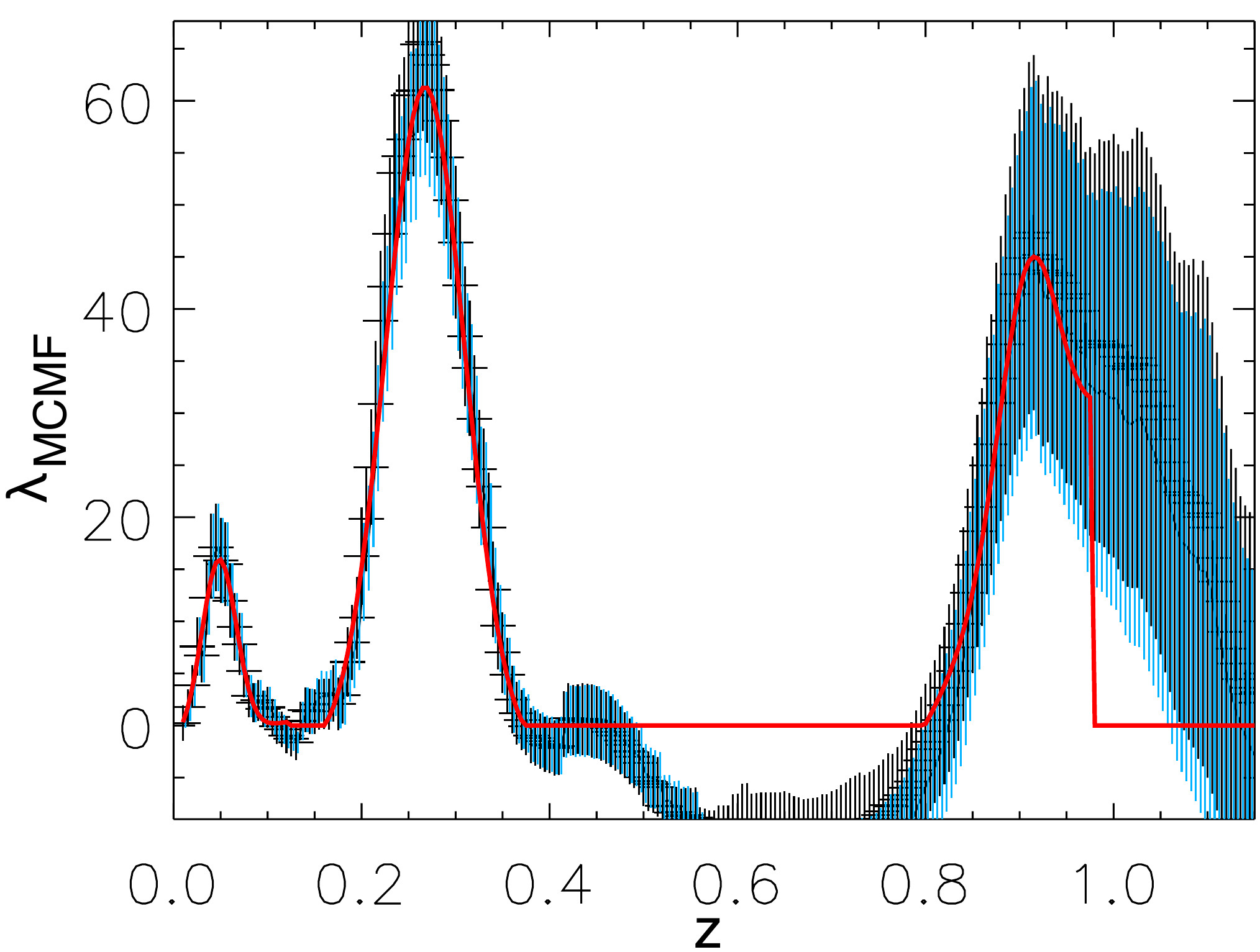}
\includegraphics[keepaspectratio=true,width=\columnwidth]{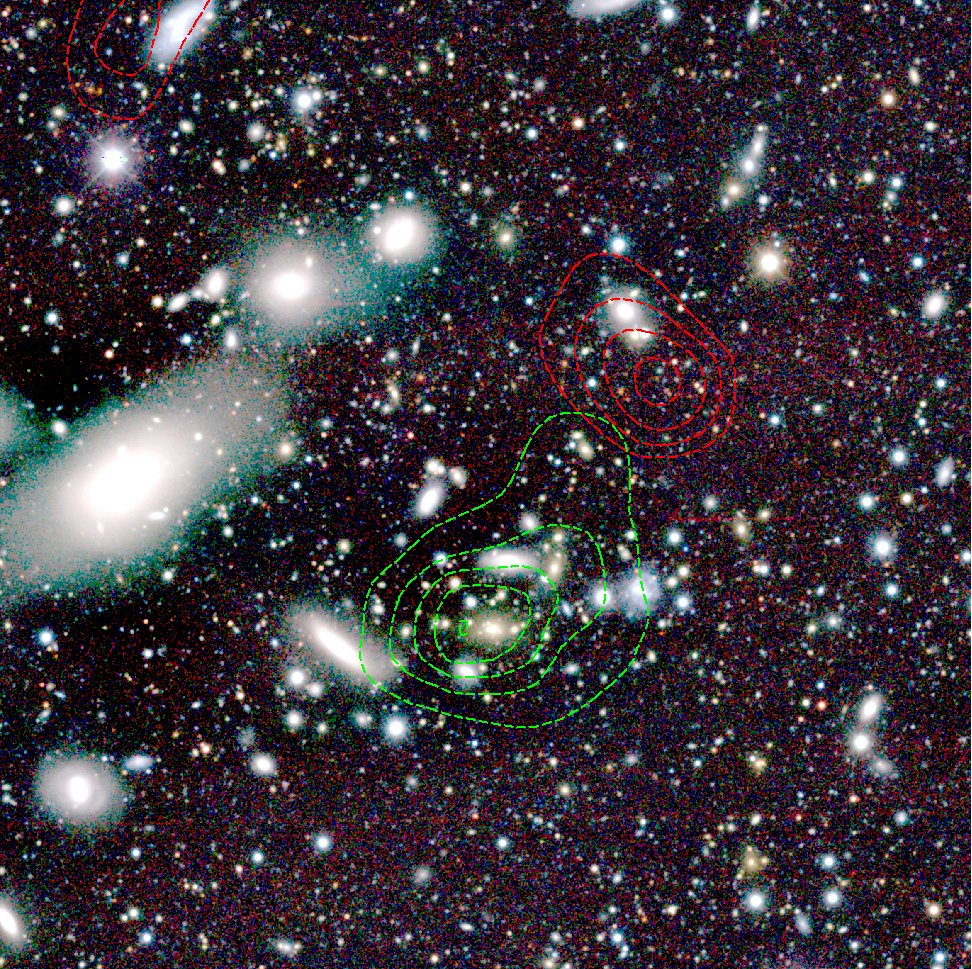}
\vskip-0.05in
\caption{RASS position with three significant peaks in redshift at z=0.05, 0.24 and 0.87. The top panel shows the richness versus redshift plot, with Gaussian fits to the three peaks. Masking by the two clusters at lower redshifts results in an apparent under density at redshifts z>0.5. Bottom panel: $grz$ pseudo-color image of the 10$\times$10\,arcmin region around the RASS position. Contours show the galaxy number density corresponding to the two redshift peaks with $z>0.2$ identified in the upper panel.}
\label{fig:triple}
\vskip-0.15in
\end{figure}

\section{Cluster Confirmation Method}
\label{sec:method}

The optical counterparts of X-ray selected galaxy clusters can be identified in several ways, depending on the availability of optical pass bands and data quality.  Methods are based on galaxy clustering, photometric or spectroscopic redshifts or on their appearance in color-magnitude space. The latter method is called the red-sequence (RS) method \citep[][]{gladders00} and uses the fact that galaxy clusters contain many elliptical galaxies, which are dominated by a stellar population passively evolving since $2<z<5$ \citep[e.g.,][]{bower92,ellis97,depropris99,lin06}.  Those galaxies therefore show a strong 4000\,\AA\ break, resulting in a tight color-magnitude relation for pass bands that include this break at a given redshift.  This color-magnitude relation is called the RS and gives the method its name. We use the RS method together with the galaxy clustering information as the basis for constructing a matched filter for cluster verification and redshift estimation.

Our aim is the identification of X-ray selected galaxy clusters from the 2RXS catalog \citep{boller16}. We therefore search for cluster candidates in the optical data at the location of every X-ray source that lies in the footprint of our SPT-East dataset.
In this search we use the X-ray flux as a redshift dependent mass constraint, allowing us to identify the physically relevant radius within which to search for optical counterparts.  
This is done by calculating the cluster richness $\lambda$, which is the excess of the weighted sum of galaxies within this radius with respect to the expected number galaxies in the absence of a cluster. The weights can be seen as filter functions that follow the expected behaviour of cluster galaxies in color, magnitude and angular space and therefore maximize the chance to detect the cluster at the correct redshift.
The final step to quantify the probability of a chance superposition of the X-ray and optical sources involves comparing the measured richness and redshift of the optical counterpart to the distributions of richness and redshift determined along random lines of sight where there is no 2RXS detection but similar X-ray exposure time.

The details of the different components of this multi-component cluster matched filter confirmation tool (MCMF) are explained in the following subsections.

\subsection{X-ray Luminosity}\label{sec:XL}

To define the region in which we are searching for counterparts, we need to estimate the X-ray luminosity under the assumption that the observed X-ray source is a galaxy cluster.
The basis of the X-ray luminosity estimate for each source is the X-ray count rate in the 2RXS catalog. The count rate is obtained within a minimum aperture of 5$'$ radius around each 2RXS position. If the extent of the source is found to be larger than this, the extraction radius is increased. In the SPT-East region, all but two of the sources have extraction radii equal to this minimum value.

Because fits to the X-ray spectra are only possible for a tiny fraction of the sources, we have to assume a spectral model with given temperature to obtain luminosities from the 2RXS catalog. We can convert the X-ray count rate to an X-ray luminosity in the 0.1-2.4 keV band using an APEC plasma model \citep{smith01} of a certain temperature and metal abundance, a specific redshift and the appropriate neutral hydrogen column density.  We create for this task a look up table for different redshifts, metal abundance, plasma temperatures and neutral hydrogen column densities.
 
For this work we fix temperature and metal abundance to 5\,keV and 0.4 solar metallicity, respectively. We further assume that our fiducial luminosity $L_{X}$, corresponds closely to $L_{500}$, the luminosity in the 0.1-2.4 keV band within a radius within which the mean density is 500 times the critical density at the assumed cluster redshift. Although our fiducial luminosity is derived from count rates within a fixed aperture, we do not expect this inconsistency to be an important source of scatter given the large intrinsic scatter in the cluster $L_{X}$--mass relation.  It would result in a mass and redshift dependent bias.  Assuming the X-ray surface brightness profile following a beta profile with $\beta\approx2/3$, the flux differs less than 6\,percent for radii between 0.65 to 2 times $r_{500}$ from the value at $r_{500}$.  Such a small bias is not important for the analysis presented here.
A more sophisticated iterative approach using a luminosity--temperature relation is planned in combination with re-centering and re-extraction of the X-ray count rate for the upcoming full DES analysis of the 2RXS catalog. All luminosities are presented in units of $10^{44}$\,erg\,s$^{-1}$.  

\subsection{Cluster Mass and Followup Region of Interest}

We measure the cluster matched filter richness \LamMCMF\ as a function of redshift along the line of sight toward each candidate.  \LamMCMF\ is extracted within a radius $r_{500}$.  We derive this radius using the estimated luminosity at that redshift and a $L_{X}$--mass scaling relation.  For this analysis we use the RASS based scaling relation given in \citet{mantz10b}, which has the form:
\begin{eqnarray}
\label{eq:nominalMLT}
\expectation{\ell(m)} &=& \beta_0 + \beta_1 m. 
\end{eqnarray}
With the terms
\begin{eqnarray}
\label{eq:MLTdefs}
\ell &=& \logTen\left(\frac{L_{X}}{E(z)10^{44}\erg\second^{-1}}\right), \nonumber\\
  m &=& \logTen\left(\frac{E(z)M_{500}}{10^{15}\Msun}\right), 
\end{eqnarray}
which include factors of the normalized Hubble parameter, $E(z)=H(z)/H_0$.  As scaling parameter $\beta_0$ and $\beta_1$ we take 0.8 and 1.34  \citep[as given in Table~7,][]{mantz10b}, assuming our fiducial X-ray luminosity is a reasonable estimate of $L_{500}$.
The exact choice of the scaling relation used at this stage of the project is not crucial, because we are mainly using these mass estimates to determine the region of interest within which we search for optical counterparts.
The region of interest, $r_{500}$, is finally derived from $M_{500}$ using our fiducial cosmology and the redshift.

 \subsection{Radial Filter}
 
To use the clustering information in our code, we apply a radial weighting $\Sigma(R)$ based on a Navarro, Frenk and White (NFW) profile \citep{navarro97}.  The corresponding projected NFW that we use as a spatial weighting function is \citep{bartelmann96}
\begin{equation}
\Sigma(R) \propto \frac{1}{(R/R_s)^2-1}f(R/R_s),
\label{eqn:radfilter}
\end{equation}
where $R_s$ is the characteristic scale radius, and
\begin{equation}
f(x) = 
\begin{cases}
1-\frac{2}{\sqrt{x^2-1}}\arctan\sqrt{ \frac{x-1}{x+1} } & (x>1) \cr
1-\frac{2}{\sqrt{1-x^2}}\arctanh\sqrt{ \frac{1-x}{x+1} } & (x<1) .
\end{cases}
\end{equation}
We use a scale radius $\mathrm{R}_\mathrm{s}= R_{500}/3$, which is consistent with the typical concentration of RS galaxies found in massive clusters extending to redshift $z\sim1$ \citep{hennig16}.  We adopt a minimum radius of 0.1$h^{-1}$Mpc, below which we set the radial weight to be constant \citep{rykoff14}.

The profile $\Sigma(R)$ is truncated at the cluster radius $R_{500}(z)$, and the correction term $C_{\mathrm{rad}}$ is chosen such that the radial filter is normalized as
\begin{equation}
1 = C_{\mathrm{rad}}\int_0^{R_{500}(z)} dR\ 2\pi R\Sigma(R).
\label{eq:radnorm}
\end{equation}
The radial weight assigned to a galaxy $i$ at a given radial distance $R_i$ from the assumed center is then simply,
\begin{equation}
n_i (z)= C_{\mathrm{rad}}(z) 2\pi R_i \Sigma(R_i).
\end{equation}

\citet{rykoff12} showed that their matched filter richness is only weakly dependent on the shape of the radial filter. They found that the ratio of the intrinsic scatter of the $L_X-\lambda$ relation using a flat radial weight over that using an NFW based radial filter is only $\sigma_{L_X|\lambda_{flat}}/\sigma_{L_X|\lambda_{NFW}}=1.03\pm0.015$. The precise choice of the parameter values entering the NFW profile are therefore likely of low relevance and will be explored in future work, once a larger cluster sample is available.

\subsection{Color-Magnitude Filter}\label{sec:colmag}

To obtain a clean selection of RS galaxies, we use information from all the available bands rather than only using the bands that bracket the 4,000\,\AA\ break.  The basic requirement, then, is that a cluster galaxy candidate has to be consistent with a passively evolving galaxy in all colors, even if there is no tight RS visible at a given redshift and color.

As a baseline RS model we use the same models as described in \citet{hennig16}. Those models have been used for a DES study of galaxy populations of massive, Sunyaev-Zel'dovich Effect selected clusters up to a redshift of 1.1 \citep[see also][]{zenteno11,song12b,liu15,zenteno16}. Because the models are meant to describe the RS in color-magnitude space, where the filter combination includes the 4,000\,\AA\ break and the next reddest band, it is not guaranteed that the model performs well for other filter combinations.  We therefore use two galaxy cluster subsamples from the SPT survey with spectroscopic cluster redshifts \citep{ruel14,bayliss16} to calibrate the RS models and estimate the width of the RS for a given redshift and color.  In this tuning process we keep the slope fixed to that from the model and just allow the normalization to vary.  Furthermore, we estimate the total scatter (observational plus intrinsic) of the RS in each color.  Ultimately, we may return to the underlying RS model and tune it in a less {\it ad hoc} manner once we extend the analysis to the full DES dataset.

We define all galaxies that lie within three times the standard deviation from the RS model in color-magnitude space in all of the following color combinations $(c_1,c_2,c_3)=(g-r,r-i,i-z)$ to be candidate cluster members.  We weight each cluster galaxy candidate $i$ with a redshift $z$ dependent value
\begin{equation}
 w_i(z)=\frac{\prod\limits_{j=1}^{3}{G\left(c_{i,j}-\left<c(z)\right>_j,\sigma_{c_j}(z)\right)}}{N(\sigma_{c_1}(z),\sigma_{c_2}(z),\sigma_{c_3}(z))},
 \label{eq:color_weight}
\end{equation}
where $G\left(\Delta c_{i,j},\sigma_{c_j}(z)\right)$ is the value of a normalized Gaussian function at color offset $\Delta c_{i,j}=c_{i,j}-\left<c(z)\right>_j$, where for band $j$ and redshift $z$ the expected RS color and measurement standard deviation are $\left<c(z)\right>_j$ and $\sigma_{c_j}(z)$, respectively.  We normalize the weights by $N(\sigma_{c_1}(z),\sigma_{c_2}(z),\sigma_{c_3}(z))$, which is the average weight of a population of galaxies that follows the expected distribution of a cluster at the investigated redshift.

\subsection{Luminosity Cut and Incompleteness Correction}

We do not apply a luminosity based weight to the galaxies, but we limit the number of galaxies investigated at a given redshift by selecting only galaxies that are brighter than $i\leqq m^*(z)+1.25$, where $m^*(z)$ is the expected characteristic magnitude for a cluster at redshift $z$.  This magnitude cutoff can exceed the completeness limit  $c_\mathrm{lim}$ of the data at higher redshift in some locations within the SPT-East region.  These clusters would be still detectable, but the estimated richness would be biased low.  We account for this effect by rescaling the measured richness to $c_\mathrm{lim}$ by using the correction factor
\begin{equation}
 C_{\mathrm{cmp}}=\frac{\int_{m^*(z)-4.6}^{m^*(z)+1.25} S(m^*(z),m,\alpha) dm}{\int_{m^*(z)-4.6}^{c_\mathrm{lim}} S(m^*(z),m,\alpha) dm},
 \label{eq:richcor}
\end{equation}
where $S(m^*(z),m,\alpha)$ is the Schechter function \citep{schechter76}, in which $m^*(z)$ is the characteristic magnitude expected at redshift $z$.  The faint end slope $\alpha$ is set to  $\alpha=-1$ in our analysis, but it can be adjusted with redshift and mass to match any measured trends \citep[e.g.][]{zenteno16}.  Assuming the galaxy population follows the Schechter function, equation~(\ref{eq:richcor}) gives the fraction of galaxies brighter than the completeness limit compared to the number of galaxies that would be brighter than $m^*(z)+1.25$.  When $m^*(z)+1.25<c_\mathrm{lim}$ we set $C_{\mathrm{cmp}}$=1.

\subsection{Masking and Background Estimation}

At each RASS position we select all sources within a distance of 0.5\,degrees. We create a smoothed source density map with a pixel scale of 2$''$, using a box-car smoothing with box size that contains on average 16 sources. We use square and rectangular boxes to further improve the detection and the modeling of masked regions.

This robustly masks regions around very bright sources, accounting for the complex edges of the DES-SV area and non-uniformities in the shapes of very bright sources.  The number of unmasked pixels gives the available area for the background and cluster counts. These mask images are also used to adapt the normalization of the radial weight, accounting for masked regions.

We use two different approaches to estimate the background correction for a cluster at a given redshift. The global background approach uses a merged source catalog of tiles with similar or higher completeness limits than that estimated locally.  This approach takes advantage of the increased statistics that comes with larger area but is also not sensitive to variations in local image properties such as image depth variation or stellar density. 
The local background approach uses all sources within the range $r_{500}<r<0.5$ degrees. This follows the local conditions more closely, which becomes increasingly important at the higher redshifts probed by our method.

\subsection{Identifying Cluster Candidates and Estimating Redshifts}
We define our filtered richness \LamMCMF\ as 
\begin{align}
\label{eq:richn}
 \lambda_{\mathrm{MCMF}}(z)=&\frac{C_{\mathrm{cmp}}(z)A_{\mathrm{tcl}}(z)}{A_{\mathrm{cl}}(z)}\left( \sum_{i} w_i(z) n_i(z) \right. \nonumber \\
 &\left. - \frac{A_{\mathrm{cl}}(z)}{A_{\mathrm{BG}}(z)} \sum_{j} w_j(z)\right) ,
\end{align}
the sum of the color and the radial weight over all cluster candidates minus the scaled background, where $j$ runs over all background galaxies that fulfill the same color and magnitude cuts as for the cluster candidates. Here the elements $A_{\mathrm{cl}}$ and $A_{\mathrm{BG}}$ correspond to the unmasked cluster and background area and $A_{\mathrm{tcl}}$ to the total area within $r_{500}(z)$.  \LamMCMF\ is calculated for redshifts between 0.01<z<1.15 in steps of $\delta z=0.005$.   For each \LamMCMF\ estimate, we calculate the uncertainty $\Delta$\LamMCMF\ assuming Poisson statistics. We call the ratio of the \LamMCMF\ measurement to the Poisson uncertainty the signal to noise ratio $S= \lambda_{\mathrm{MCMF}}/\Delta \lambda_{\mathrm{MCMF}}$ of the detection.

The distribution of \LamMCMF\ versus redshift is searched for up to five peaks. These peaks are then subsequently fitted with Gaussian functions, and the three most significant peaks are kept for further analysis.  To improve the robustness of the fits and to avoid confusion with multiple peaks, we restrict the fitting range to be $|z-z_{\mathrm{peak}}|<0.08$ for $z<0.2$, $z_{\mathrm{peak}}<0.10$ for $z<0.2$ and $z_{\mathrm{peak}}<0.15$ for $z>0.4$. The best fit provides the richness and redshift estimate for each peak.  No deblending of nearby peaks is performed.  The search for peaks is performed for the global and the local background approach independently; however, we generally use the local approach for redshifts above $z=0.15$ and the global approach at lower redshifts. Fig.~\ref{fig:clusterexamples} shows three examples of clusters identified with MCMF, and Fig.~\ref{fig:triple} shows an example of a RASS candidate with multiple optical counterpart peaks along the line of sight.

To obtain optical positions for each of the identified peaks, a weighted galaxy density map is created by using a Voronoi Tessellation method and then smoothed with a Gaussian kernel of 250\,kpc size. The weighting is performed by using the derived color weights from the richness estimator at the redshift of the identified peak.  This efficiently suppresses the signal from structures at other redshifts and reduces the median offset between SPT centers and galaxy density centers by 0.1$'$ or 16\,percent compared to the unweighted estimates \citep[for discussion of offsets between SPT and optical positions see][]{song12b,saro15}.  Finally, we use {\tt SExtractor} \citep{bertin96} to identify the galaxy density peak that lies nearest the X-ray candidate position.

\subsection{Quantifying Probability of Random Superposition}
\label{sec:random_superposition}
To estimate the chance of random superpositions of unrelated optical clusters onto lines of sight corresponding to X-ray sources, we apply our method to random positions in the survey field \citep[see also][]{saro15}.  In this estimate we exclude 6$'$ radius regions around the 2RXS X-ray positions and exclude also regions with low RASS exposure times.  The lower exposure time cut ensures that the median X-ray exposure time in the region where random lines of sight are evaluated is similar to that in the region where the 2RXS followup is carried out.  For this analysis we use 2,587 random positions, and we run our code using X-ray count rates corresponding to the 25, 50, 75 and 95 percentiles of the count rate distribution of RASS sources that overlap the SPT-East region.

Then, for each peak found at a RASS position, we calculate the fraction \ProbLam\ of sources in the random catalog with $|z_{RASS}-z_{rand}|<0.075$ that have richnesses equal to or below the richness found for the RASS cluster candidate.  Because the aperture of the cluster finder is based on the X-ray count rate, we use the random catalog where the count rate is closest to that of the RASS cluster candidate.  This fraction corresponds to the probability that the optical cluster counterpart identified for a particular RASS source at a given redshift and mass is real as opposed to a random superposition of an unassociated optical system.

As an alternative approach, we also estimate the fraction \ProbS\ of sources in the random catalog with $|z_{RASS}-z_{rand}|<0.075$ that have a signal to noise ratio $S$ equal to or below that found for the RASS cluster candidate.  This estimator accounts for the local variations in imaging depth as well as the effects of masking, but at the price of partially loosening the correlation between X-ray detection probability, mass and redshift.


\section{Application to RASS and DES-SV Datasets}
\label{sec:application}

We test our method by applying it to the RASS and DES-SV datasets described in section~\ref{sec:data}.  Below we first discuss the performance of the matched cluster filter in the face of stellar contamination (section~\ref{sec:stellarcontamination}).  Then in section~\ref{sec:sample} we present the cluster sample and examine its characteristics, including the photometric redshift accuracy, the optical--X-ray scaling relations and the center offset distributions.  Finally, in section~\ref{sec:comparison} we compare our cluster sample to three other samples extracted over the same sky area.

\subsection{Sensitivity to Stellar Contamination}\label{sec:stars}
\label{sec:stellarcontamination}

Stars can interfere with our followup of galaxy clusters if they slip through our cluster filter, creating fake signals, affecting galaxy colors by reflected light or being X-ray emitters by themselves. In this section we address these issues by performing two tests.  The cluster confirmation method we present will be applied over a large area and therefore has to be robust against large changes in the stellar density.  The changes in stellar density are especially large in the DES-SV region, due to its proximity to the Magellanic clouds.  Our dataset is therefore well suited for these tests. 

\subsubsection{Stars and Star Clusters}

\begin{figure}
\vskip-0.05in
\includegraphics[keepaspectratio=true,width=\columnwidth]{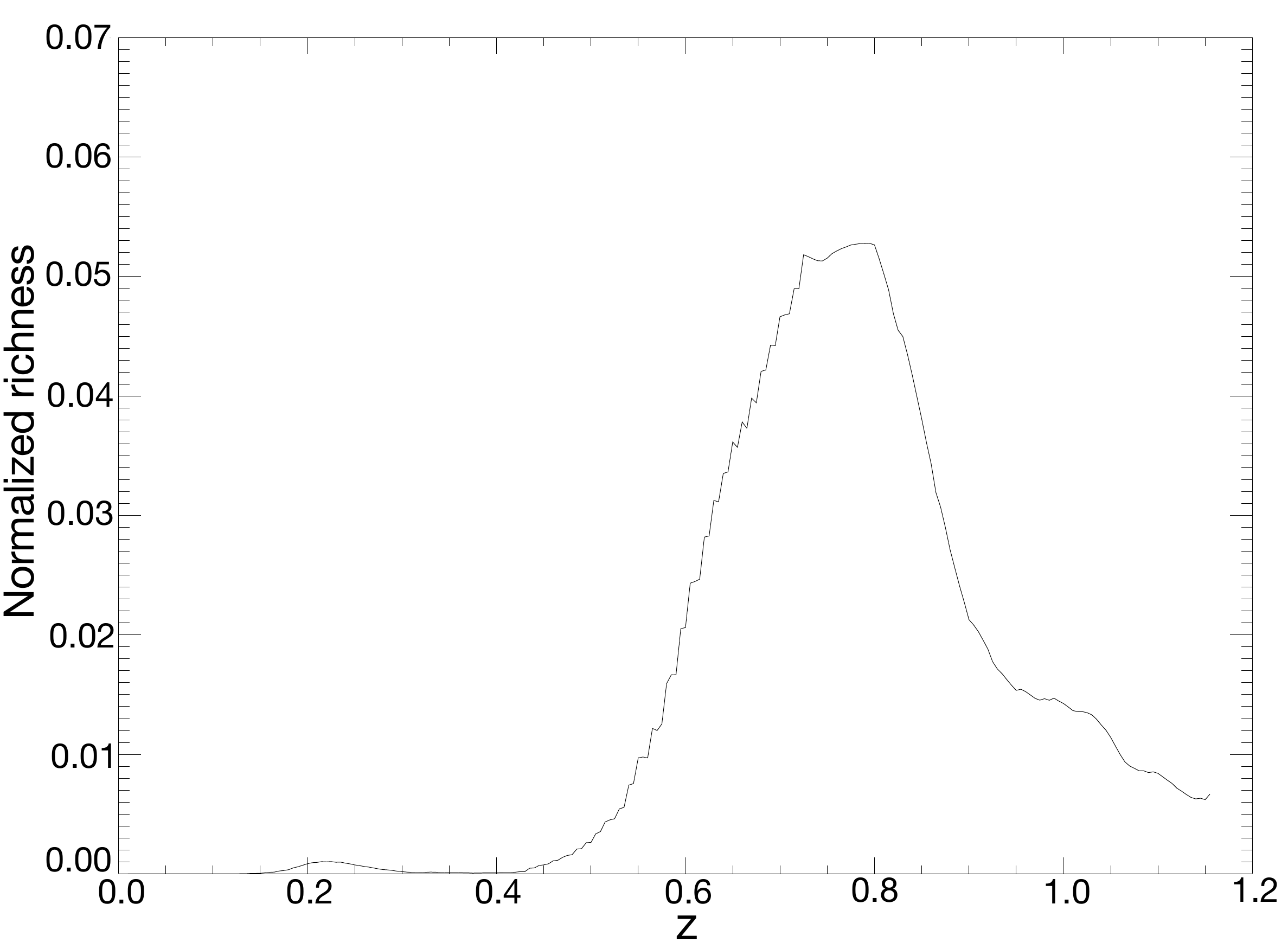}
\vskip-0.05in
\caption{The fraction of stars leaking into the galaxy sample for a completeness limit of i=22.5 mag. A maximum of 5\,percent of the input signal is reached at a redshift of z=0.75.}
\label{fig:stelcont}
\vskip-0.15in
\end{figure}

The first test uses a catalog of only stars and measures what fraction of the input signal makes it through our cluster filter.  To obtain a clean star catalog, we use the DECam observations of the COSMOS field and match the detected sources with
stars from the COSMOS photo-z catalog \citep{Ilbert09}.  No radial filtering or weighting is used in the test. Rather, the input catalog is filtered by color and magnitude, and as a last step the cut in {\tt spread\_model} is applied.  The output signal is normalized to the number of stars falling into the magnitude range used for a given redshift.  We run this test for several limiting magnitudes, and the result for the case of i=22.5 can be seen in Fig.~\ref{fig:stelcont}.  

As already mentioned, the cut in spread model is reliable to an apparent magnitude of 22.  Because of that  and because of the magnitude dependency of the used color filters we do not expect a significant signal up to redshifts of z$\approx0.5$.  This is reflected in Fig.~\ref{fig:stelcont}, where the signal stays well below 0.01 up to a redshift of 0.55.  At a limiting magnitude 22.5 the signal increases to a maximum of 5.2\,percent and for a limiting magnitude of 23 the normalized signal reaches a maximum of 6\,percent.  Our cluster followup algorithm employs a local background subtraction for redshifts greater than z=0.15. Therefore a false signal by stars can only be created by a local over density of stars within the investigated aperture of $r_{500}$. 

\begin{figure}
\includegraphics[keepaspectratio=true,width=\columnwidth]{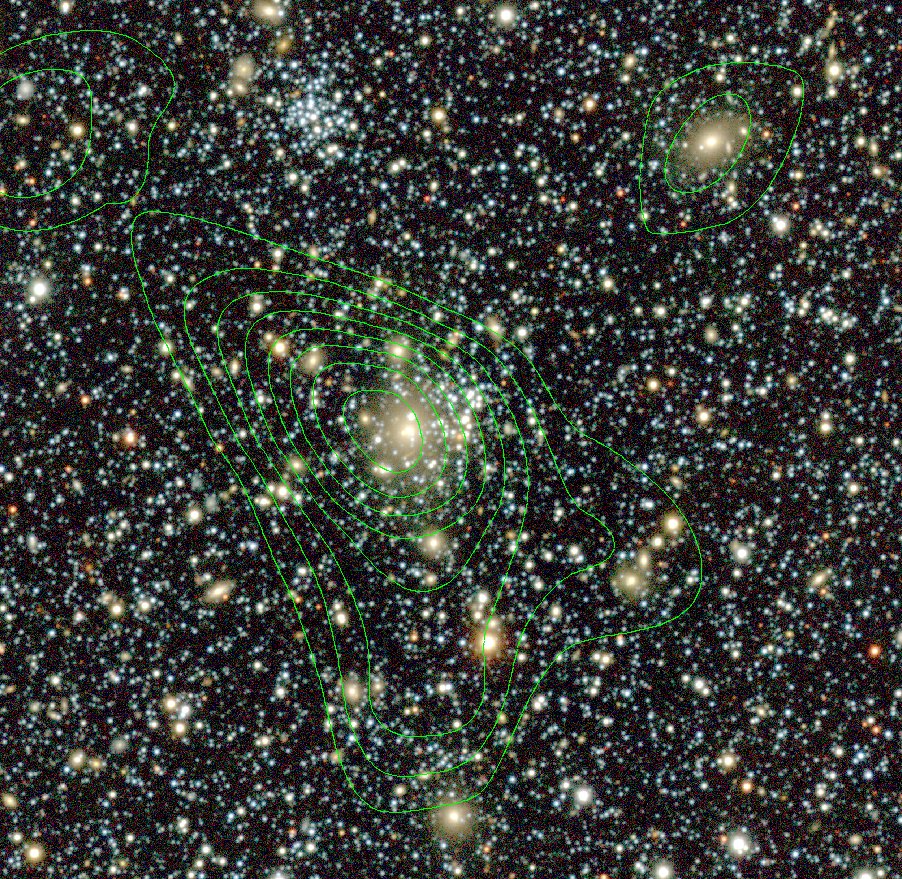}
\includegraphics[keepaspectratio=true,width=\columnwidth]{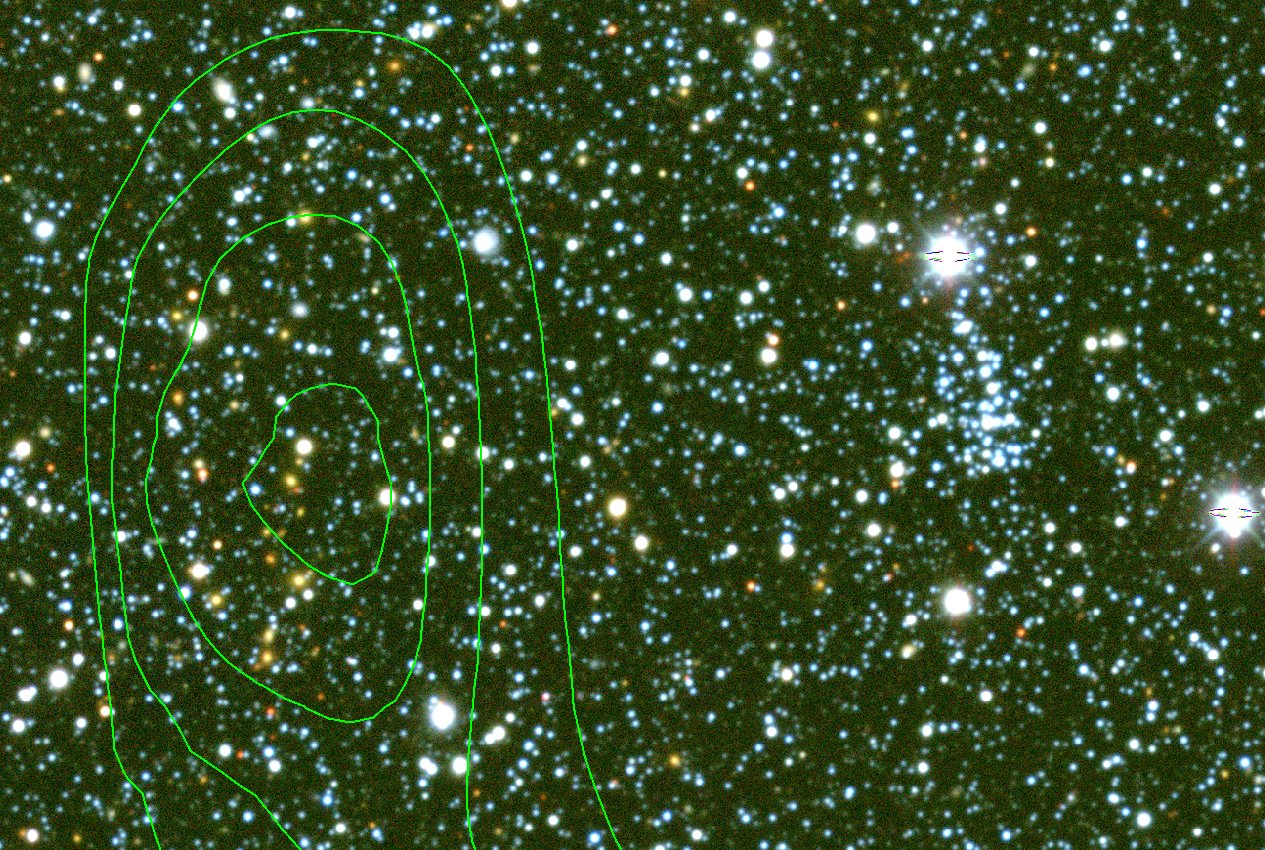}
\vskip-0.05in
\caption{$grz$ pseudo-color images around star cluster positions with significant \LamMCMF\ measurements. The image shows the presence of galaxy clusters close to the star cluster positions. The top image shows the case with to star clusters next a galaxy cluster. The galaxy density contours further highlight that the \LamMCMF\ peaks found are related to those galaxy clusters and not caused by stars leaking into the galaxy sample.}
\label{fig:stelcont2}
\vskip-0.15in
\end{figure}

To explore the impact of star clusters, we perform a second, more extreme test. We identify 29 star clusters in our field and use their locations as an input catalog for our cluster finder. We use the observed medium luminosity from our RASS sources to define our aperture and radial filter.  In one out of 29 positions the filter returns a \LamMCMF\ that fulfills our selection criteria to be an optical counterpart of an X-ray source.  Two other positions have a \LamMCMF\ that exceed three times the Poisson based measurement error.  An investigation of the optical images and galaxy density maps for those three cases reveal two real galaxy clusters next to the position of the star clusters. 
In fact, two of three star clusters are very close to one another, and thus it is the same galaxy cluster that is found in both cases.
Pseudo-color images showing the star and galaxy clusters are shown in Fig.~\ref{fig:stelcont2}. The redshifts for the clusters are  z=0.16 and z=0.48 and therefore below the redshift range where we expect contamination by stars.
We note that due to the vicinity to the LMC a significant fraction of the investigated star clusters consist of young stars associated to the LMC. A repetition of this test with globular clusters is therefore planned once a greater number is observed within the DES.

Finally, we investigate the impact of scattered light caused by bright stars by searching the UCAC4 catalog \citep{Zacharias12} for bright (V<11\,mag) sources within 2$'$ of 2XRS X-ray positions. We investigate the images and galaxy density maps for all 33 cluster candidates that fulfill our standard quality cuts and have a bright source near them.  For moderately bright stars we do not observe any effects on the galaxy density contours, indicating that if there are regions affected, these are smaller than the typical smoothing scale. For the brightest sources the region around the star gets masked due to a lack of identified objects in the photometric catalog.  Thus, the net impact of bright stars is to reduce the area over which optical followup of X-ray sources is possible.

\subsubsection{X-ray Emitting Stars} \label{sec:xrstars}

\begin{figure}
\includegraphics[keepaspectratio=true,angle=90,width=\columnwidth]{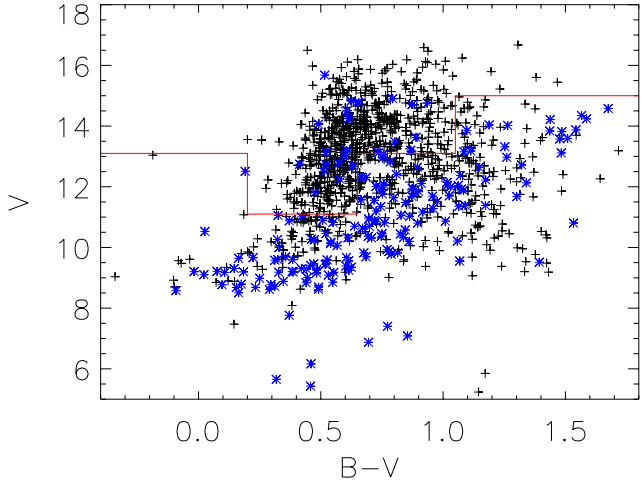}
\vskip-0.05in
\caption{Color-magnitude distribution of UCAC4 sources within 3$'$ from the X-ray positions. Sources outside 1$'$ are plotted as black '+' symbols, sources within 0.5$'$ are shown as blue '*' symbols. The separation between X-ray and non X-ray sources is indicated as red line. The radial distribution of both samples can be found in Fig.\ref{fig:STraddist}}
\label{fig:STcolormag}
\vskip-0.15in
\end{figure}

\begin{figure}
\includegraphics[keepaspectratio=true,angle=90,width=\columnwidth]{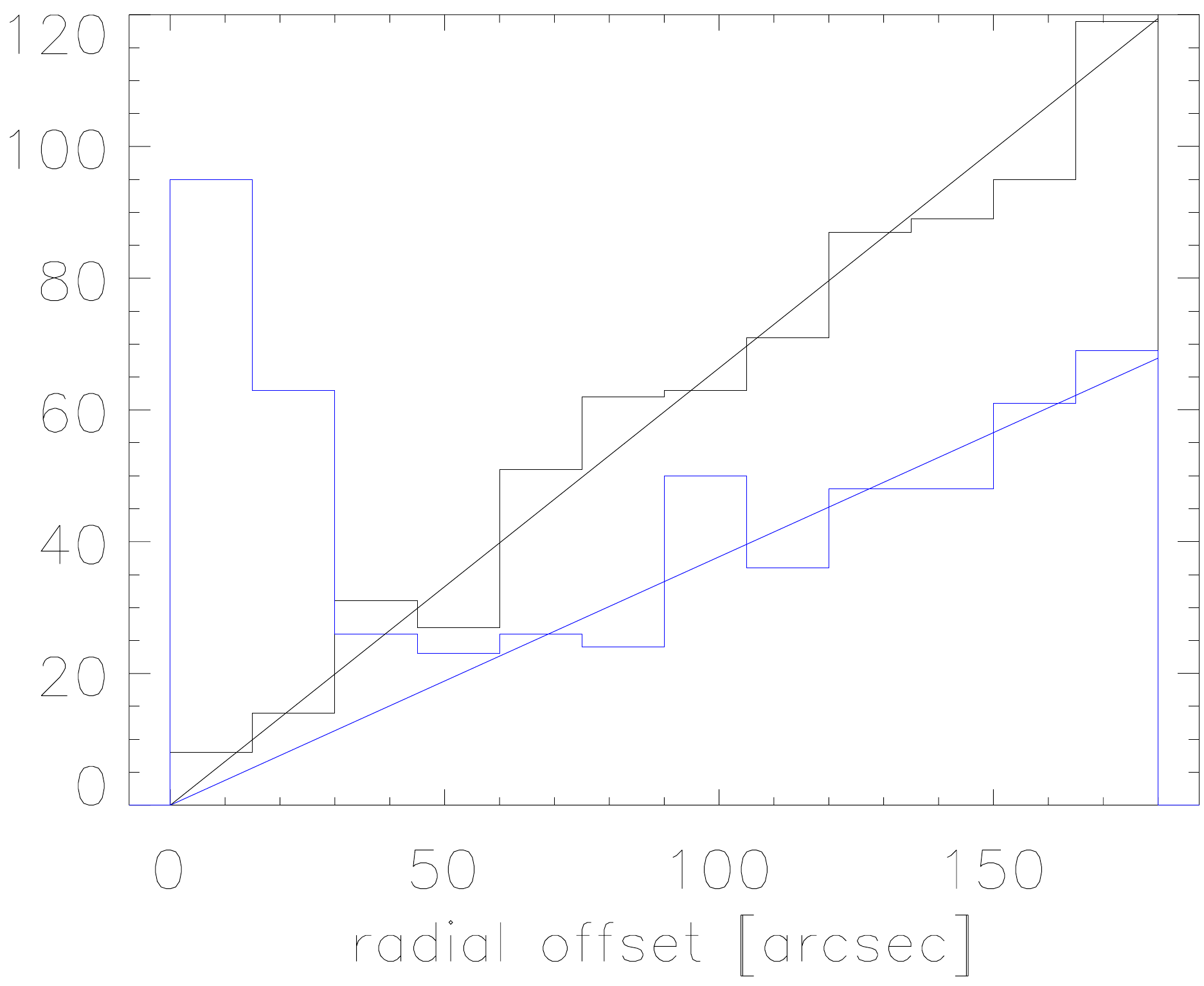}
\vskip-0.1in
\caption{Histogram of separations of UCAC4 sources around 2RXS sources for the two source populations indicated in Fig.\ref{fig:STcolormag}. Sources with colors and magnitudes falling above the red lines of Fig.~\ref{fig:STcolormag} are used for the black histogram, sources falling below the line are shown in the blue histogram. The blue histogram shows a strong excess of sources below 1$'$ over the expected number of chance super positions, whose indicated as a blue line. The black histogram is consistent with random superposition, whose expected distribution is shown as a straight line.}
\vskip-0.15in
\label{fig:STraddist}
\end{figure}

X-ray emitting stars tend to be extremely bright in the optical, and therefore we can use this information to remove any random superpositions of X-ray emitting stars with optical clusters of galaxies.  The majority of X-ray emitting stars detected in RASS are too bright to have a valid DES measurement; we therefore cross match our 2RXS catalog with the UCAC4 catalog. In Fig.~\ref{fig:STcolormag} we show the distribution of UCAC4 sources above 1$'$ and below 0.5$'$ offset to the X-ray positions. UCAC4 sources with small offsets and therefore high probability of being an optical counter part of the X-ray source seem to cover a distinct region in color magnitude space. We split the sources into a non X-ray emitting and in a potentially X-ray emitting population based on their position in color magnitude space. The distribution of radial offsets to the X-ray position for the two subsamples are shown in Fig.~\ref{fig:STraddist}. The distribution of non X-ray emitters in black is consistent with a constant source density, indicated as a black line. The potential X-ray emitters are shown in blue. An excess of sources above the expectation of a constant source density is seen for small offsets within 1$'$. We find 160 (153) sources in excess of the background for offsets below 1$'$ (0.75$'$). We use this distribution to assign a probability of a source being an X-ray emitting UCAC4 source \ProbStar\, for all sources within 1$'$ offset to the X-ray position and below the red line in color-magnitude space. 
The probabilities are calculated in bins of 15$"$, and are the ratio of excess counts over total counts. The expected number of chance super positions is obtained from a fit to the histogram at distances greater 80$"$ and forcing to be zero at zero distance and is indicated as a blue line in Fig.~\ref{fig:STraddist}. We find for the first four bins values of 0.970, 0.865, 0.456 and 0.140.
Applied to 2RXS catalog within the DES-SV area, we note that only four out of ninety cluster candidates with \ProbLam$>0.985$ have \ProbStar$>0.86$. The correlation between bright UCAC4 and 2RXS sources is not included in the creation of the random catalogs. Including those sources would result in a contamination of 3 to 5$\%$ additional to that expected from \ProbLam alone.

\begin{figure}
\includegraphics[angle=0,keepaspectratio=true,width=\columnwidth]{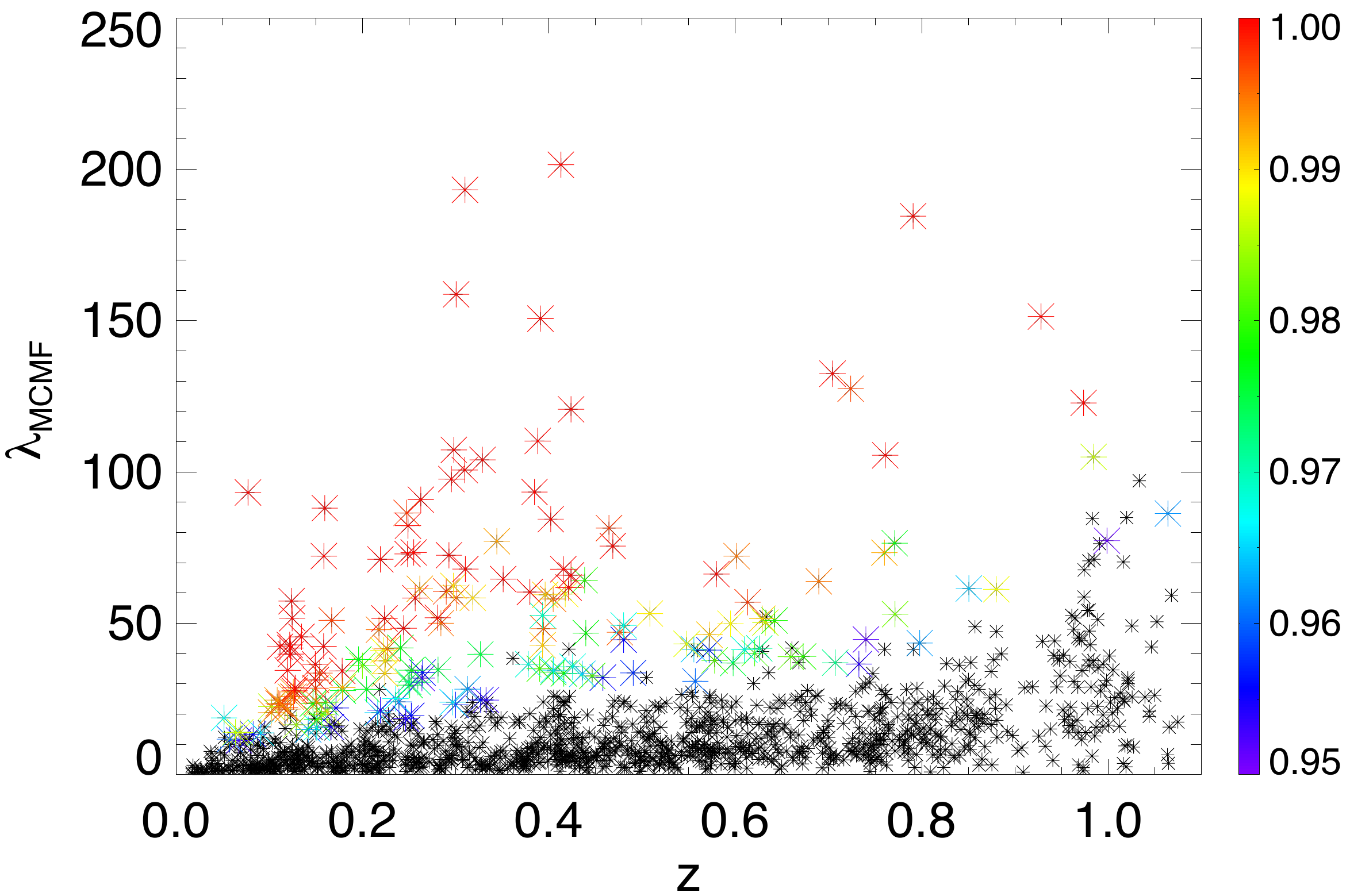}
\vskip-0.05in
\caption{Richness versus redshift distribution for the followup sample.  Those RASS sources with \ProbLam$<0.95$, corresponding to more than 5\,percent chance of being a false superposition, are shown in black, and those with \ProbLam$>0.95$ are color coded according to \ProbLam.}
\label{fig:richnessdist}
\vskip-0.15in
\end{figure}

\subsection{Galaxy Cluster Sample}
\label{sec:sample}

For each 2RXS position, we obtain for up to three peaks the probability that the source is not a random superposition \ProbLam, the probability of the detection signal to noise \ProbS\ being greater than null, the signal to noise $S$, the photometric redshift $z$, the photometric redshift uncertainty $\sigma_z$, the richness measurement \LamMCMF, the \LamMCMF\ uncertainty $\sigma_{\lambda}$ and the galaxy density based positions and their offsets from the X-ray position.  Additionally, we obtain the X-ray luminosity $L_{X}$ and the mass $M_{500}$ for each peak using the X-ray count rate, the optically derived redshift and the scaling relations as previously described. Further, we estimate \ProbStar\ for all 2RXS postitions.  The distribution of the 1241 RASS sources in \LamMCMF\ and redshift is shown in Fig.~\ref{fig:richnessdist}.  Those sources that have \ProbLam$<0.95$ are shown in black; given their measured richnesses and redshifts these sources are less then 2\,$\sigma$ inconsistent with the distribution of \LamMCMF\ extracted from 1000 random lines of sight.  Those above this threshold are color coded according to their significance in comparison with the random lines of sight. 

\begin{figure}
\includegraphics[angle=90,keepaspectratio=true,width=\columnwidth]{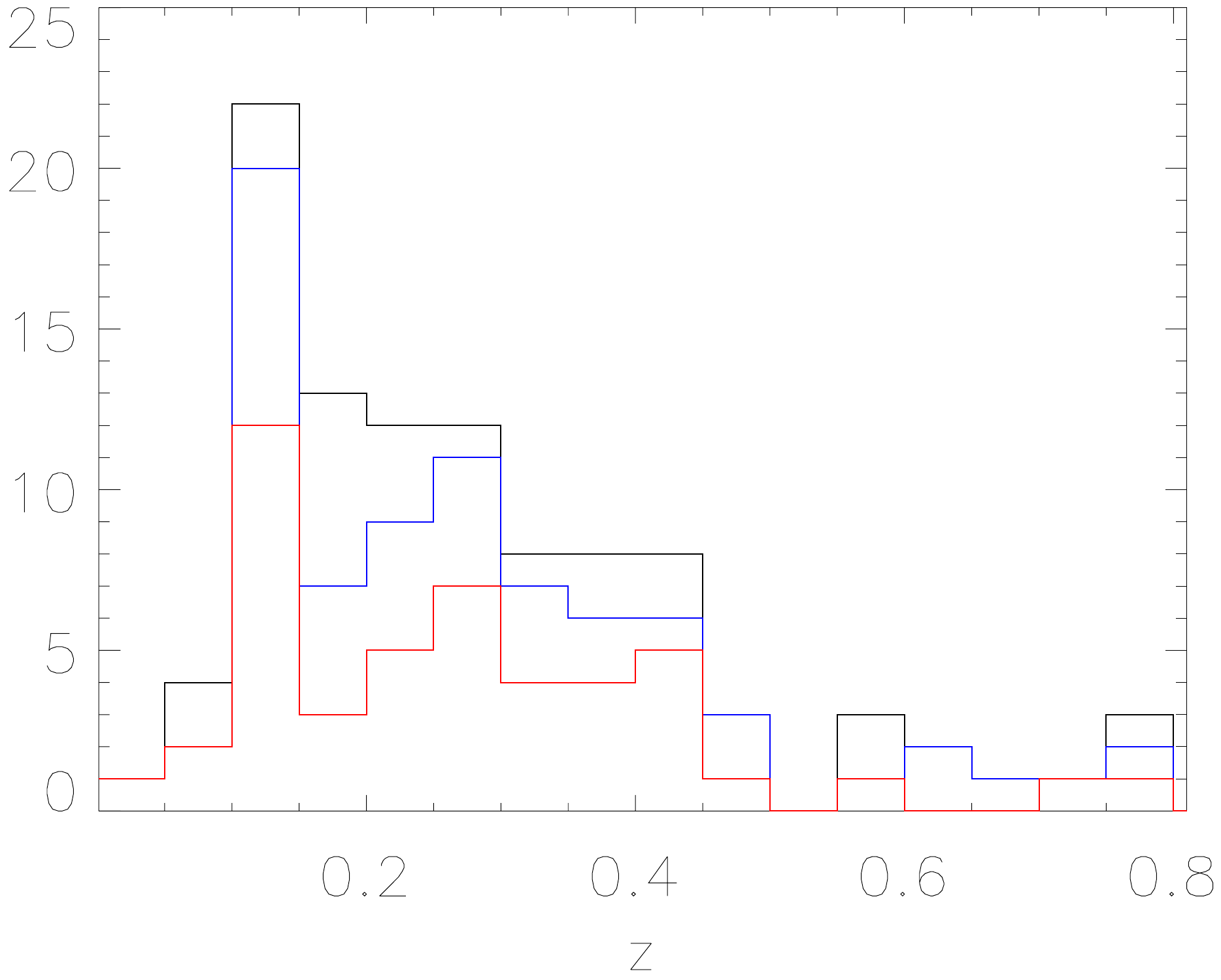}
\vskip-0.1in
\caption{Redshift distribution of RASS clusters with cuts of \ProbLam\ and \ProbS$>0.98$ in black, \ProbLam\ and \ProbS$>0.99$ in blue and \ProbLam\ and \ProbS$>0.999$ in red.}
\label{fig:redshiftdist}
\vskip-0.15in
\end{figure}

Depending on the applied cuts in \ProbLam\ and \ProbS\ we find up to 100 cluster candidates out of 1241 RASS sources in the DES-SV footprint. The redshifts range from $z=0.05$ to $z=0.79$.  Fig.~\ref{fig:redshiftdist} shows the distribution of RASS sources in redshift for different cuts in \ProbLam\ and \ProbS.  Independent of the cut the cluster with the highest redshift stays in the sample, indicating that this is a robust cluster detection.  A prominent feature in the redshift distribution is visible at redshifts  $0.1<z<0.15$. This is likely a combination of the lower luminosity (mass) threshold corresponding to the 2XRS flux limit at low redshifts and true large scale structure, which is also visible in Fig.~\ref{fig:footprint}. RASS candidates exceeding \ProbLam\ and \ProbS$>0.98$ and \ProbStar$<0.1$ can be found in Table~\ref{tab:main_table}. 

We find that the cluster candidate list for 2RXS contains some neighbouring sources that share the same optical counterpart.  Thus, we carry out an additional filtering step where we examine all 2RXS sources within 5$'$ of other sources and whose optical counterparts exhibit small differences in redshift $\Delta z<0.04$. Fourteen cases of multiple 2RXS sources associated with the same cluster are excluded, while two cases that may refer to distinct substructures are kept. Additionally, requiring the center of the optical counterpart to be within 1.5$'$ selects 12 of the 14 multiple associations while correctly keeping 2RXS sources associated with distinct substructures.  Further testing is needed to finalize the automation of this step, which will be important in studying the 2RXS catalog over very large solid angle. 

\begin{figure}
\vskip-0.1in
\includegraphics[keepaspectratio=true,width=\columnwidth]{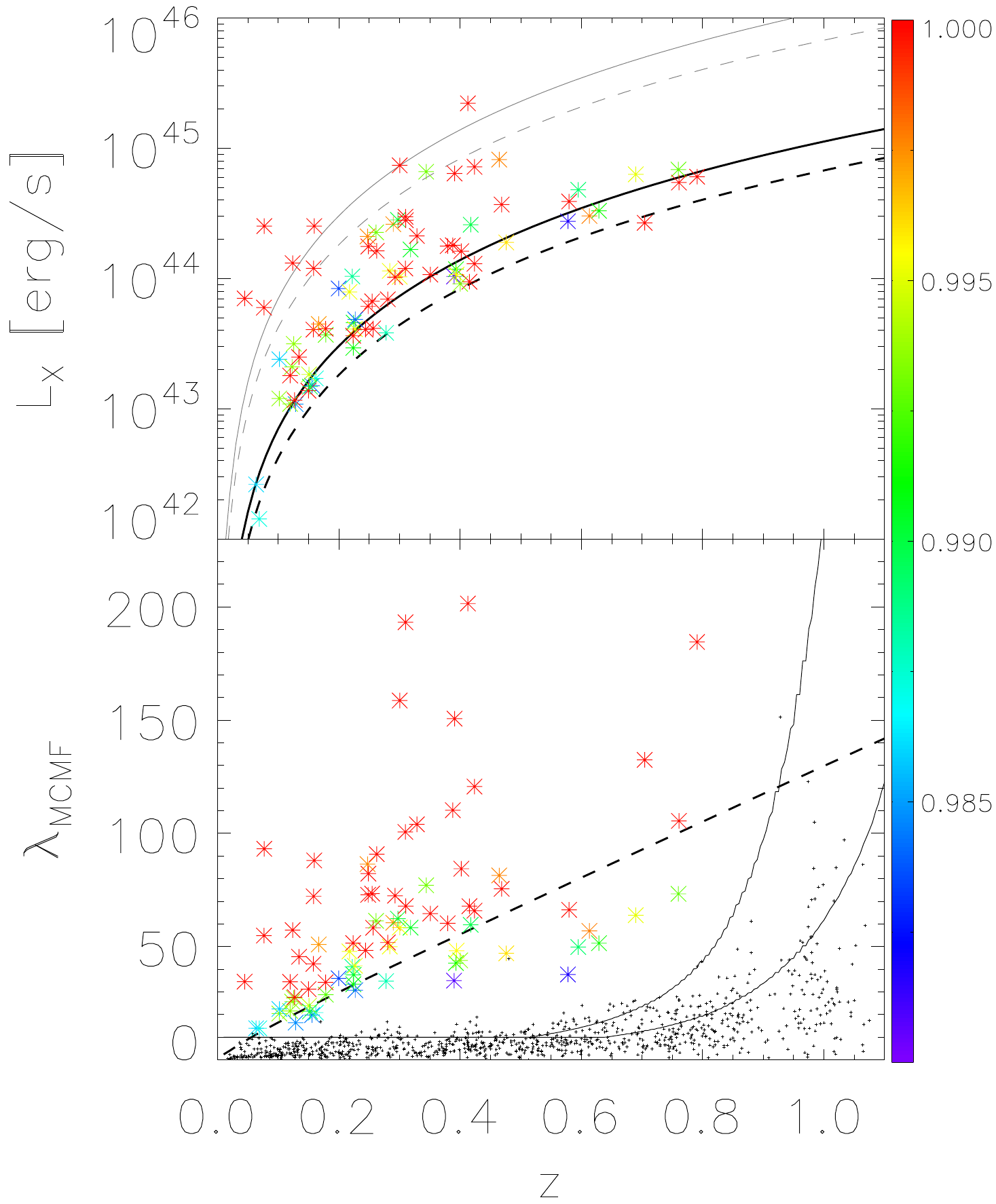}
\vskip-0.07in
\caption{The top panel contains the distribution of RASS+DES-SV clusters in luminosity--redshift space.  The thin continuous (dashed) grey lines shows the flux limit of REFLEX (REFLEX II). The thick black lines show a flux limit 1/10 of these, indicating that our sample pushes to these flux limits.  Below is the distribution of RASS+DES-SV clusters in richness--redshift space. The dashed line indicates 1/10 REFLEX II flux limit converted in \LamMCMF\ using the luminosity \LamMCMF\ relation with \ProbCut$=0.99$.  The solid curves indicate the $S=3$ limit for i=21.5 mag and i=22.1 mag. Black dots are 2RXS sources with $S<3$.  }
\label{fig:limflux}
\vskip-0.15in
\end{figure}

As previously noted, the 2RXS catalog contains only sources that have detection likelihood parameter of at least 6.5, the additional cut of at least six source photons as in 1RXS was dropped.  This likelihood or signal to noise threshold translates into a position dependent flux limit, depending on the local exposure time, galactic hydrogen column and observed background.  Given the large variation in exposure times in the region we study there is a correspondingly large variation in the effective flux limit of the survey. This causes a luminosity-redshift distribution (Fig.~\ref{fig:limflux}) without a hard cut at a limiting flux. A flux cut, if preferred, can be applied as a post processing step.

\subsubsection{Photo-z Performance}
\begin{figure}
\vskip-0.15in
\includegraphics[angle=90,keepaspectratio=true,width=1.0\columnwidth]{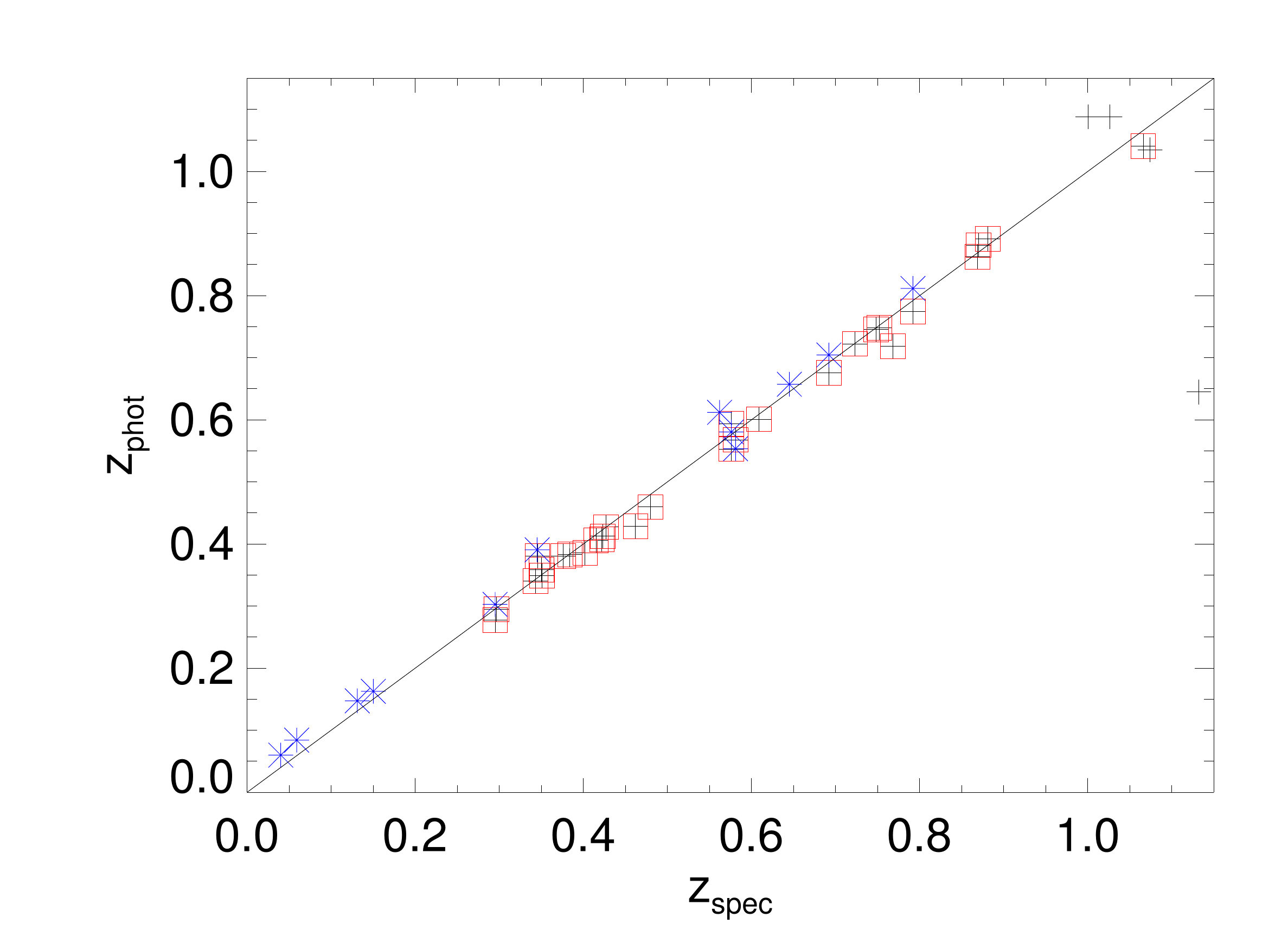}
\vskip-0.1in
\caption{Cluster photometric versus spectroscopic redshifts are plotted (black pluses) for the test sample. Clusters with \LamMCMF\ three times larger than the Poisson uncertainty are marked with red squares, and those in the DES-SV area are marked with blue points.
 } \label{fig:photperf}
 \vskip-0.15in
\end{figure}

To test and calibrate our photometric redshifts we used two approaches. We first use the same spectroscopic sample as in section~\ref{sec:colmag} to estimate a possible redshift dependent bias in our photometric redshifts and to measure the scatter of the redshift estimates about the true redshifts. Fig~\ref{fig:photperf} contains a plot of our photometric redshifts against the spectroscopic redshifts for this sample.  To be as close as possible to the measurement method in the DES-SV area, we use the masses listed in \citet{bleem15} and the luminosity mass relation to create a virtual count rate, that we then use in our pipeline. Using the SPT based masses has the advantage of providing a homogeneous mass proxy over that sample. This allows us to probe the photo-z performance of the MCMF beyond the limitation of the depth of the RASS catalog. We find and account for a slight redshift dependent offset between spectroscopic and photometric redshifts of $z_{spec}=z_{phot}*0.941+0.01$ for this sample of 29 clusters with spectroscopic redshifts. The redshift dependency seen here of $0.941\pm0.017$ might be surprising, given that our RS models are tuned using the same sample. But this redshift dependency can be partially explained due to the fact that the aperture, the used magnitude limit as well as the width of the RS model, change at each step in redshift. The aperture is connected to the assumed luminosity at the given redshift, which increases with redshift. Similarly, the allowed magnitude and color range of the sources considered in our matched filter increases with redshift due to the evolution on $m^{*}$ and measurement errors. All this can lead to the inclusion of faint cluster sources that were fainter or have colors that were slightly off from the RS model. All these effects cause a tendency for \LamMCMF\ to peak at a redshift that is slightly higher than the redshift of the cluster, even though the RS model is correct at the true cluster redshift. 

We apply this correction and probe richnesses up to a redshift of $z=1.1$, expecting to get reliable redshifts out to $z\approx1$.  The latter restriction comes from the difficulty to find and fit a peak if there are not enough points probing the drop in \LamMCMF\ expected at redshifts beyond that of the cluster.  Using this sample of 29 clusters with $S>3$, we obtain a characteristic scatter $\sigma_{\Delta z/(1+z)}=0.010$, using an outlier resistant standard deviation 
\footnote{We use the IDL function robust\_sigma.pro from the IDL Astronomy Users Library, \url{http://www.idlastro.gsfc.nasa.gov}}. We note from our data that $S>2.5$ appears to be sufficient to obtain good photo-z estimates, but at this lower signal to noise it appears there are also outliers at redshifts $z>1$. 

We also check whether known clusters in the SPT-East field give similar performance.  We search the MCXC \citep{piffaretti11}, the BAX \citep{Sadat04} and the Abell \citep{abell58} catalogs and match the positions with our RASS catalog, using the position and count rate from 2RXS as input.  All told, we find twelve clusters that have signal to noise $S>2.5$. For those clusters we find a characteristic scatter of $\sigma_{\Delta z/(1+z)}=0.014$.

\begin{figure}
\includegraphics[angle=0,keepaspectratio=true,width=\columnwidth]{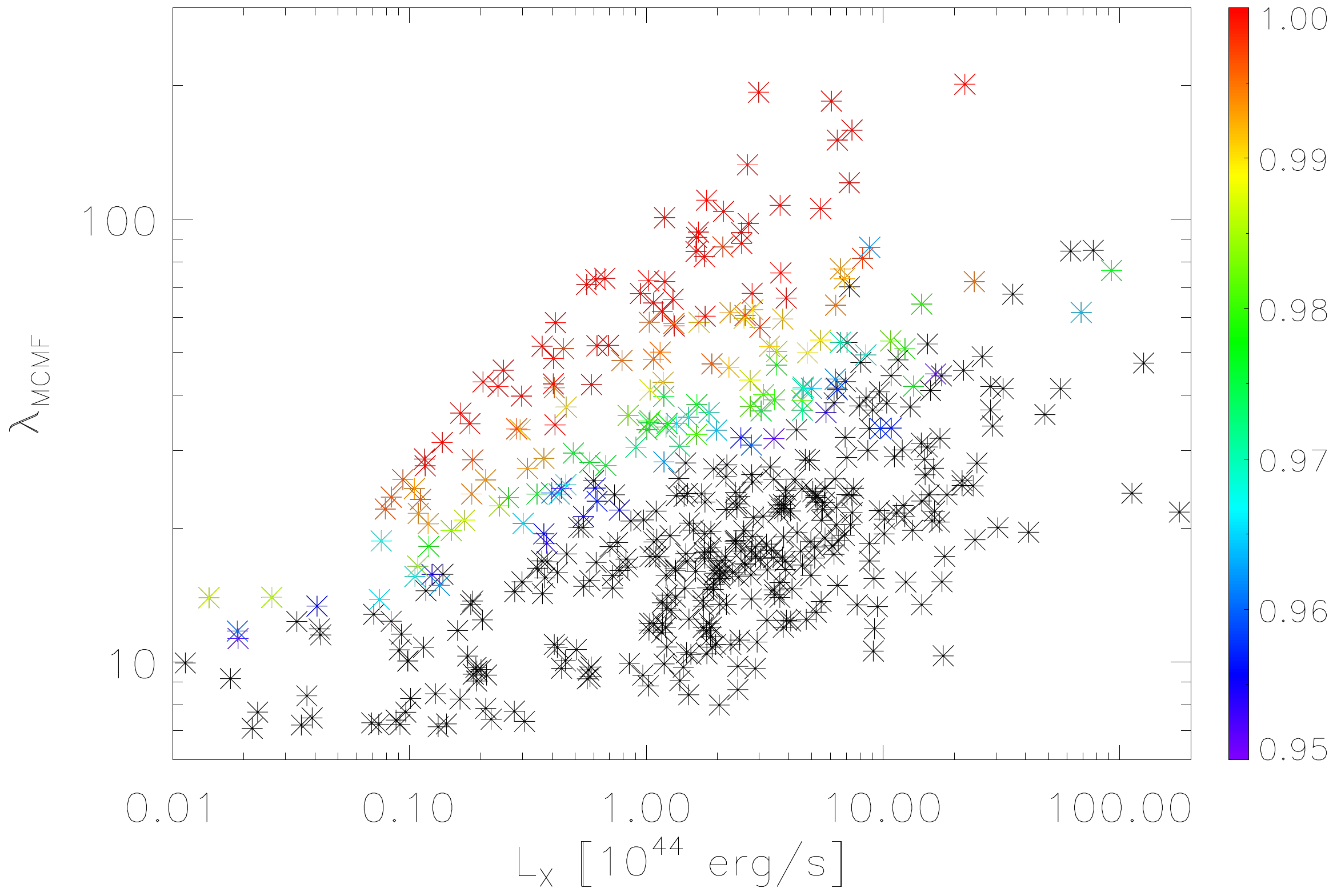}
\vskip-0.1in
\caption{Richness versus luminosity. All RASS sources with \LamMCMF\ based detection probability \ProbLam$<0.95$ and statistical signal to noise $S<2.5$ are shown in black. Those sources with \ProbLam$>0.95$ and $S>2.5$ are color coded according to \ProbLam.}
\label{fig:lumrichnessdist}
\vskip-0.15in
\end{figure}
%

 \begin{table*}
 	\centering
 	\caption{List of all 2RXS clusters with at least one peak in redshift exceeding \ProbLam\ and \ProbS$>0.98$ and \ProbStar$<0.1$. Multiple X-ray associations with the same optical cluster are excluded. Only the parameters related to the two peaks with highest value in \ProbLam\ are shown.}
 	\label{tab:main_table}
 	\setlength\tabcolsep{3.5pt}
 	\begin{tabular}{l c c c c c c c c c c c r r r r} 
 		\hline
 		Name & RA & DEC & \ProbLam$_1$ & \ProbLam$_2$ & \ProbS$_1$ &  \ProbS$_2$ & \ProbStar & $z_1$ & $z_2$ & $\sigma_{z,1}$ & $\sigma_{z,2}$ & \LamMCMF$_1$ & \LamMCMF$_2$ & $\sigma_{\lambda1}$  & $\sigma_{\lambda2}$ \\
 		\hline
2RXC J0418.3-5850 & 64.6230 & -58.8490 & 0.988 & 0.938 & 0.986 & 0.531 & 0.00 & 0.278 & 1.017 & 0.013 & 0.020 & 34.8 & 70.3 & 6.3 & 45.3 \\ 
2RXC J0420.6-5246 & 65.2350 & -52.7810 & 0.991 & 0.593 & 0.991 & 0.540 & 0.00 & 0.318 & 0.596 & 0.013 & 0.016 & 58.4 & 14.9 & 8.2 & 6.6 \\ 
2RXC J0423.5-5432 & 65.9700 & -54.5460 & 1.000 & 0.951 & 1.000 & 0.946 & 0.00 & 0.255 & 0.428 & 0.013 & 0.014 & 73.4 & 30.8 & 9.1 & 6.3 \\ 
2RXC J0424.3-5557 & 66.1130 & -55.9550 & 0.995 & 0.685 & 0.995 & 0.683 & 0.00 & 0.226 & 0.458 & 0.012 & 0.015 & 41.6 & 13.3 & 6.8 & 4.7 \\ 
2RXC J0425.4-6143 & 66.4340 & -61.7310 & 1.000 & 0.000 & 1.000 & 0.000 & 0.00 & 0.704 & - & 0.017 & - & 132.5 & - & 14.7 & - \\ 
2RXC J0426.2-5517 & 66.5980 & -55.3000 & 0.990 & 0.723 & 0.984 & 0.774 & 0.00 & 0.417 & 0.053 & 0.014 & 0.011 & 59.6 & 5.3 & 8.6 & 2.3 \\ 
2RXC J0426.4-4545 & 66.6830 & -45.7570 & 1.000 & 0.000 & 1.000 & 0.000 & 0.00 & 0.281 & - & 0.013 & - & 51.8 & - & 7.8 & - \\ 
2RXC J0426.5-6003 & 66.6960 & -60.0500 & 0.986 & 0.315 & 0.981 & 0.338 & 0.00 & 0.064 & 0.461 & 0.011 & 0.015 & 14.0 & 6.2 & 4.0 & 3.5 \\ 
2RXC J0428.4-5349 & 67.1750 & -53.8280 & 0.993 & 0.950 & 0.993 & 0.965 & 0.00 & 0.261 & 0.048 & 0.013 & 0.010 & 61.4 & 16.2 & 8.4 & 4.1 \\ 
2RXC J0428.6-6019 & 67.2480 & -60.3210 & 1.000 & 0.664 & 1.000 & 0.678 & 0.00 & 0.219 & 0.531 & 0.012 & 0.015 & 71.1 & 13.0 & 8.6 & 5.1 \\ 
2RXC J0429.1-6020 & 67.3120 & -60.3490 & 1.000 & 0.425 & 1.000 & 0.439 & 0.00 & 0.223 & 0.531 & 0.012 & 0.015 & 51.6 & 9.1 & 7.4 & 4.5 \\ 
2RXC J0430.2-6127 & 67.5820 & -61.4530 & 1.000 & 0.977 & 1.000 & 0.985 & 0.00 & 0.076 & 0.771 & 0.011 & 0.018 & 54.9 & 76.5 & 7.6 & 14.7 \\ 
2RXC J0430.4-5336 & 67.6660 & -53.6150 & 1.000 & 0.900 & 1.000 & 0.749 & 0.00 & 0.045 & 0.991 & 0.010 & 0.020 & 34.6 & 76.3 & 6.2 & 39.9 \\ 
2RXC J0430.5-5737 & 67.7010 & -57.6250 & 0.995 & 0.076 & 0.993 & 0.066 & 0.00 & 0.284 & 0.077 & 0.013 & 0.011 & 50.0 & 0.6 & 7.6 & 1.2 \\ 
2RXC J0431.1-6246 & 67.8060 & -62.7710 & 0.991 & 0.505 & 0.993 & 0.573 & 0.00 & 0.224 & 0.057 & 0.012 & 0.011 & 37.6 & 2.6 & 6.3 & 1.6 \\ 
2RXC J0431.1-6033 & 67.8090 & -60.5550 & 0.998 & 0.755 & 0.998 & 0.947 & 0.00 & 0.613 & 0.894 & 0.016 & 0.019 & 56.9 & 25.6 & 8.6 & 8.3 \\ 
2RXC J0431.3-6126 & 67.8650 & -61.4350 & 1.000 & 0.928 & 1.000 & 0.903 & 0.00 & 0.077 & 0.129 & 0.011 & 0.011 & 93.2 & 18.5 & 10.1 & 5.4 \\ 
2RXC J0432.2-5942 & 68.0650 & -59.7040 & 0.984 & 0.552 & 0.990 & 0.562 & 0.00 & 0.156 & 0.291 & 0.012 & 0.013 & 19.8 & 6.3 & 4.6 & 3.1 \\ 
2RXC J0432.6-4549 & 68.2430 & -45.8330 & 1.000 & 0.770 & 1.000 & 0.794 & 0.00 & 0.248 & 0.136 & 0.012 & 0.011 & 72.9 & 7.5 & 9.1 & 2.9 \\ 
2RXC J0433.3-5511 & 68.3580 & -55.1870 & 0.995 & 0.549 & 0.995 & 0.575 & 0.00 & 0.690 & 0.496 & 0.017 & 0.015 & 63.9 & 10.9 & 10.6 & 4.4 \\ 
2RXC J0434.1-4551 & 68.5280 & -45.8530 & 0.987 & 0.000 & 0.985 & 0.000 & 0.00 & 0.162 & - & 0.012 & - & 20.9 & - & 4.9 & - \\ 
2RXC J0434.1-4943 & 68.5440 & -49.7310 & 1.000 & 0.532 & 1.000 & 0.792 & 0.00 & 0.244 & 0.033 & 0.012 & 0.010 & 48.4 & 2.0 & 7.0 & 1.0 \\ 
2RXC J0434.6-4726 & 68.7290 & -47.4430 & 1.000 & 0.102 & 0.998 & 0.066 & 0.00 & 0.310 & 0.077 & 0.013 & 0.011 & 67.9 & 1.0 & 9.1 & 1.9 \\ 
2RXC J0435.4-5801 & 68.9090 & -58.0220 & 0.989 & 0.813 & 0.989 & 0.763 & 0.00 & 0.222 & 0.457 & 0.012 & 0.015 & 41.1 & 19.8 & 6.9 & 6.1 \\ 
2RXC J0435.5-4540 & 68.9560 & -45.6710 & 0.998 & 0.153 & 0.998 & 0.167 & 0.00 & 0.167 & 0.027 & 0.012 & 0.010 & 51.0 & 0.8 & 7.6 & 1.0 \\ 
2RXC J0436.3-6032 & 69.1270 & -60.5380 & 0.996 & 0.762 & 0.996 & 0.932 & 0.00 & 0.476 & 0.781 & 0.015 & 0.018 & 47.0 & 20.0 & 7.3 & 6.5 \\ 
2RXC J0437.1-4731 & 69.3000 & -47.5280 & 0.990 & 0.622 & 0.990 & 0.654 & 0.00 & 0.297 & 0.179 & 0.013 & 0.012 & 62.4 & 10.0 & 8.8 & 3.9 \\ 
2RXC J0437.3-5753 & 69.3620 & -57.8910 & 1.000 & 0.897 & 0.995 & 0.892 & 0.00 & 0.379 & 0.598 & 0.014 & 0.016 & 60.3 & 25.9 & 8.4 & 7.0 \\ 
2RXC J0438.1-4559 & 69.5210 & -45.9840 & 0.997 & 0.000 & 0.992 & 0.000 & 0.00 & 0.154 & - & 0.012 & - & 33.6 & - & 6.5 & - \\ 
2RXC J0438.1-4858 & 69.5220 & -48.9770 & 0.991 & 0.470 & 0.989 & 0.495 & 0.00 & 0.224 & 0.544 & 0.012 & 0.015 & 33.4 & 8.9 & 6.2 & 4.2 \\ 
2RXC J0438.2-5419 & 69.5650 & -54.3270 & 1.000 & 0.707 & 1.000 & 0.733 & 0.00 & 0.424 & 0.575 & 0.014 & 0.016 & 120.7 & 24.5 & 11.9 & 7.8 \\ 
2RXC J0438.2-4555 & 69.5910 & -45.9210 & 0.995 & 0.415 & 0.987 & 0.185 & 0.00 & 0.151 & 1.006 & 0.012 & 0.020 & 23.9 & 22.9 & 5.8 & 21.1 \\ 
2RXC J0438.3-4906 & 69.6170 & -49.1090 & 0.998 & 0.000 & 0.998 & 0.000 & 0.00 & 0.247 & - & 0.012 & - & 86.5 & - & 10.0 & - \\ 
2RXC J0439.1-4600 & 69.8060 & -46.0140 & 0.993 & 0.560 & 0.986 & 0.619 & 0.00 & 0.344 & 0.128 & 0.013 & 0.011 & 77.1 & 6.5 & 10.0 & 3.0 \\ 
2RXC J0439.6-4842 & 70.0000 & -48.7120 & 1.000 & 0.347 & 1.000 & 0.611 & 0.00 & 0.256 & 0.040 & 0.013 & 0.010 & 58.3 & 1.5 & 7.9 & 1.0 \\ 
2RXC J0440.5-4657 & 70.2080 & -46.9630 & 1.000 & 0.752 & 1.000 & 0.667 & 0.00 & 0.329 & 0.570 & 0.013 & 0.016 & 104.0 & 19.3 & 10.1 & 7.0 \\ 
2RXC J0440.5-4743 & 70.2180 & -47.7210 & 1.000 & 0.107 & 1.000 & 0.059 & 0.00 & 0.310 & 0.567 & 0.013 & 0.016 & 193.2 & 5.2 & 10.8 & 6.0 \\ 
2RXC J0440.6-4510 & 70.2480 & -45.1710 & 1.000 & 0.408 & 0.997 & 0.405 & 0.00 & 0.158 & 0.703 & 0.012 & 0.017 & 42.4 & 12.7 & 7.2 & 7.4 \\ 
2RXC J0441.0-4830 & 70.2620 & -48.5070 & 1.000 & 0.508 & 1.000 & 0.508 & 0.00 & 0.249 & 0.121 & 0.012 & 0.011 & 82.2 & 4.8 & 9.7 & 2.6 \\ 
2RXC J0441.1-4502 & 70.3080 & -45.0490 & 0.998 & 0.825 & 0.998 & 0.614 & 0.00 & 0.165 & 0.968 & 0.012 & 0.020 & 42.3 & 45.5 & 7.2 & 29.7 \\ 
2RXC J0446.1-5142 & 71.5330 & -51.7130 & 0.998 & 0.504 & 0.998 & 0.485 & 0.00 & 0.464 & 0.312 & 0.015 & 0.013 & 81.4 & 9.8 & 10.2 & 4.5 \\ 
2RXC J0446.6-4834 & 71.7390 & -48.5680 & 0.993 & 0.464 & 0.988 & 0.451 & 0.00 & 0.760 & 1.024 & 0.018 & 0.020 & 73.3 & 26.1 & 13.9 & 19.0 \\ 
2RXC J0447.2-5055 & 71.8510 & -50.9230 & 1.000 & 0.599 & 1.000 & 0.647 & 0.00 & 0.402 & 0.235 & 0.014 & 0.012 & 84.4 & 7.3 & 9.6 & 3.2 \\ 
2RXC J0447.3-5145 & 71.8910 & -51.7540 & 1.000 & 0.912 & 1.000 & 0.902 & 0.00 & 0.415 & 0.556 & 0.014 & 0.016 & 67.9 & 22.9 & 8.7 & 6.2 \\ 
2RXC J0447.5-5045 & 71.9410 & -50.7570 & 1.000 & 0.554 & 1.000 & 0.515 & 0.00 & 0.580 & 0.795 & 0.016 & 0.018 & 66.3 & 15.7 & 9.5 & 9.3 \\ 
2RXC J0448.3-4540 & 72.1250 & -45.6700 & 0.995 & 0.889 & 0.995 & 0.971 & 0.00 & 0.300 & 0.535 & 0.013 & 0.015 & 58.4 & 26.5 & 8.3 & 5.2 \\ 
2RXC J0449.1-4900 & 72.2960 & -49.0120 & 1.000 & 0.257 & 1.000 & 0.323 & 0.00 & 0.791 & 0.235 & 0.018 & 0.012 & 184.5 & 3.5 & 22.8 & 2.3 \\ 
2RXC J0449.2-4815 & 72.3490 & -48.2650 & 0.982 & 0.185 & 0.988 & 0.109 & 0.00 & 0.578 & 0.727 & 0.016 & 0.017 & 37.7 & 7.1 & 7.1 & 6.6 \\ 
2RXC J0449.3-5407 & 72.3560 & -54.1260 & 0.991 & 0.937 & 0.991 & 0.912 & 0.00 & 0.629 & 0.209 & 0.016 & 0.012 & 51.5 & 15.5 & 8.7 & 4.6 \\ 
2RXC J0449.6-4440 & 72.4800 & -44.6790 & 1.000 & 0.484 & 1.000 & 0.378 & 0.00 & 0.159 & 0.605 & 0.012 & 0.016 & 88.1 & 17.7 & 9.9 & 8.7 \\ 
2RXC J0451.5-5058 & 72.9620 & -50.9710 & 1.000 & 0.717 & 0.998 & 0.720 & 0.00 & 0.761 & 0.398 & 0.018 & 0.014 & 105.5 & 12.6 & 14.7 & 4.3 \\ 
2RXC J0451.5-4521 & 72.9690 & -45.3550 & 0.984 & 0.828 & 0.984 & 0.817 & 0.00 & 0.227 & 0.497 & 0.012 & 0.015 & 30.8 & 18.4 & 5.8 & 5.4 \\ 
2RXC J0500.5-5115 & 75.2290 & -51.2620 & 1.000 & 0.000 & 1.000 & 0.000 & 0.00 & 0.158 & - & 0.012 & - & 72.2 & - & 8.8 & - \\ 
2RXC J0500.6-6346 & 75.2420 & -63.7670 & 0.981 & 0.000 & 0.983 & 0.000 & 0.00 & 0.390 & - & 0.014 & - & 35.0 & - & 6.2 & - \\ 
2RXC J0503.4-5658 & 75.9030 & -56.9750 & 1.000 & 0.649 & 1.000 & 0.704 & 0.00 & 0.134 & 0.433 & 0.011 & 0.014 & 45.5 & 13.6 & 6.7 & 4.5 \\ 
2RXC J0504.0-4929 & 76.0100 & -49.4900 & 0.995 & 0.579 & 0.991 & 0.678 & 0.00 & 0.218 & 0.364 & 0.012 & 0.014 & 47.9 & 10.1 & 7.6 & 3.7 \\ 
2RXC J0505.3-6145 & 76.3610 & -61.7500 & 1.000 & 0.287 & 1.000 & 0.200 & 0.00 & 0.262 & 0.734 & 0.013 & 0.017 & 90.8 & 11.0 & 9.7 & 8.5 \\ 
 		\hline
 	\end{tabular}
 \end{table*} 

\begin{table*}
 	\centering
 	\contcaption{List of all 2RXS clusters with at least one peak in redshift exceeding \ProbLam\ and \ProbS$>0.98$ and \ProbStar$<0.1$. Multiple X-ray associations with the same optical cluster are excluded. Only the parameters related to the two peaks with highest value in \ProbLam\ are shown.}
 	\label{tab:example_table}
 	\setlength\tabcolsep{3.5pt}
 	\begin{tabular}{ l c c c c c c c c c c c r r r r} 
 		\hline
 		Name & RA & DEC & \ProbLam$_1$ & \ProbLam$_2$ & \ProbS$_1$ &  \ProbS$_2$ & \ProbStar\ & $z_1$ & $z_2$ & $\sigma_{z,1}$ & $\sigma_{z,2}$ & \LamMCMF$_1$ & \LamMCMF$_2$ & $\sigma_{\lambda1}$  & $\sigma_{\lambda2}$ \\
 		\hline
2RXC J0506.1-6310 & 76.5260 & -63.1700 & 1.000 & 0.923 & 1.000 & 0.907 & 0.00 & 0.293 & 0.171 & 0.013 & 0.012 & 72.5 & 13.6 & 8.6 & 4.2 \\ 
2RXC J0509.3-5640 & 77.3690 & -56.6740 & 0.989 & 0.605 & 0.996 & 0.603 & 0.00 & 0.595 & 0.279 & 0.016 & 0.013 & 49.8 & 7.8 & 8.4 & 3.5 \\ 
2RXC J0510.0-6118 & 77.5050 & -61.3050 & 1.000 & 0.572 & 1.000 & 0.667 & 0.00 & 0.388 & 0.503 & 0.014 & 0.015 & 110.2 & 11.2 & 10.5 & 4.2 \\ 
2RXC J0515.1-6159 & 78.8030 & -61.9950 & 0.994 & 0.904 & 0.994 & 0.891 & 0.00 & 0.126 & 0.230 & 0.011 & 0.012 & 27.3 & 19.2 & 5.6 & 5.2 \\ 
2RXC J0516.3-5431 & 79.1450 & -54.5180 & 1.000 & 0.127 & 1.000 & 0.146 & 0.00 & 0.300 & 0.010 & 0.013 & 0.010 & 158.7 & 0.6 & 13.1 & 1.0 \\ 
2RXC J0517.4-5602 & 79.4140 & -56.0410 & 0.993 & 0.792 & 0.988 & 0.805 & 0.00 & 0.178 & 0.090 & 0.012 & 0.011 & 28.8 & 6.7 & 6.0 & 2.7 \\ 
2RXC J0517.6-5834 & 79.4890 & -58.5740 & 1.000 & 0.754 & 1.000 & 0.397 & 0.00 & 0.111 & 0.120 & 0.011 & 0.011 & 42.2 & 9.7 & 6.6 & 6.2 \\ 
2RXC J0518.2-5601 & 79.6020 & -56.0190 & 1.000 & 0.477 & 1.000 & 0.609 & 0.00 & 0.351 & 0.515 & 0.014 & 0.015 & 64.6 & 8.5 & 8.4 & 3.6 \\ 
2RXC J0518.3-5616 & 79.6170 & -56.2700 & 0.994 & 0.000 & 0.990 & 0.000 & 0.00 & 0.123 & - & 0.011 & - & 25.7 & - & 5.4 & - \\ 
2RXC J0518.5-5720 & 79.6910 & -57.3450 & 0.998 & 0.341 & 1.000 & 0.211 & 0.00 & 0.290 & 0.949 & 0.013 & 0.019 & 60.5 & 19.0 & 8.1 & 16.9 \\ 
2RXC J0519.2-6109 & 79.8430 & -61.1510 & 0.986 & 0.000 & 0.989 & 0.000 & 0.00 & 0.102 & - & 0.011 & - & 22.5 & - & 4.8 & - \\ 
2RXC J0524.5-5550 & 81.1990 & -55.8420 & 0.983 & 0.383 & 0.983 & 0.492 & 0.00 & 0.200 & 0.391 & 0.012 & 0.014 & 36.0 & 7.9 & 6.6 & 3.5 \\ 
2RXC J0524.6-5814 & 81.2430 & -58.2380 & 1.000 & 0.082 & 1.000 & 0.082 & 0.00 & 0.310 & 0.741 & 0.013 & 0.017 & 100.6 & 5.2 & 10.4 & 6.3 \\ 
2RXC J0525.6-6212 & 81.4800 & -62.2110 & 1.000 & 0.189 & 1.000 & 0.176 & 0.00 & 0.150 & 0.556 & 0.011 & 0.016 & 31.3 & 5.5 & 5.9 & 4.3 \\ 
2RXC J0528.4-6218 & 82.1720 & -62.3080 & 0.997 & 0.000 & 0.997 & 0.000 & 0.00 & 0.117 & - & 0.011 & - & 22.1 & - & 4.9 & - \\ 
2RXC J0528.6-6214 & 82.2420 & -62.2420 & 0.985 & 0.068 & 0.991 & 0.082 & 0.00 & 0.129 & 0.235 & 0.011 & 0.012 & 16.4 & 1.3 & 4.1 & 2.0 \\ 
2RXC J0534.3-6238 & 83.6320 & -62.6490 & 1.000 & 0.124 & 1.000 & 0.032 & 0.00 & 0.128 & 0.131 & 0.011 & 0.011 & 28.7 & 1.4 & 5.4 & 5.4 \\ 
2RXC J0536.2-5847 & 84.0960 & -58.7860 & 0.996 & 0.538 & 0.994 & 0.540 & 0.00 & 0.394 & 0.706 & 0.014 & 0.017 & 48.2 & 12.3 & 7.3 & 6.5 \\ 
2RXC J0536.4-6251 & 84.1590 & -62.8570 & 1.000 & 0.857 & 1.000 & 0.901 & 0.00 & 0.178 & 0.084 & 0.012 & 0.011 & 34.3 & 7.6 & 6.2 & 2.6 \\ 
2RXC J0538.3-6031 & 84.6190 & -60.5260 & 0.994 & 0.962 & 0.989 & 0.127 & 0.00 & 0.400 & 1.064 & 0.014 & 0.021 & 43.9 & 86.3 & 7.0 & 80.4 \\ 
2RXC J0538.4-6046 & 84.6460 & -60.7720 & 0.988 & 0.434 & 0.986 & 0.217 & 0.87 & 0.548 & 0.826 & 0.015 & 0.018 & 43.2 & 13.0 & 8.2 & 11.3 \\ 
2RXC J0540.5-6144 & 85.2200 & -61.7380 & 1.000 & 0.000 & 1.000 & 0.000 & 0.00 & 0.124 & - & 0.011 & - & 51.7 & - & 7.4 & - \\ 
2RXC J0541.0-6251 & 85.2530 & -62.8670 & 1.000 & 0.087 & 1.000 & 0.082 & 0.00 & 0.123 & 0.810 & 0.011 & 0.018 & 39.8 & 7.3 & 6.2 & 8.4 \\ 
2RXC J0541.1-6414 & 85.2860 & -64.2340 & 0.991 & 0.226 & 0.989 & 0.307 & 0.00 & 0.393 & 0.542 & 0.014 & 0.015 & 42.7 & 5.8 & 6.9 & 3.5 \\ 
2RXC J0541.2-6254 & 85.3350 & -62.9150 & 0.997 & 0.000 & 0.997 & 0.000 & 0.14 & 0.110 & - & 0.011 & - & 23.5 & - & 4.9 & - \\ 
2RXC J0542.4-6154 & 85.6510 & -61.9040 & 1.000 & 0.167 & 1.000 & 0.167 & 0.00 & 0.124 & 0.367 & 0.011 & 0.014 & 57.3 & 5.5 & 7.6 & 4.1 \\ 
2RXC J0543.0-6219 & 85.7650 & -62.3200 & 1.000 & 0.646 & 1.000 & 0.681 & 0.00 & 0.468 & 0.064 & 0.015 & 0.011 & 75.5 & 3.7 & 9.1 & 2.1 \\ 
2RXC J0549.2-6205 & 87.3300 & -62.0870 & 1.000 & 0.772 & 1.000 & 0.717 & 0.00 & 0.413 & 0.135 & 0.014 & 0.011 & 201.5 & 10.7 & 15.1 & 4.3 \\ 
2RXC J0552.0-6243 & 88.0050 & -62.7200 & 0.990 & 0.928 & 0.983 & 0.903 & 0.00 & 0.152 & 0.417 & 0.012 & 0.014 & 21.4 & 23.7 & 5.3 & 5.9 \\ 
2RXC J0552.3-6427 & 88.1120 & -64.4670 & 1.000 & 0.533 & 1.000 & 0.572 & 0.00 & 0.424 & 0.261 & 0.014 & 0.013 & 65.9 & 5.6 & 8.9 & 2.8 \\ 
2RXC J0555.3-6406 & 88.8670 & -64.1050 & 1.000 & 0.970 & 1.000 & 0.970 & 0.00 & 0.391 & 0.205 & 0.014 & 0.012 & 150.7 & 26.7 & 12.7 & 5.6 \\ 
2RXC J0602.1-6445 & 90.5350 & -64.7560 & 0.987 & 0.499 & 0.987 & 0.618 & 0.00 & 0.069 & 0.422 & 0.011 & 0.014 & 13.9 & 8.2 & 3.7 & 3.2 \\ 
 		\hline
 	\end{tabular}
 \end{table*} 
 
\subsubsection{Scaling Relations}

We use the scaling relation between \LamMCMF\ and luminosity or mass as a tool to understand selection effects and to test the basic properties of the sample. Contamination of the cluster sample by non-clusters will affect intrinsic scatter and other basic scaling relation parameters. With increasing cut in \ProbLam and with the corresponding increasing in sample purity, we expect the intrinsic scatter to fall and scaling relation parameters to stabilize once a clean sample is obtained.

A well behaved scaling between \LamMCMF\ and mass proxies can be expected by construction of our richness estimator. However this is not crucial for the success of our confirmation process since \ProbLam\, the main parameter for confirmation, completely works self consistent in \LamMCMF\ space. This makes \ProbLam\ more robust against changes in the scaling relations such as redshift evolution, which would affect all \LamMCMF\ in the same way, leaving \ProbLam\ unchanged. Due to its design as a conformation tool, there will always be other mass proxies available that are based on the method the original catalog is drawn from. Such proxies, like the X-ray luminosity in this work, will be considered as the main mass proxy for future analysis. The richness \LamMCMF\ might be used as a secondary mass proxy if correlations are taken into account. 

We stress here that the luminosities given here are based on a simplistic model with a fixed temperature and the X-ray fluxes typically within a fixed aperture of 5\,arcmin diameter.  Masses are derived from these luminosities using the \citet{mantz10b} scaling relation assuming that the measured flux corresponds to a measurement within $r_{500}$.  Masses of the SPT cluster sample are taken from \citet{bleem15}, while the \LamMCMF\ is derived using a mock X-ray count rate as input. This count rate is derived from the mass, using the spectroscopic redshift and the mass-luminosity scaling relation with fixed temperature. These assumptions have to be taken into account when comparing to other publications, but in general these have only a small impact on MCMF performance. 
 
 We use here the assumption that the scaling relations follow a simple power law of the form
\begin{equation}
	\expectation{\lambda_\mathrm{MCMF}'(x)}=Ax^B,
\end{equation}
where $\lambda_\mathrm{MCMF}'$  is the scaled richness, $A$ is the normalization, $B$ is the power law index and $x$ is either the scaled luminosity or mass.  For the pivot of the scaling relations we choose $10^{44}$\,erg/s for the luminosity and $4\times10^{14}$\,M$_\odot$ for the mass based scaling relation. We scale our richness by dividing $\lambda_\mathrm{MCMF}$ by 90. 

We fit our scaling relation model to the data using a maximum-likelihood method described in \citet{Kelly07} to find the best fit power law to our data, including log-normal intrinsic scatter, accounting for measurement errors in the x and y direction and the Malmquist-bias. We used the implementation described in \citet{Sommer14} that directly fits a power law to the data instead of transferring the problem to log-log space to use the original implementation given in \citet{Kelly07}. The individual fit results for different selection cuts are listed in Table~\ref{tab:scaling}. Intrinsic scatter is given in log-space using the natural logarithm.

Fig.~\ref{fig:lumrichnessdist} contains a plot of the \LamMCMF\ and X-ray luminosity distribution of the sample, where we have included not only the color-coded systems with high probability of being real \ProbLam$>0.95$ but also (in black) those systems that include contamination from false superpositions.  There is no clear separation in the distribution, although there are many systems piled up in the low significance black points.  This indicates that the scaling relation parameters will be sensitive to the significance cuts applied to the data.

\begin{figure}
\includegraphics[keepaspectratio=true,width=\columnwidth]{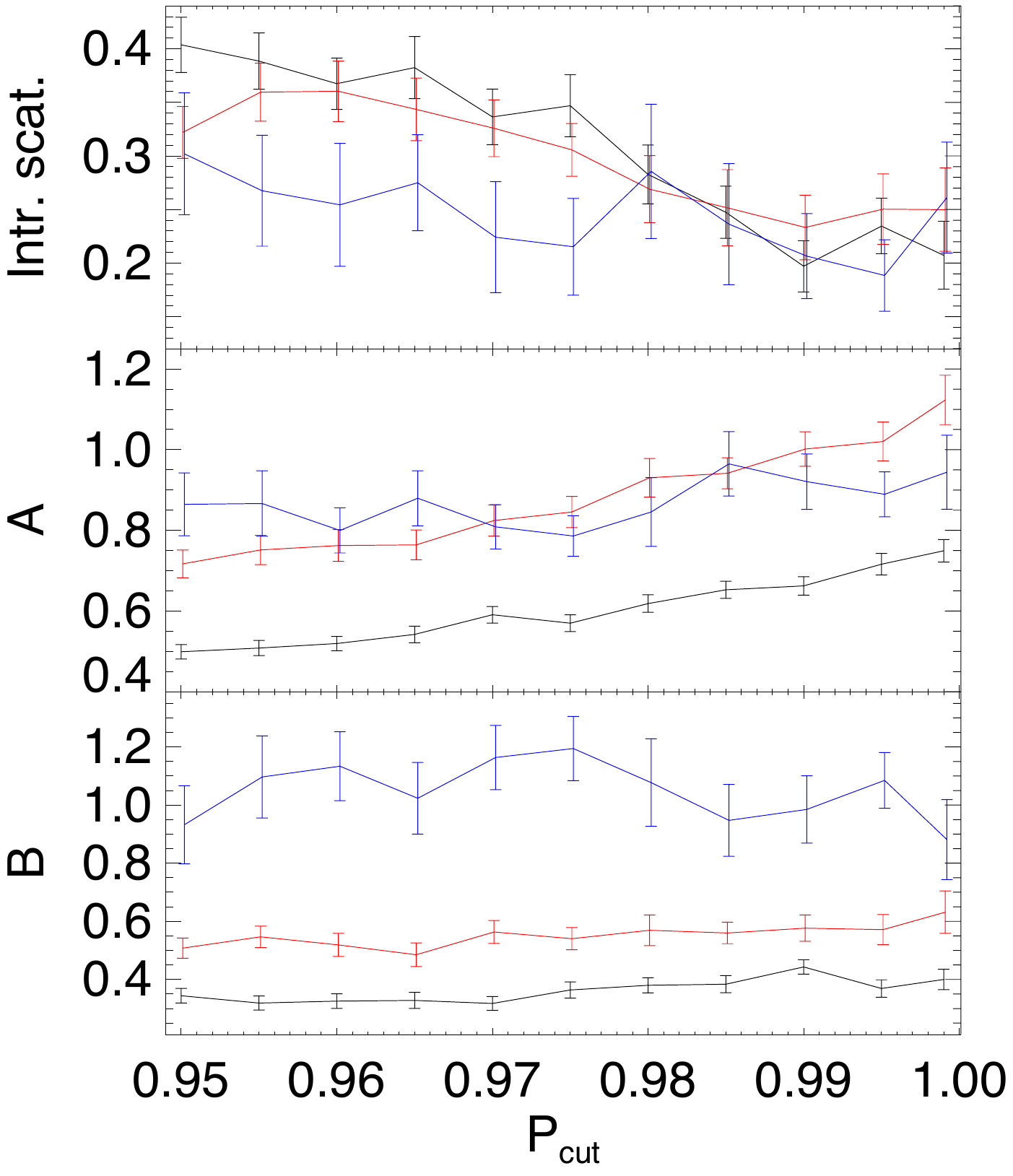}
\vskip-0.1in
\caption{Intrinsic scatter, normalization (A) and power law index (B) of the L-$\lambda_\mathrm{MCMF}'$ (black) and M-$\lambda_\mathrm{MCMF}'$ (red) scaling relation of RASS cluster candidates and M-$\lambda_\mathrm{MCMF}'$ scaling relation of SPT clusters (blue) against cut in \ProbLam\ and \ProbS. In case of SPT we adopt a lower cut in S corresponding to the minimum S found for RASS clusters at this cut in  \ProbLam\ and \ProbS.}
\label{fig:intrscat}
\vskip-0.15in
\end{figure}

In Fig.~\ref{fig:intrscat} is a plot of the dependency of the best fit scaling relation parameters on the selection parameter \ProbCut, which is the minimum value in either \ProbLam\ or \ProbS\ that is applied in the sample selection.  There is a suggestion that the intrinsic scatter shrinks with increasing \ProbCut, which is increasing sample purity.  For the \LamMCMF--mass relation the intrinsic scatter becomes constant at \ProbCut$=0.985$. For the \LamMCMF--luminosity relation the intrinsic scatter is constantly decreasing with increasing \ProbCut, although the trend may not be significant for cut values above 0.985.  But broadly speaking this analysis suggests that with sufficiently high \ProbCut\ values one obtains a sample that exhibits scaling relations that are insensitive to the exact value of the cut.

\begin{figure} 
\includegraphics[keepaspectratio=true,width=\columnwidth]{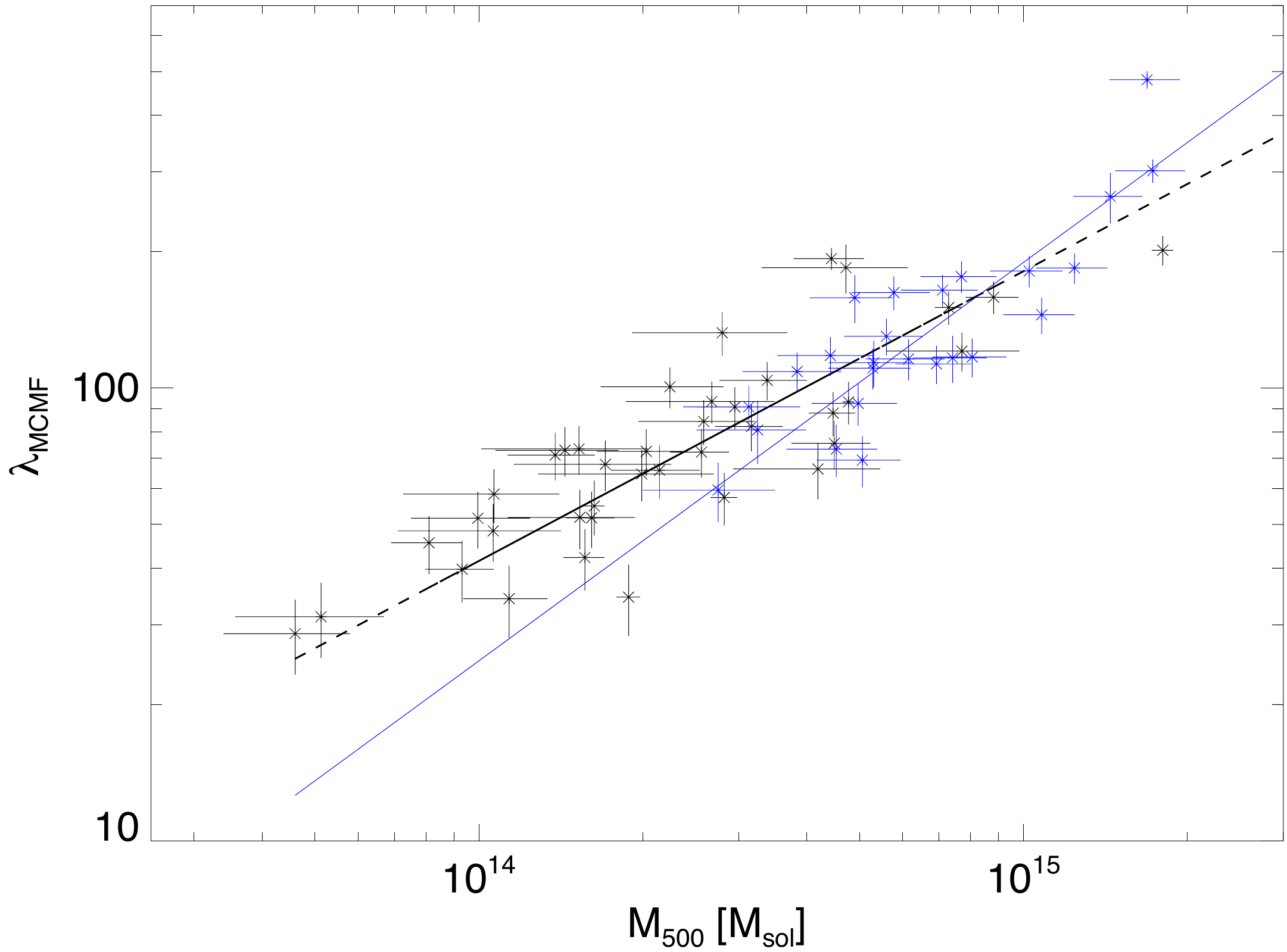}
\vskip-0.1in
\caption{Richness-mass relation: Distribution of 2RXS sources with \ProbLam\ and \ProbS\  $>0.999$ are shown in black, SPT clusters from the spectroscopic sample are shown in blue. Best fit scaling relations for the corresponding cut are shown as black dashed and blue continuous lines.}
\label{fig:scaling}
\vskip-0.15in
\end{figure}

While the slopes of the scaling relations seem to be robust against the chosen \ProbCut, there is a somewhat different behavior exhibited by the SPT and RASS based \LamMCMF--mass relations.  Fig.~\ref{fig:scaling} shows these samples plotted, providing some insights into the slope differences.
Overall, the RASS sample seems to be a reasonable extension of the SPT sample towards lower masses.

\begin{table*}
\centering
\caption{Dependency of scaling relation parameters on the lower limit \ProbCut\ applied to \ProbLam\ and \ProbS, for the \LamMCMF--luminosity relation (L), \LamMCMF--mass relation (M) of 2RXS sources and for the \LamMCMF--mass relation for SPT clusters with spectroscopic redshifts (S). $\mathrm{S}_{\mathrm{min}}$ lists the lowest value of S in that sample, N the number of cluster candidates and Cont. the expected contamination by random superpositions based on Eq.~\ref{eq:contamination}}.
\label{tab:scaling}
\setlength\tabcolsep{3.0pt}
\begin{tabular}{c c c c c c c c c c c c c c c c c c c c c c} 
 \hline
  & \multicolumn{6}{c}{\LamMCMF--luminosity} & \multicolumn{6}{c}{\LamMCMF--mass 2RXS} & \multicolumn{6}{c}{\LamMCMF--mass SPT} \\
\ProbCut & $A_{\mathrm{L}}$ & $eA_{\mathrm{L}}$ & $B_{\mathrm{L}}$ & $eB_{\mathrm{L}}$ & $\sigma_{\mathrm{int,L}}$ & $e\sigma_{\mathrm{int,L}}$ & $A_{\mathrm{M}}$ & $eA_{\mathrm{M}}$ & $B_{\mathrm{M}}$ & $eB_{\mathrm{M}}$ & $\sigma_{\mathrm{int,M}}$ & $e\sigma_{\mathrm{int,M}}$ & $A_{\mathrm{S}}$ & $eA_{\mathrm{S}}$ & $B_{\mathrm{S}}$ & $eB_{\mathrm{S}}$ & $\sigma_{\mathrm{int,S}}$ & $e\sigma_{\mathrm{int,S}}$ & $\mathrm{S}_{\mathrm{min}}$ & N & Cont. \\
\hline
0.950 & 0.50 & 0.02 & 0.34 & 0.03 & 0.40 & 0.03 & 0.72 & 0.03 & 0.51 & 0.03 & 0.32 & 0.02 & 0.86 & 0.08 & 0.93 & 0.13 & 0.30 & 0.06 & 3.5 & 134 & 43\% \\
0.955 & 0.51 & 0.02 & 0.32 & 0.02 & 0.39 & 0.03 & 0.75 & 0.04 & 0.55 & 0.04 & 0.36 & 0.03 & 0.87 & 0.08 & 1.10 & 0.14 & 0.27 & 0.05 & 3.5 & 128 & 41\% \\
0.960 & 0.52 & 0.02 & 0.33 & 0.03 & 0.37 & 0.02 & 0.76 & 0.04 & 0.52 & 0.04 & 0.36 & 0.03 & 0.80 & 0.06 & 1.13 & 0.12 & 0.25 & 0.06 & 3.5 & 121 & 39\% \\
0.965 & 0.54 & 0.02 & 0.33 & 0.03 & 0.38 & 0.03 & 0.76 & 0.04 & 0.49 & 0.04 & 0.34 & 0.03 & 0.88 & 0.07 & 1.02 & 0.12 & 0.28 & 0.04 & 3.5 & 113  & 36\% \\
0.970 & 0.59 & 0.02 & 0.32 & 0.02 & 0.34 & 0.03 & 0.82 & 0.04 & 0.56 & 0.04 & 0.33 & 0.03 & 0.81 & 0.05 & 1.16 & 0.11 & 0.22 & 0.05 & 3.5 & 101 & 35\% \\
0.975 & 0.57 & 0.02 & 0.36 & 0.03 & 0.35 & 0.03 & 0.85 & 0.04 & 0.54 & 0.04 & 0.31 & 0.02 & 0.79 & 0.05 & 1.19 & 0.11 & 0.22 & 0.05 & 3.5 & 98 & 30\% \\
0.980 & 0.62 & 0.02 & 0.38 & 0.03 & 0.28 & 0.03 & 0.93 & 0.05 & 0.57 & 0.05 & 0.27 & 0.03 & 0.85 & 0.09 & 1.08 & 0.15 & 0.29 & 0.06 & 3.5 & 88 & 27\% \\ 
0.985 & 0.65 & 0.02 & 0.38 & 0.03 & 0.25 & 0.02 & 0.94 & 0.04 & 0.56 & 0.04 & 0.25 & 0.04 & 0.96 & 0.08 & 0.95 & 0.12 & 0.24 & 0.06  & 3.7 & 78 & 23\% \\
0.990 & 0.66 & 0.02 & 0.44 & 0.02 & 0.20 & 0.02 & 1.00 & 0.04 & 0.58 & 0.05 & 0.23 & 0.03 & 0.92 & 0.07 & 0.99 & 0.12 & 0.21 & 0.04 & 4.2 & 66 & 18\% \\
0.995 & 0.72 & 0.03 & 0.37 & 0.03 & 0.23 & 0.03 & 1.02 & 0.05 & 0.57 & 0.05 & 0.25 & 0.03 & 0.89 & 0.06 & 1.09 & 0.10 & 0.19 & 0.03 & 4.5 & 54 & 11\% \\
0.999 & 0.75 & 0.03 & 0.40 & 0.03 & 0.21 & 0.03 & 1.12 & 0.06 & 0.63 & 0.07 & 0.25 & 0.04 & 0.94 & 0.09 & 0.88 & 0.14 & 0.26 & 0.05 & 5.3 & 39 & 3\% \\
 \hline
\end{tabular}
 \end{table*} 

\subsubsection{Cluster Catalog Contamination}

With the information from MCMF, it is straightforward to characterize the contamination due to chance superposition of a non-cluster X-ray source with an optical cluster counterpart.  As described in Section~\ref{sec:random_superposition}, we use followup of a large catalog of random positions to characterize the probability \ProbLam\ of obtaining a particular richness \LamMCMF\ as a function of redshift for our sample.  The process of following up a candidate catalog and imposing a threshold value \ProbCut\ in \ProbLam\ of \ProbCut$=0.97$, for example, then implies that 3~percent of the candidate sources that are not clusters will be matched to what are in actuality random superpositions.  Thus, the final contamination fraction $f_\mathrm{c}$ of the resulting MCMF cluster catalog depends on the threshold \ProbCut\ and the contamination in the original candidate catalog.
\begin{equation}
f_\mathrm{c}={N_\mathrm{cont}\over N_\mathrm{cat}}={(1-P_\mathrm{cut})\over P_\mathrm{cut}} {(N_\mathrm{cand}-N_\mathrm{cat})\over N_\mathrm{cat}}
\label{eq:contamination}
\end{equation}
where the input candidate catalog contains $N_\mathrm{cand}$ members of which $N_\mathrm{cont}$ are non-cluster, and the final cluster catalog has $N_\mathrm{cat}$ members (including any contamination that has slipped through).

The 2RXS candidate catalog is highly contaminated, with only a small fraction of sources corresponding to real clusters.  Approximately 22~percent of the sources are spurious, and roughly 90~percent of the real sources are either AGN or stars.  Nevertheless, with an appropriate threshold in \ProbLam\ it is possible to produce cluster catalogs with very low contamination.  As an example, if we adopt \ProbCut$>0.999$ applied to the 1241 2RXS sources overlapping DES-SV, we find 39 clusters.  Given these numbers, equation~(\ref{eq:contamination}) indicated that we would expect a contamination fraction $f_\mathrm{c}=3$~percent or that $\sim1.2$ of those 39 clusters in the output catalog are random superpositions.
In Table~\ref{tab:scaling} we list the expected contamination for the different selections used to investigate the dependency of the scaling relation on those cuts.

For eROSITA we expect the situation to be much better, indeed.  The PSF size for the eROSITA survey is expected to be $\sim25"$~half energy width, which is comparable to the PSF within the inner ring of   ROSAT PSPC pointed observations.  \citet{vikhlinin98} used the extent of X-ray sources in his serendipitous survey to identify clusters, demonstrating a contamination of between $\sim2$ and 10~percent (depending on the flux limit) in a sample of somewhat more than 200 clusters.  The application of MCMF to a input candidate catalog with 10~percent contamination would result in a final cluster catalog contamination fraction that is one to two orders of magnitude below the contaminations listed in Table~\ref{tab:scaling}.

\subsubsection{Centering}

For each cluster candidate we estimate a center based on the galaxy density maps of RS galaxies at the cluster redshift. The maps are created using Voronoi Tessellation and smoothed with a 250\,kpc kernel. The center position is the average between the position of the nearest density peak and the barycenter of the same peak, found by {\tt SExtractor}.

\begin{figure}
\includegraphics[keepaspectratio=true,width=\columnwidth]{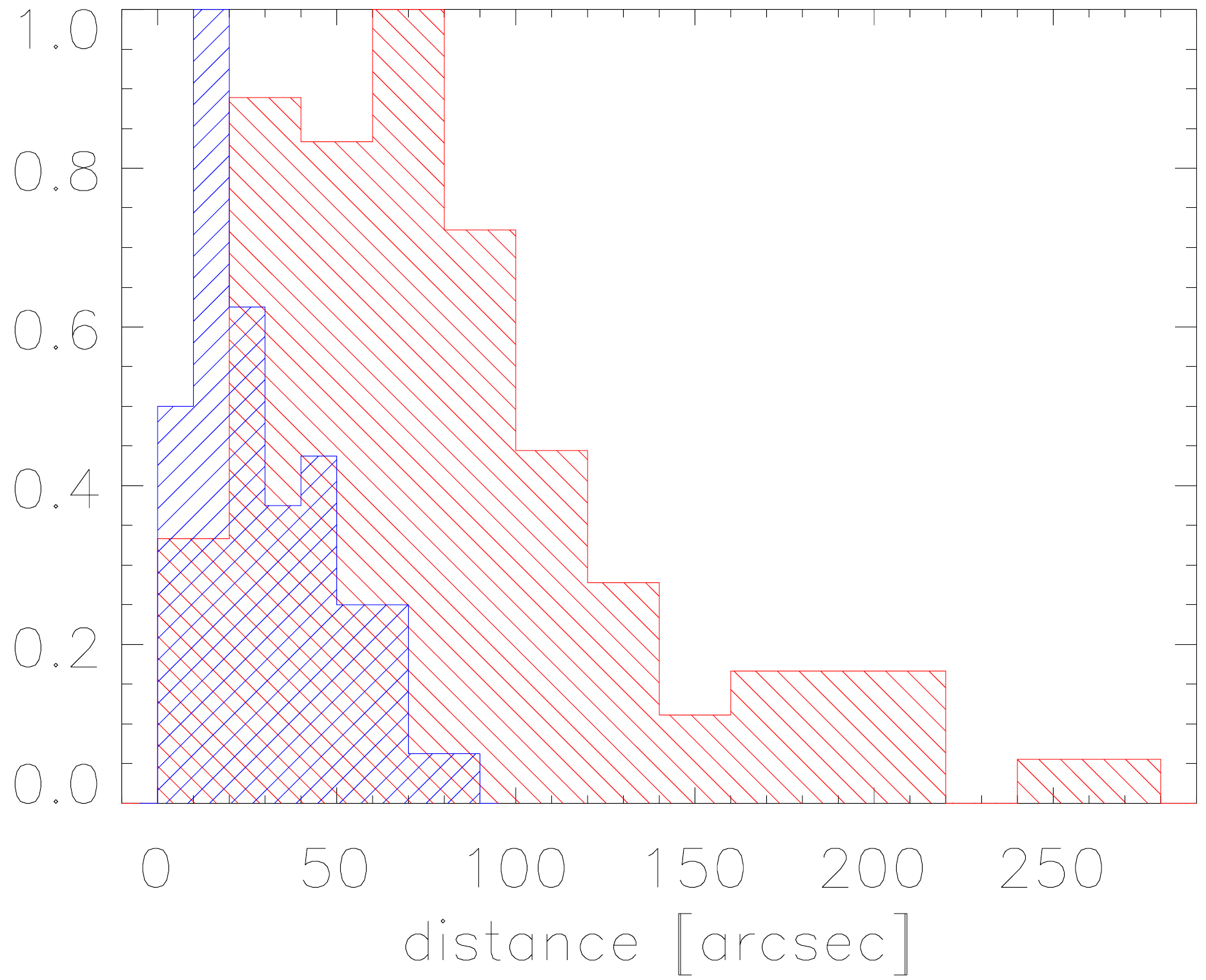}
\includegraphics[keepaspectratio=true,width=\columnwidth]{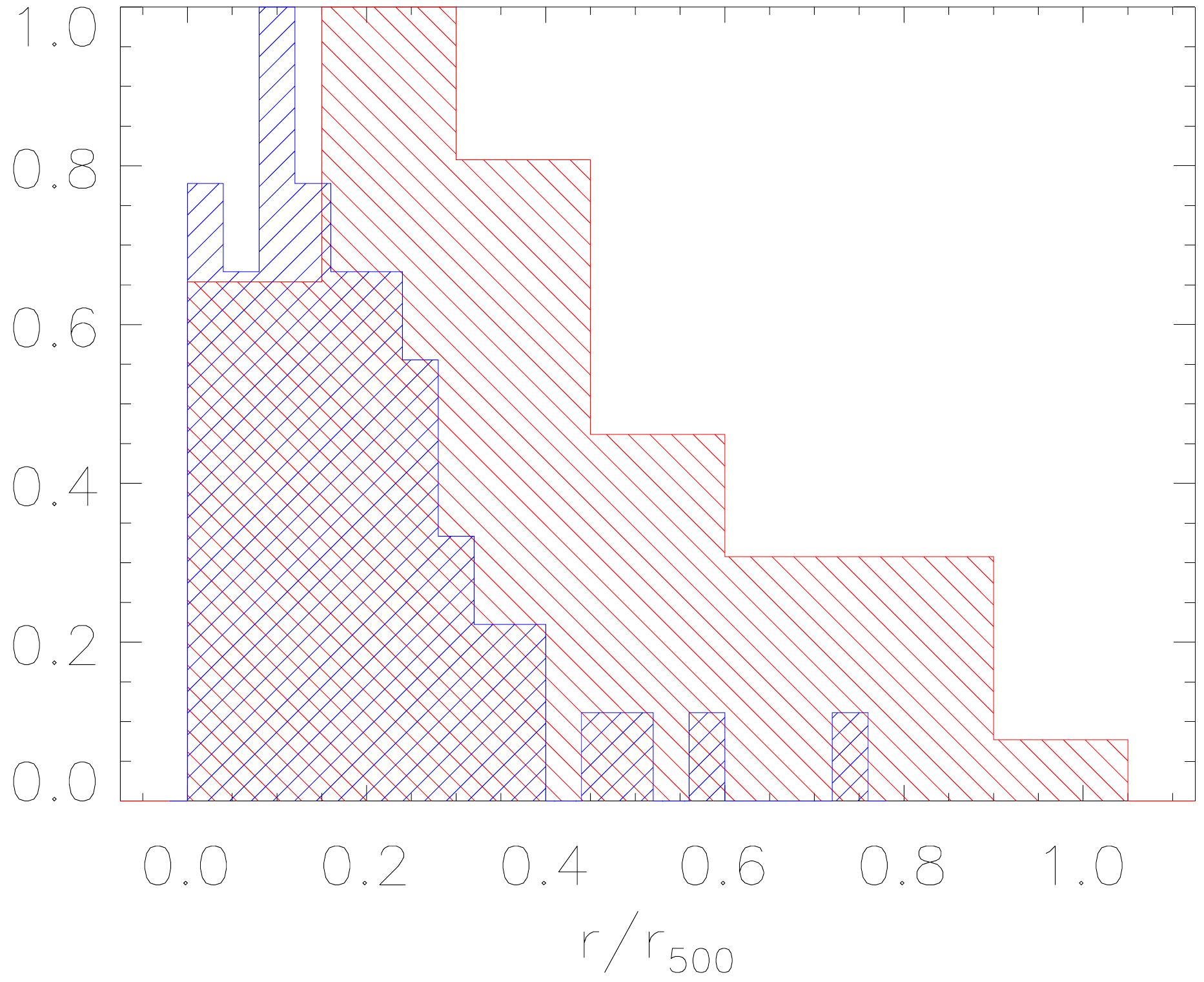}
\vskip-0.1in
\caption{Distribution of offsets between X-ray and galaxy density center (red) for the \ProbCut\ >0.98 sample and between galaxy density center and SPT center (blue) for the SPT cluster sample. The peak of the distributions are normalized to one.}
\label{fig:centering}
\vskip-0.15in
\end{figure}

The positional accuracy for point sources is measured for the RASS to be of 0.3$'$ \citep{voges00}. Despite the fact that most of the RASS clusters appear to be point like due to the large survey PSF of the RASS, the true surface brightness distribution is extended and can be of complex shape. Fig.~\ref{fig:centering} shows the distribution of offsets between galaxy based and X-ray based centers for the RASS cluster sample as well as the offset distribution between Sunyaev-Zel'dovich effect \citep[SZE;][]{sunyaev70} based and galaxy density based centers for the SPT cluster sample. The distribution of RASS center offsets is significantly broader than that of the SPT sample or that expected for RASS point sources.

%


\subsection{Comparison to Other Cluster Catalogs}
\label{sec:comparison}

The DES-SV field used for this work is overlapped by the SPT survey area and the MCXC cluster catalog.  Moreover, these DES-SV data have been used to construct a RedMaPPer cluster catalog.  In this section we compare our RASS+DES-SV catalog to these other catalogs. 

\subsubsection{SPT-SZ Clusters}

We find 56 SZE selected clusters from SPT in the investigated area, and 22 of these clusters have a RASS detection within 4$'$. These numbers will significantly increase once we extend our analysis to the full 2,500\,deg$^2$ SPT-SZ survey region.  Eight clusters fall below our selection criteria of \ProbLam$>0.98$ and \ProbS$>0.98$, but only four have \ProbLam$<0.97$.  All four of these SPT clusters have redshifts  $z>0.55$ and all except one have offsets from the corresponding 2RXS source larger than 2$'$.  Of these four systems, all but one is well detected in our optical cluster finder, but the \LamMCMF\ falls in a range where there is a greater than 3\,percent chance of a random superposition. The second lowest significance cluster out of these four clusters has an X-ray to SZE center offset of 1.3$'$, an SZE based mass estimate of M$_{500}=3.76\times10^{14}$ M$_\odot$ and a photo-z estimate of z$=0.76\pm0.03$.  Our results for this cluster include a photo-z $z=0.706$ and an X-ray based mass of M$_{500}=4.0\times10^{14}$\,M$_\odot$.  Thus, we find consistent cluster parameters, although this falls near the lower limit for optical confirmation at that redshift.

Only one cluster match, SPT-CLJ0432-6150, falls significantly below the standard selection criteria and shows a small X-ray to SZE positional offset of 0.5$'$. The signal to noise of the optical counterpart is $S=2.4$, and it exhibits significances of \ProbLam$=0.64$ and \ProbS$=0.65$ at a redshift of $z=0.63$. The SPT based mass estimate is $M_{500}=(2.39\pm0.65)\times10^{14}\mathrm{M}_{\odot}$, and the photometric redshift is given as z=$0.98\pm0.07$. Using the RASS X-ray count rate and the redshift from SPT, we obtain an X-ray based mass of $M_{500}=3.5\times10^{14}\mathrm{M}_{\odot}$, which is consistent with the SZE based mass estimate. Based on the flux limit we see in Fig.~\ref{fig:limflux} and using the \citet{mantz10b} scaling relation we expect a mass limit at z=0.98 of about M$_{500}=5.2\times10^{14}\mathrm{M}_{\odot}$, which is clearly above the mass estimates from X-ray and SZE.  So given the low significances of the $z=0.63$ counterpart, if we had  sufficiently deep optical data at this location, we would have likely identified the higher redshift counterpart and rejected the counterpart at $z=0.63$.

To further clarify whether the cluster is a misidentified lower redshift cluster or a missed high-z cluster, we utilize the galaxy density map automatically created for each peak in \LamMCMF\ and overlaid on the $grz$ pseudo-color image. Because the photometric redshift suggested in \citet{bleem15} would place the optical counterpart near our adopted depth limits, we relax our standard depth setting, effectively changing our magnitude limit from $i=21.8$ to $i=22.1$. 

\begin{figure}
\centering
\includegraphics[keepaspectratio=true,width=\columnwidth]{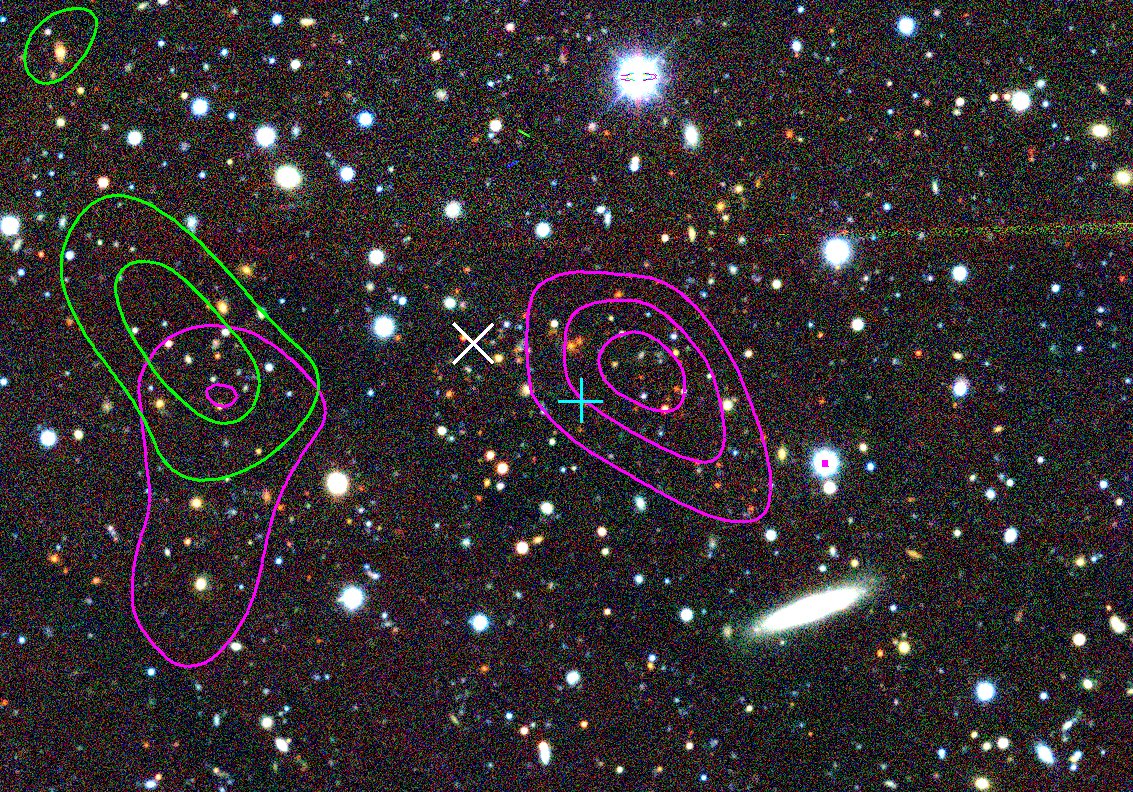}
\includegraphics[keepaspectratio=true,width=\columnwidth]{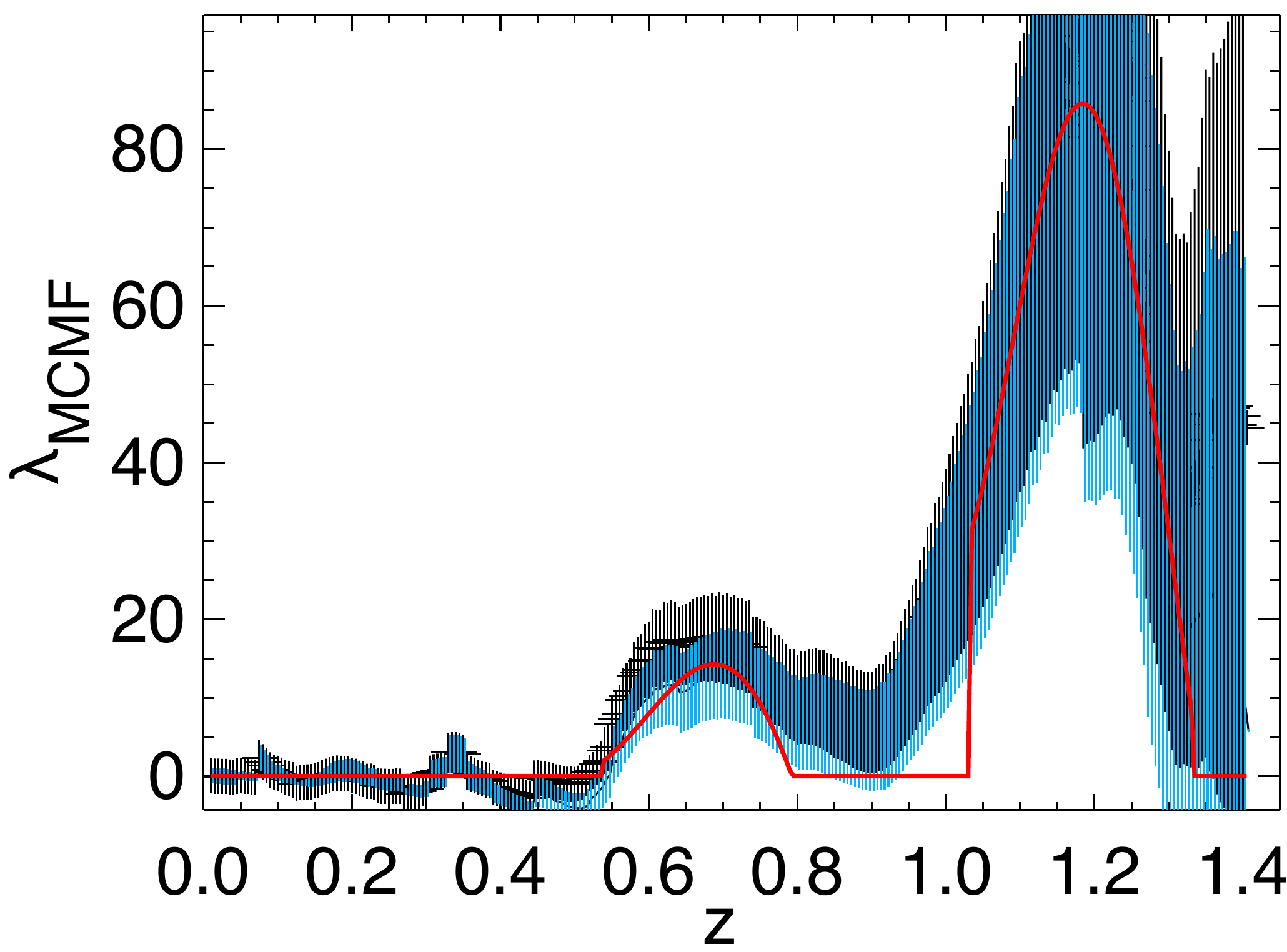}
\vskip-0.10in
\caption{Top panel: g, r, z pseudo-color image of the central 5x3.5 arcmin region around the RASS detection close to cluster SPT-CLJ0432-6150, not identified with our standard settings. Cyan cross: RASS position, White cross: SPT position. Green contours: Galaxy density for z=0.63 galaxies. Magenta contours: Galaxy density for z=1.09 galaxies.  Lower panel: Richness vs redshift plot, with modifications to explore higher redshifts.}
\label{fig:highZspt}
\vskip-0.15in
 \end{figure}

The \LamMCMF\ versus redshift plot using this high-z extension can be seen in Fig.~\ref{fig:highZspt}. As one can see, the modified settings do indeed show a cluster candidate at a redshift of $z\sim1.09$ and a \LamMCMF\ of $86\pm40$.  For this redshift we obtain an X-ray based mass of $5.1\times10^{14}\mathrm{M}_{\odot}$, which is nearer to the flux limit induced mass of $5.7\times10^{14}\mathrm{M}_{\odot}$. However the \LamMCMF\ supports a lower mass estimate of $3.2\times10^{14}\mathrm{M}_{\odot}$, closer to the SZE based mass estimate.
The galaxy density contours of the peak at z=1.09 are in agreement with the X-ray and SZE positions, supporting the high redshift cluster and disfavoring the z=0.63 structure. The differences between X-ray, SZE and \LamMCMF\ based masses suggest that the X-ray flux is likely boosted by other sources, such as AGN or the observed structure at z=0.63. Our photometric redshift seems to be consistent with that presented in \citet{bleem15} within 2$\sigma$, while suggesting a slightly higher redshift for that cluster.  This example illustrates the importance of having a strong positional prior (and to a lesser extent mass prior) when identifying optical counterparts of X-ray or SZE selected clusters.

As a further consistency check, we search for the highest mass RASS clusters missing in the SPT cluster sample. The cluster with the highest X-ray mass that does not have an SPT counterpart is at $z=0.76$ and has a mass of M$_{500}=4.6\times10^{14}\mathrm{M}_{\odot}$ and a \LamMCMF\ based mass estimate of M$_{500}=4.1\times10^{14}\mathrm{M}_{\odot}$. This makes it also the second richest cluster that is missing. The richest cluster missing has a \LamMCMF\ of 136, corresponding to mass of  M$_{500}=5.8\times10^{14}\mathrm{M}_{\odot}$ using our best fit scaling relation. The X-ray based mass estimate is M$_{500}=2.75\times10^{14}\mathrm{M}_{\odot}$, suggesting a significantly lower mass.  Judging from Fig.~7 of \citep{bleem15}, the cluster masses are in the range where the SPT cluster sample is only $\sim$50\,percent complete. Finding a missing cluster is therefore not in contradiction with the expectations.  The larger sample enabled by the extension of this work to the full survey will enable a detailed quantitative consistency test of the sample as applied in \citet{saro15}.
 
\subsubsection{MCXC Clusters}
 
The MCXC \citep{piffaretti11} catalog is a meta-catalog of X-ray detected clusters of galaxies and combines various publicly available RASS based X-ray catalogs, such as NORAS \citep{2000ApJS..129..435B}, REFLEX\citep{boehringer04} and MACS \citep{ebeling01}.  We find seven matches between 2RXS and MCXC in our footprint, using a cross identification radius of 3$'$. The largest offset between 2RXS and MCXC matches is 1.5$'$, and the mean offset is 0.75$'$. We find for all except one of these matches significance values of $S>3$, \ProbLam$>0.78$ and \ProbS$>0.86$.

The only match with $S<3$ has significance values of $S=1.7$, \ProbLam$=0.35$ and \ProbS$=0.32$. This is well below the threshold required to rule out a chance superposition or to consider it even a statistically significant detection. The MCXC catalog lists a mass of $M_{500}=7\times 10^{13} M_\odot$ and a redshift of $z=0.33$. A cluster of that mass would have been detected at that location with $S>5$.  Optical investigation at the MCXC location does not show an obvious cluster counterpart.  No additional X-ray data besides those from ROSAT were found for that location. This cluster was originally published in the southern SHARC catalog \citep{Burke03}, and is the only one in that catalog with a quality flag of three.  We therefore conclude that this MCXC cluster is not a real cluster.

\begin{figure}
\includegraphics[keepaspectratio=true,width=0.98\columnwidth]{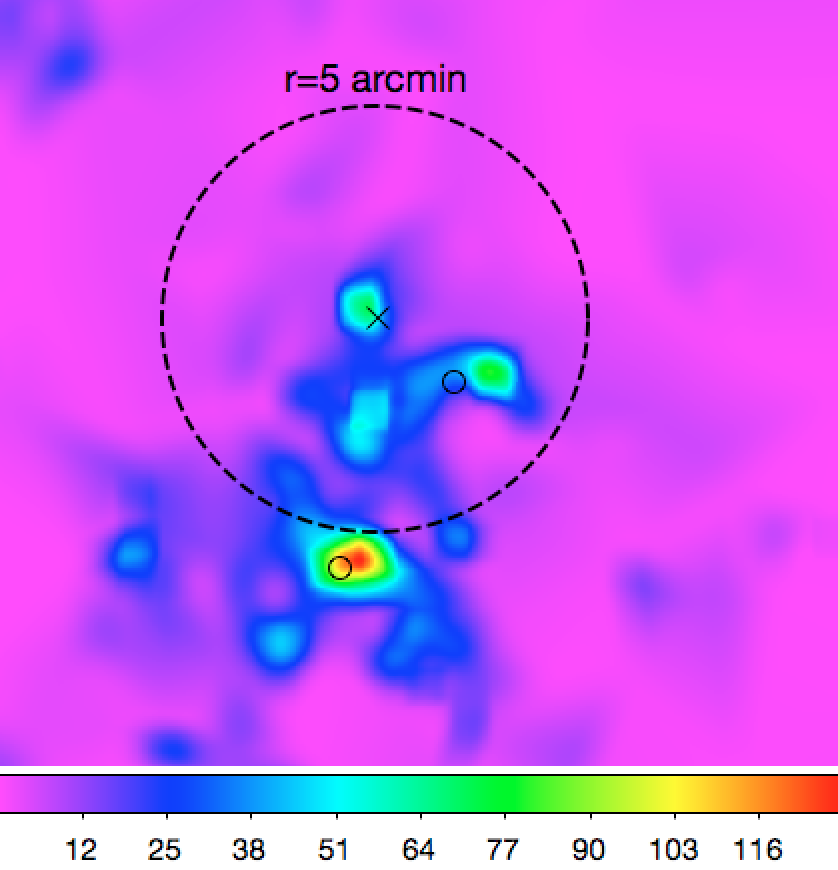}
\includegraphics[keepaspectratio=true,width=0.98\columnwidth]{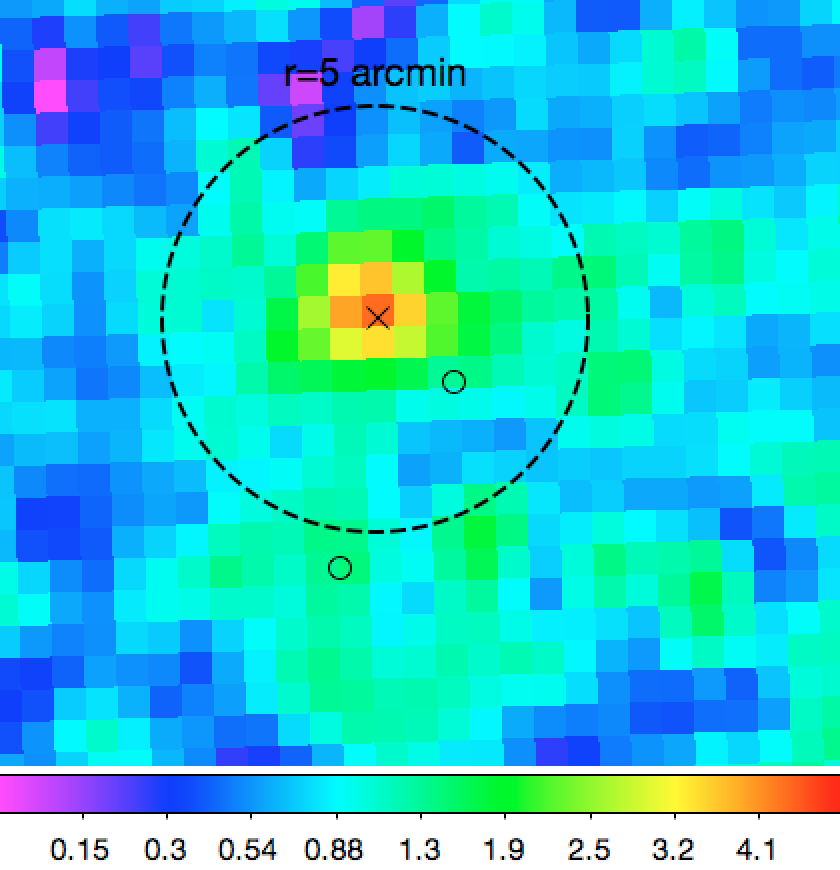}
\caption{Region around 2RXC~J0434.6-4726. Top: Galaxy density map of red sequence galaxies at z=0.31. Bottom: Smoothed RASS photon count image in the energy range 0.1-2.4 keV of the same region. The black cross makes the 2RXS position, the dashed circle marks the extraction region of the 2RXS measurment of 5 arcmin, corresponding to 1.4 Mpc at z=0.31. The small black circle close to the main galaxy density peak, marks the position of the RedMaPPer cluster of $\lambda_\mathrm{RM}=75$ and $z_\mathrm{RM}=0.316$ discussed in Sec.~\ref{susec:redmcomp}. The small black circle within 5 arcmin from the 2RXS source corresponds to a  RedMaPPer cluster of $\lambda_\mathrm{RM}=11$ and $z_\mathrm{RM}=0.29$.}
\label{fig:missingRedmap}
\end{figure}

\begin{figure}
\includegraphics[keepaspectratio=true,width=\columnwidth]{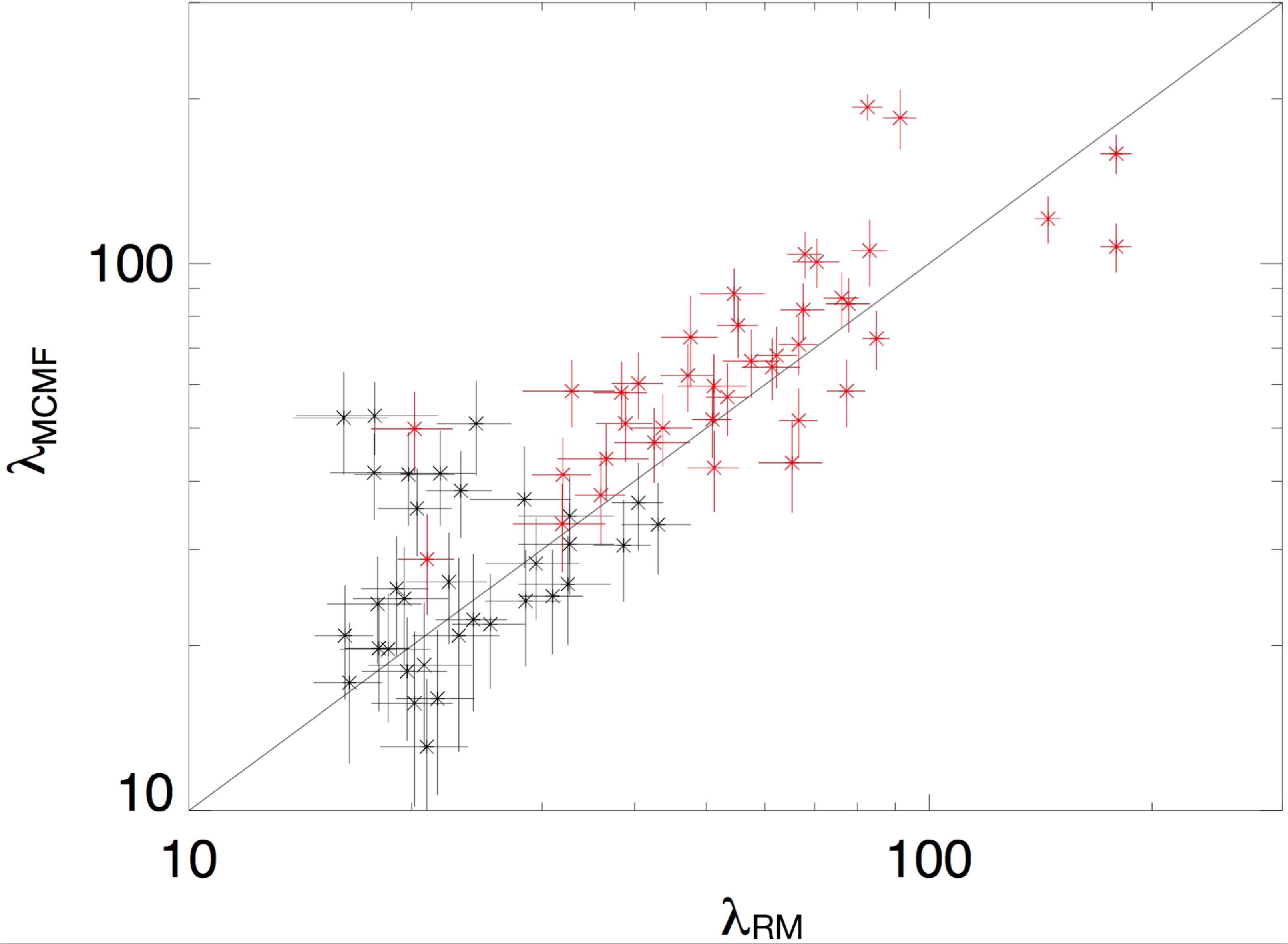}
\vskip-0.1in
\caption{Richness comparison for matches between the RASS+DES-SV and RedMaPPer catalogs. Highlighted in red are clusters with $\lambda >20$ and \ProbCut >0.98. The black line shows $\lambda_\mathrm{MCMF}=\lambda_\mathrm{RM}$.}
\label{fig:compRedmap}
\end{figure}

\subsubsection{RedMaPPer Clusters}  \label{susec:redmcomp}
 
 A natural source for comparison is the RedMaPPer catalog for the same region \citep{rykoff16}.  Although there are similarities between MCMF and RedMaPPer in the radial and color filter, there are a few things that may cause differences in the output catalogs. First, RedMaPPer is a cluster finder, while our code is tuned for cluster confirmation around X-ray positions.  Also, in MCMF there is no recentering to the optical counterpart position.  Our aperture for counting galaxies is defined by the mass implied by the X-ray luminosity. Further, we use independently derived RS models and data reduction pipelines.  In the remainder of the section we focus in turn on (1) high significance 2RXS clusters from our catalog that are missing in the RedMaPPer catalog, (2) high $\lambda_\mathrm{RM}$ RedMaPPer systems missing in our catalog, and then (3) 2RXS clusters from our catalog with RedMaPPer catalog counterparts.
  
First, we search for 2RXS cluster candidates with \ProbLam$>0.99$ (there are 60 of these) that have no RedMaPPer cluster candidate within a distance of 2.5$'$. We find 14 such clusters, four of which lie outside the footprint of the RedMaPPer cluster catalog. In three cases, the 2RXS position has a offset of $>150$ arcsec from the main galaxy density peak, which itself is consistent with a RedMaPPer cluster.  In one case, the BCG was not correctly identified by RedMaPPer. Three clusters fall close to stars, where different masking strategies may have led to the non detections in the RedMaPPer catalog. One cluster, 2RXC~J0426.5-6003 (z=0.06), falls below the lower redshift limit of z=0.1, applied to the RedMaPPer catalog. The neighbouring cluster 2RXC~J0428.4-5349 at z=0.26 is not in the RedMaPPer catalog either. Both clusters are part of the region with three distinct redshift peaks shown in Fig.~\ref{fig:triple} and might have been excluded by RedMaPPer due to masking of bright cluster members in the low-z clusters. The last missing cluster 2RXC~J0536.2-5847 was excluded from the RedMapper catalog because a large fraction of the cluster region was masked. We note that in our dataset the cluster region does not suffer from significant masking.

Second, we search for RedMaPPer clusters with $\lambda_\mathrm{RM}>60$ that do not have a 2RXS cluster candidate having \ProbLam$>0.99$ within 4$'$. This is motivated by the observed offset distribution in Fig.~\ref{fig:centering} and the limits seen in Fig.~\ref{fig:limflux}.  We find 19 RedMaPPer clusters with a median redshift of z=0.59. This high median redshift already suggests that those system may be too high in redshift and to low in mass to be detected in 2RXS. The cluster with the highest richness missed by 2RXS has $\lambda_\mathrm{RM}=107$, a redshift $z_\mathrm{RM}=0.76$ and is the fifth highest redshift cluster without a 2RXS counterpart.  A cluster of this richness and redshift is close to the detection limit of 2RXS, as can be seen in Fig.~\ref{fig:limflux}.  This is true even given the small systematic offset between $\lambda_\mathrm{RM}$ and  $\lambda_\mathrm{MCMF}$ discussed in the following paragraph and given the intrinsic scatter between luminosity and richness. The same arguments hold for all clusters down to redshift $~0.4$, noting that the second richest missing cluster has only a richness of $\lambda_\mathrm{RM}=87$ at $z_\mathrm{RM}=0.81$. Concentrating on the low redshift range, we find five missing clusters with $0.26<z_\mathrm{RM}<0.43$ with a range in richness of $60<\lambda_\mathrm{RM}<77$. Three of them are in regions with RASS exposure times below half of the median exposure time. Another cluster lies in a region with 80\% of the median exposure time. The remaining cluster that is not in a particularly low exposure time region has $\lambda_\mathrm{RM}=75$ and $z_\mathrm{RM}=0.316$. The nearest 2RXS source is 2RXC~J0434.6-4726 that is $6'$ away from the RedMaPPer position. MCMF identifies this 2RXS candidate as a cluster with redshift of $z_\mathrm{MCMF}=0.310$ and richness $\lambda_\mathrm{MCMF}=68$ but assigns the optical center to a local over density of galaxies likely associated with the main cluster.  In the current incarnation, MCMF simply searches for the nearest peak in the RS weighted galaxy density map. It does not perform any likelihood analysis to identify the most likely galaxy density peak associated with the cluster candidate given richness or luminosity.  Otherwise, the code would have shown a large X-ray to optical offset, which would have indicated a likely correlated X-ray point source. In total, we do not see strong evidence for missing RedMaPPer clusters in 2RXS that should have been detected, but the large offset of 2RXC~J0434.6-4726 is a failure mode that underscores the current limitations of MCMF.

Finally, we match the catalogs by searching for the nearest RedMaPPer counterpart that lies within 2.5$'$ of each of our clusters.  Requiring $\lambda_\mathrm{RM}>15$ and $S>2.5$, we find 73 matches.  Restricting the catalog to $\lambda_\mathrm{RM}>20$ as suggested by \citet{rykoff16}, we find 60 matches.  Note that the RedMaPPer catalog only exists for $\delta>-61^\circ$.  The corresponding redshift scatter is $\sigma_{\Delta z/(1+z)}=0.011$ for both of these catalogs.  Given that the same raw data are used in both cases, this scatter reflects the differences in the algorithms, indicating that there is good agreement between the two codes.

Fig.~\ref{fig:compRedmap} contains a comparison of the richnesses we measure \LamMCMF\ versus the RedMaPPer richnesses \LamRM.  Although we developed our code to enable a precise and quantitative selection of the optical counterpart with different radial and color weighting than the RedMaPPer algorithm, the \LamMCMF's we measure are clearly well correlated to \LamRM.  The upcoming larger RASS selected sample from the full DES dataset will enable a more extensive comparison.

Considering all matches with $\lambda_\mathrm{RM}>20$ and $|z_\mathrm{MCMF}-z_\mathrm{RM}|/(1+z_\mathrm{RM})<0.04$, we find a median offset between RedMaPPer and 2RXS positions of 1.65$'$. For the same sample we find an offset between our optical centers to RedMaPPer centers of about 0.33$'$ and between our centers and 2RXS of 1.15$'$.  If we restrict the comparison to the \ProbLam\ and \ProbS$>0.98$ sample, we find 41 matches. Here the median offset between RedMaPPer and 2RXS reduces to 1.1$'$ and is only marginally larger than the median offset between 2RXS and our centers of 1$'$.  The median offset between RedMaPPer and our centers stay at 0.33$'$ for this sub-sample.  The lower average offset of our centers to 2RXS can be explained by two effects. First our method simply searches for peaks towards the 2RXS sources, while the RedMaPPer centers are found independently.  A second argument is the potential problem of the RedMaPPer algorithm to find the BCG candidate for very low redshift systems where saturation and blending effects may play an important role. Our algorithm uses smoothed density maps, which allows us to recover the correct cluster center even when the brightest cluster members are saturated or blended.

In summary, we see very good agreement between the characteristics of the matched sample of RedMaPPer clusters and our own, three examples of 2RXS systems not making it into the RedMaPPer catalog, and a case of a 2RXS system that could be an X-ray point source that has been associated with a clump of RS galaxies that lie in the outskirts of a larger optical system identified by RedMaPPer.  The analysis of the 2RXS cluster sample extracted from the full DES region will allow more precise cross checks with RedMaPPer.


\section{Conclusions}
\label{sec:conclusions}

In this work we present our multi-component matched filter cluster confirmation method MCMF and apply it to the 2RXS X-ray source catalog \citep{boller16} using the DES-SV optical dataset.  We identify optical counterparts using the overdensity of galaxies having colors consistent with the RS, extracting richnesses \LamMCMF\ and photometric redshifts, and then quantifying the probability that the identified counterpart is a random superposition of an unassociated optical system.  We present a catalog of 88 RASS selected, MCMF confirmed clusters that cover a redshift range from $0.05<z<0.8$ and a mass range of $2\times10^{13}$ to $2\times10^{15}\,\mathrm{M}_{\odot}$.  With our MCMF method we follow up sources that are as much as ten times fainter than the typical sources previously used to identify RASS galaxy cluster samples. When restricting to more conservative cuts of \ProbCut\ >0.999, we find 39 clusters with an expected contamination of 3\%, which is about ten times the number of identified REFLEX clusters within the same footprint. The contamination of the cluster catalog by random superpositions can be directly derived from the catalog, given the adopted \ProbCut, which enables one to use MCMF to create cluster catalogs with the desired contamination.

In addition to following up RASS selected sources, we follow up SPT selected clusters, creating mock X-ray count rates using the SZE based masses presented for these clusters \citep{bocquet15,bleem15}.  This enables a test of the photometric redshifts from our method using a sample of 29 spectroscopically confirmed SPT selected clusters.
Photometric redshifts for our cluster candidates reach a characteristic accuracy of $<\sigma_z/(1+z)> \approx 0.010$.  This performance is comparable to that of other cluster finding codes such as RedMaPPer.  Moreover, the \LamMCMF--mass distribution of SPT clusters nicely follows the behavior of the mostly lower mass systems identified using our MCMF confirmed RASS selected clusters.

We compare our MCMF confirmed 2RXS cluster sample to several other existing catalogs over the same portion of the sky.  These include SPT, MCXC and RedMaPPer.  We could find no clear evidence of SPT clusters missing in the 2RXS sample or high mass 2RXS clusters missing in the SPT sample.  All MCXC systems were confirmed, save for one system that was flagged as a problematic cluster candidate in the original SHARC survey.  A direct comparison between the RedMaPPer sample and ours enables another test.  We find a single 2RXS+MCMF system that appears to be an X-ray point source associated with RS galaxies in the outskirts of a larger optical system identified by RedMaPPer.  The cross-matched 2RXS+MCMF and RedMaPPer catalog allow further tests of  our photometric redshifts.  The two sets of redshifts are in good agreement, exhibiting scatter similar in scale to that seen when we compare to spectroscopic redshifts.  Because in this case the same raw optical data are used for both catalogs, this good agreement is a test only of the differences between the methods and therefore measures only one component of the photometric redshift error.  Finally, there is a strong correlation between the RedMaPPer richnesses \LamRM\ and \LamMCMF.


Using the best fit luminosity--\LamMCMF\ scaling relation, we show that the cluster sample is primarily RASS limited out to a redshift $z\sim0.9$ within the DES-SV dataset.  Given that DES-SV does not reach the full expected DES depth, this implies that a complete RASS cluster confirmation of the DES area does not require the final DES dataset.  Using the average source density of RASS and our cuts for optical counterpart significance, we estimate that we will obtain a cluster sample of 1500 to 2500 clusters over the full DES area.  The sample would be one of the largest homogeneously X-ray selected cluster samples in the southern hemisphere and likely remain so until the forthcoming launch of eROSITA.  A cosmological analysis of our MCMF sample would be improved by a re-extraction of X-ray properties and improved calibration of the relevant mass-observable relations. 


\section*{Acknowledgements}

We thank Th. Boller, M. Freyberg and H. Brunner from the MPE high energy group for helpful conversations.  We acknowledge the support of the Max Planck Gemeinschaft Faculty Fellowship program and the High Energy Group at MPE.  Further, we acknowledge the support of the DFG Cluster of Excellence ``Origin and Structure of the Universe'', the Transregio program TR33 ``The Dark Universe'', the Ludwig-Maximilians-Universit\"at and the DLR supported Euclid development project. The data processing has been carried out on the computing facilities of the Computational Center for Particle and Astrophysics (C2PAP), located at the Leibniz Supercomputer Center (LRZ).  

Funding for the DES Projects has been provided by the U.S. Department of Energy, the U.S. National Science Foundation, the Ministry of Science and Education of Spain, the Science and Technology Facilities Council of the United Kingdom, the Higher Education Funding Council for England, the National Center for Supercomputing  Applications at the University of Illinois at Urbana-Champaign, the Kavli Institute of Cosmological Physics at the University of Chicago, the Center for Cosmology and Astro-Particle Physics at the Ohio State University, the Mitchell Institute for Fundamental Physics and Astronomy at Texas A\&M University, Financiadora de Estudos e Projetos, Funda{\c c}{\~a}o Carlos Chagas Filho de Amparo {\`a} Pesquisa do Estado do Rio de Janeiro, Conselho Nacional de Desenvolvimento Cient{\'i}fico e Tecnol{\'o}gico and the Minist{\'e}rio da Ci{\^e}ncia, Tecnologia e Inova{\c c}{\~a}o, the Deutsche Forschungsgemeinschaft and the Collaborating Institutions in the Dark Energy Survey.   The Collaborating Institutions are Argonne National Laboratory, the University of California at Santa Cruz, the University of Cambridge, Centro de Investigaciones Energ{\'e}ticas, Medioambientales y Tecnol{\'o}gicas-Madrid, the University of Chicago, University College London, the DES-Brazil Consortium, the University of Edinburgh, the Eidgen{\"o}ssische Technische Hochschule (ETH) Z{\"u}rich, Fermi National Accelerator Laboratory, the University of Illinois at Urbana-Champaign, the Institut de Ci{\`e}ncies de l'Espai (IEEC/CSIC), the Institut de F{\'i}sica d'Altes Energies, Lawrence Berkeley National Laboratory, the Ludwig-Maximilians Universit{\"a}t M{\"u}nchen and the associated Excellence Cluster Universe, the University of Michigan, the National Optical Astronomy Observatory, the University of Nottingham, The Ohio State University, the University of Pennsylvania, the University of Portsmouth, SLAC National Accelerator Laboratory, Stanford University, the University of Sussex, Texas A\&M University, and the OzDES Membership Consortium.  The DES data management system is supported by the National Science Foundation under Grant Number AST-1138766. The DES participants from Spanish institutions are partially supported by MINECO under grants AYA2012-39559, ESP2013-48274, FPA2013-47986, and Centro de Excelencia Severo Ochoa SEV-2012-0234.  Research leading to these results has received funding from the European Research Council under the European Union's Seventh Framework Programme (FP7/2007-2013) including ERC grant agreements  240672, 291329, and 306478.

\section*{Affiliations}
$^{1}$ Faculty of Physics, Ludwig-Maximilians-Universit\"at, Scheinerstr. 1, 81679 Munich, Germany\\
$^{2}$ Max Planck Institute for Extraterrestrial Physics, Giessenbachstrasse, 85748 Garching, Germany\\
$^{3}$ Excellence Cluster Universe, Boltzmannstr.\ 2, 85748 Garching, Germany\\
$^{4}$ Department of Physics, IIT Hyderabad, Kandi, Telangana 502285, India\\
$^{5}$ Fermi National Accelerator Laboratory, P. O. Box 500, Batavia, IL 60510, USA\\
$^{6}$ CNRS, UMR 7095, Institut d'Astrophysique de Paris, F-75014, Paris, France\\
$^{7}$ Department of Physics \& Astronomy, University College London, Gower Street, London, WC1E 6BT, UK\\
$^{8}$ Sorbonne Universit\'es, UPMC Univ Paris 06, UMR 7095, Institut d'Astrophysique de Paris, F-75014, Paris, France\\
$^{9}$ Laborat\'orio Interinstitucional de e-Astronomia - LIneA, Rua Gal. Jos\'e Cristino 77, Rio de Janeiro, RJ - 20921-400, Brazil\\
$^{10}$ Observat\'orio Nacional, Rua Gal. Jos\'e Cristino 77, Rio de Janeiro, RJ - 20921-400, Brazil\\
$^{11}$ Department of Astronomy, University of Illinois, 1002 W. Green Street, Urbana, IL 61801, USA\\
$^{12}$ National Center for Supercomputing Applications, 1205 West Clark St., Urbana, IL 61801, USA\\
$^{13}$ Kavli Institute for Particle Astrophysics \& Cosmology, P. O. Box 2450, Stanford University, Stanford, CA 94305, USA\\
$^{14}$ Jet Propulsion Laboratory, California Institute of Technology, 4800 Oak Grove Dr., Pasadena, CA 91109, USA\\
$^{15}$ Department of Astronomy, University of Michigan, Ann Arbor, MI 48109, USA\\
$^{16}$ Department of Physics, University of Michigan, Ann Arbor, MI 48109, USA\\
$^{17}$ Kavli Institute for Cosmological Physics, University of Chicago, Chicago, IL 60637, USA\\
$^{18}$ SLAC National Accelerator Laboratory, Menlo Park, CA 94025, USA\\
$^{19}$ Center for Cosmology and Astro-Particle Physics, The Ohio State University, Columbus, OH 43210, USA\\
$^{20}$ Department of Physics, The Ohio State University, Columbus, OH 43210, USA\\
$^{21}$ Astronomy Department, University of Washington, Box 351580, Seattle, WA 98195, USA\\
$^{22}$ Cerro Tololo Inter-American Observatory, National Optical Astronomy Observatory, Casilla 603, La Serena, Chile\\
$^{23}$ Australian Astronomical Observatory, North Ryde, NSW 2113, Australia\\
$^{24}$ Departamento de F\'{\i}sica Matem\'atica,  Instituto de F\'{\i}sica, Universidade de S\~ao Paulo,  CP 66318, CEP 05314-970, S\~ao Paulo, SP,  Brazil\\
$^{25}$ Department of Physics and Astronomy, University of Pennsylvania, Philadelphia, PA 19104, USA\\
$^{26}$ Department of Astrophysical Sciences, Princeton University, Peyton Hall, Princeton, NJ 08544, USA\\
$^{27}$ Instituci\'o Catalana de Recerca i Estudis Avan\c{c}ats, E-08010 Barcelona, Spain\\
$^{28}$ Institut de F\'{\i}sica d'Altes Energies (IFAE), The Barcelona Institute of Science and Technology, Campus UAB, 08193 Bellaterra (Barcelona) Spain\\
$^{29}$ Department of Physics and Astronomy, Pevensey Building, University of Sussex, Brighton, BN1 9QH, UK\\
$^{30}$ Centro de Investigaciones Energ\'eticas, Medioambientales y Tecnol\'ogicas (CIEMAT), Madrid, Spain\\
$^{31}$ Instituto de F\'\i sica, UFRGS, Caixa Postal 15051, Porto Alegre, RS - 91501-970, Brazil\\
$^{32}$ School of Physics and Astronomy, University of Southampton,  Southampton, SO17 1BJ, UK\\
$^{33}$ Universidade Federal do ABC, Centro de Ci\^encias Naturais e Humanas, Av. dos Estados, 5001, Santo Andr\'e, SP, Brazil, 09210-580\\
$^{34}$ Computer Science and Mathematics Division, Oak Ridge National Laboratory, Oak Ridge, TN 37831

\label{lastpage}
\end{document}